\documentclass[fleqn,usenatbib,useAMS]{mnras}
\pdfoutput=1 
\usepackage{graphicx}	% Including figure files
\usepackage{amsmath}	% Advanced maths commands
\usepackage{amssymb}	% Extra maths symbols
\usepackage{multicol}        % Multi-column entries in tables
\usepackage{bm}		% Bold maths symbols, including upright Greek

\usepackage[T1]{fontenc}
\usepackage{ae,aecompl}
\usepackage{newtxtext,newtxmath}

\usepackage[figure,figure*]{hypcap}

\newcommand{\msun}{M$_{\odot}$}
\newcommand{\ccunit}{cm$^{-3}$}

\title[Feedback with Individual Stars]{Simulating an Isolated Dwarf Galaxy with Multi-Channel Feedback and Chemical Yields from Individual Stars}

\author[A. Emerick et. al.]{
Andrew Emerick,$^{1,2}$
Greg L. Bryan,$^{1,3}$
and Mordecai-Mark Mac Low$^{2,1,3}$
\\
$^{1}$Department of Astronomy, Columbia University, New York, NY, 10027, USA\\
$^{2}$Department of Astrophysics, American Museum of Natural History, New York, NY, USA\\
$^{3}$Center for Computational Astrophysics, Flatiron Institute, 162 5th Ave, New York, NY, 10003, U.S.A}

\begin{document}

\label{firstpage}
\maketitle
%\title{Simulating an Isolated Dwarf Galaxy with Energetic Feedback and Chemical Yields from Individual Stars}

%\author{Andrew Emerick}
%\affiliation{Department of Astronomy, Columbia University, New York, NY, 10027, USA}
%\affiliation{Department of Astrophysics, American Museum of Natural History, New York, NY, USA}
%\author{Greg L. Bryan}
%\affiliation{Department of Astronomy, Columbia University, New York, NY, 10027, USA}
%\affiliation{Center for Computational Astrophysics, Flatiron Institute, 162 5th Ave, New York, NY, 10003, U.S.A}
%\author{Mordecai-Mark Mac Low}
%\affiliation{Department of Astrophysics, American Museum of Natural History, New York, NY, USA}
%\affiliation{Department of Astronomy, Columbia University, New York, NY, 10027, USA}

%\affiliation{Institut f{\"u}r Theoretische Astrophysik, Zentrum f{\"u}r Astronomie der Universit{\"a}t Heidelberg, Heidelberg, Germany}

\begin{abstract}
In order to better understand the relationship between feedback and galactic chemical evolution, we have developed a new model for stellar feedback at grid resolutions of only a few parsecs in global disk simulations, using the adaptive mesh refinement hydrodynamics code \textsc{Enzo}. For the first time in galaxy scale simulations, we simulate detailed stellar feedback from individual stars including asymptotic giant branch winds, photoelectric heating, Lyman-Werner radiation, ionizing radiation tracked through an adaptive ray-tracing radiative transfer method, and core collapse and Type Ia supernovae. We furthermore follow the star-by-star chemical yields using tracer fields for 15 metal species: C, N, O, Na, Mg, Si, S, Ca, Mn, Fe, Ni, As, Sr, Y, and Ba. We include the yields ejected in massive stellar winds, but greatly reduce the winds' velocities due to computational constraints. We describe these methods in detail in this work and present the first results from 500~Myr of evolution of an isolated dwarf galaxy with properties similar to a Local Group, low-mass dwarf galaxy. We demonstrate that our physics and feedback model is capable of producing a dwarf galaxy whose evolution is consistent with observations in both the Kennicutt-Schmidt relationship and extended Schmidt relationship. Effective feedback drives outflows with a greater metallicity than the ISM, leading to low metal retention fractions consistent with observations. Finally, we demonstrate that these simulations yield valuable information on the variation in mixing behavior of individual metal species within the multi-phase interstellar medium.
\end{abstract}


\section{Introduction}

Detailed interstellar medium (ISM) and chemical abundance properties of galaxies are sensitive tests of the underlying physical processes that govern galaxy evolution. Examining these in more detail in galaxy scale simulations is an important and exciting new discriminator between models. There is a considerable body of work studying the chemodynamical evolution of galaxies using cosmological hydrodynamics simulations \citep[e.g.][]{OppenheimerDave2008,Wiersma2009,Shen2010,Simpson2013,Snaith2015,OWLS,EAGLE,FIRE}.
% editor asked us to reduce citation count here:{Lia2002,KawataGibson2003,Kobayashi2004,Tornatore2004,Romeo2005,OppenheimerDave2008,Wiersma2009,Shen2010,MUGS2010,ErisSimulation,Simpson2013,Brook2014,Snaith2015,Oppenheimer2016,OWLS,EAGLE,FIRE}.
These simulations, coupled with additional attention to feedback processes, have made remarkable progress in reproducing global galaxy trends such as evolution of the mass-metallicity relationship \citep[e.g.][]{Obreja2014, Ma2016, Dave2017, Torrey2017} and more detailed quantities such as metallicity distribution functions (MDFs) and the evolution of individual species abundances \citep{Marcolini2008,Revaz2009,Sawala2010,RevazJablonka2012,Jeon2017,Hirai2017} . 

However, much of this work has been done with Lagrangian smoothed particle hydrodynamics schemes, with a few recent exceptions  \citep{Few2012,Simpson2013,Few2014,Vorobyov2015,Corlies2018}. In its original form, this scheme does not capture mixing between chemically inhomogeneous particles, as necessary for chemical evolution. Mixing can be modeled with sub-grid models of turbulent metal diffusion \citep[e.g.][]{Shen2010,Shen2013,Brook2014,Su2017a,Escala2018}, but there are many possible models and each is not necessarily applicable in every regime \citep[see ][]{Revaz2016}. While mixing occurs in Eulerian codes even without sub-grid models, numerical diffusion tends to result in over-mixing in simulations without sufficiently high spatial resolution. Molecular diffusion or even turbulent mixing is certainly not resolved in any galaxy-scale simulation with either method, requiring additional sub-grid models; this can be particularly important for understanding the initial pollution of otherwise pristine gas \citep[see ][ and references therein]{PanScannapiecoScalo2013,Sarmento2017}. Moreover, metal mixing efficiencies may vary species-by-species \citep[e.g.][]{Cohen2013, Roederer2014, FrebelNorris2015, Hirai2017, Cote2018, KrumholzTing2018}. Mixing behavior is tied critically to the feedback source (stellar winds, supernovae, and possibly more exotic sources) that inject metals into different phases of 
the ISM with different energies and on different timescales; the observational effect of this is poorly understood, however. The variations in how different methods handle sub-grid metal injection and metal mixing schemes can lead to uncertainties in connecting models to observations and the fundamental physics that drives galaxy evolution.

Increasing physical resolution reduces reliance on sub-grid physics for mixing.  However, at high particle mass resolution ($M \lesssim 10^3$ M$_{\odot}$) standard schemes for modeling stars as simple stellar populations lose validity  \citep[as studied in detail by][]{Revaz2016}. Below 10$^4$ M$_{\odot}$, such schemes do not fully sample the initial mass function (IMF), and cannot be considered average representations of stellar populations. This is acutely problematic at low star formation rate densities with star particle masses comparable to or below the mass of the most massive individual star ($\sim 100$ M$_{\odot}$). At high star formation rate densities, an undersampled IMF in a single low mass star particle can be compensated for by having many adjacent star particles. Various approaches exist to address this issue \citep[e.g.][]{Kobayashi2000,WeidnerKroupa2004,Pflamm-AltenburgKroupa2006,RevazJablonka2012,Kroupa2013,Rosdahl2015,Su2018}, but none are without caveats \citep{Revaz2016}, save for schemes which begin to track the star-by-star information within a given particle by directly sampling the IMF at formation time \citep[e.g.][]{Hu2017}. The most straightforward solution is to remove the single stellar population formalism entirely and simply track stars as individual particles. 
 
We introduce a new method for studying galactic chemical evolution that follows stars as individual star particles implemented in the adaptive mesh refinement code \texttt{Enzo}, designed for high resolution simulations of isolated galaxies. The relative simplicity of idealized, isolated galaxy evolution simulations allows for a focused, first-principles approach to studying multi-channel feedback mechanisms. We follow recent work using low mass dwarf galaxies as laboratories to study in detail how feedback governs galaxy evolution \citep{Forbes2016,Hu2016,Hu2017}.
%For example, \citet{Forbes2016} found that photoelectric heating can be a significant source of feedback in small galaxies. \citet{Hu2016} and \citet{Hu2017} examined the relative importance of various feedback physics in regulating ISM properties of dwarf galaxies. 
Our work builds upon our current understanding of feedback and galactic chemodynamics while making three notable advances: 1) direct star-by-star modeling, 2) stellar winds from both massive and asymptotic giant branch (AGB) stars, and 3) using an adaptive ray tracing method to follow stellar ionizing radiation. We also include core collapse and Type Ia supernova feedback, photoelectric heating from stellar far ultra-violet (FUV) radiation, and Lyman-Werner dissociation from stellar radiation. 

Using star-by-star modeling, we capture in more detail the stellar yields from individual stars released over their lifetime. This includes yields from massive and AGB stellar winds, and supernovae (SNe). In addition to better capturing how individual metal species enrich the ISM, this allows us to chemically tag individual stars. This ability opens an exciting new channel for testing models of galaxy evolution by leveraging current and ongoing observations probing the detailed distributions of chemical abundances of stars in the Milky Way and Local Group, such as APOGEE and APOGEE2 \citep{APOGEE2010,APOGEE}, the Gaia-ESO survey \citep{Gaia}, and GALAH \citep{GALAH}. This paper is the first in a series examining in detail the role that individual components of multi-channel stellar feedback play in galaxy dynamical and chemical evolution. In \cite{Emerick2018b} we investigate the importance of ionizing radiation in regulating star formation and driving outflows in our galaxy. In \cite{Emerick2018c} we explore how individual metal species mix within the ISM and are ejected via galactic winds.

We describe each mode of our multi-channel feedback 
in detail in Section~\ref{sec:methods}, describe their implementation in an isolated dwarf galaxy simulation in Section~\ref{sec:IC}, show results from this simulation in Section~\ref{sec:results}, discuss the results in Section~\ref{sec:discussion}, and conclude in Section~\ref{sec:conclusion}. Readers who may want to only briefly skim (or skip) over the details of our included physics are advised to just read the beginning of Section~\ref{sec:methods}, which contains a complete---yet brief---summary of the included physics.

\section{Methods}
\label{sec:methods}
We produce high-resolution, galaxy-scale simulations tracking stars not as single stellar populations, but as individual stars sampled from an assumed IMF.
This allows us to follow star-by-star variations in feedback physics and stellar yields in detail. To properly model the ISM, we track non-equilibrium, primordial chemistry (including molecular hydrogen) using \texttt{Grackle} \citep{GrackleMethod}, with heating and approximate self-shielding from a metagalactic ultraviolet (UV) background. We assume collisional ionization equilibrium for all other elements and use updated \texttt{Cloudy} metal-line cooling tables consistent with our self-shielding approximation (see Appendix~\ref{appendix:cooling}). We also include an updated observationally motivated dust model for the low metallicity regimes studied here ($Z \lesssim 0.1$ Z$_{\odot}$). Each star is assigned basic properties including surface gravity, effective temperature, radius, and lifetime from tabulated stellar evolution models, which inform how the stars deposit their feedback. We directly track ionizing radiation from massive stars using an adaptive ray tracing radiative transfer method that includes the effects of radiation pressure on HI. In addition, we follow the optically thin, non-ionizing UV radiation from these stars that cause photoelectric heating and Lyman-Werner dissociation of molecular hydrogen. We track the stellar wind feedback and SNe from these stars, depositing individual metal yields from both. We include AGB wind feedback and yields for lower mass stars, and track these directly as Type Ia SN progenitors. We follow yields for 15 individual metal species (C, N, O, Na, Mg, Si, S, Ca, Mn, Fe, Ni, As, Sr, Y, and Ba), chemically tagging each star as it forms with the associated local gas abundances for each species. In addition, we track a total metal density field which is the sum of all metals, including those not directly tracked. This field is used to inform the heating/cooling physics, and determines the metallicity of each star at birth. These methods are discussed in full detail below.

\subsection{Hydrodynamics and Gravity}
\label{sec:hydro}

We use the adaptive mesh refinement hydrodynamics and N-body code \texttt{Enzo}\footnote{http://www.enzo-project.org} to simulate the chemodynamical evolution and detailed feedback physics in a set of high resolution, isolated, low-mass dwarf galaxies. \texttt{Enzo} is an open-source code that is undergoing continual, active development by many researchers across several institutions. We use a substantially modified version of the current development version of \texttt{Enzo} (version 2.X) in this work.\footnote{This version is contained in a publicly available fork of the main repository: https://bitbucket.org/aemerick/enzo-emerick. Specifically, simulations presented here were conducted at changeset 7001d99.} We solve the equations of hydrodynamics using a direct-Eulerian piecewise parabolic method \citep{ColellaWoodward1984, Bryan1995} and a two-shock approximate Riemann solver with progressive fallback to more diffusive Riemann solvers in the event that higher order methods produce negative densities or energies. We compute the total gravitational potential from gas self-gravity, stars, and a static background dark matter potential (see Section~\ref{sec:IC}). Self-gravity is computed with a multigrid Poisson solver. The collisionless star particles are evolved with an adaptive particle-mesh N-body solver at an effective force resolution of $\sim 2 \Delta x$, where $\Delta x$ is the local cell size. 

We refine the mesh whenever the thermal Jeans length is no longer resolved by a minimum of 8 cells, continually refining a given region until this criterion is met or until the region reaches the maximum resolution (1.8~pc). At maximum resolution, the Jeans length can become under-resolved, leading to artificial numerical fragmentation. \citet{Truelove1997} showed that resolving the Jeans length by at least 4 cells is required to suppress this fragmentation.

We set the star formation density threshold to the value at which the Jeans length becomes resolved by only 4 cells in sub-200 K gas, or about 200 \ccunit (as discussed further in Section~\ref{sec:star formation}). Forming stars from this gas will reduce the local density, ensuring the Jeans length is resolved. However, since star formation is not instantaneous, we employ a pressure floor to support gas against artificial fragmentation and collapse. A non-thermal pressure term is added to cells once their thermal Jeans length becomes unresolved. This prevents dense, self-gravitating gas from rapidly reaching densities significantly above our resolution limit. The use of a pressure floor is common in galaxy scale simulations with limited dynamic range \citep[e.g][]{Machacek2001, 2008ApJ...680.1083R}.

Due to computational constraints we found it necessary to institute a temperature ceiling in low density, diffuse gas. These high temperatures, typically well above 10$^{7}$~K and up to 10$^{8}$~K, would place an onerous constraint on the limiting time step at high spatial resolution. At these temperatures, with typical velocities up to $\sim$10$^{3}$~km~s$^{-1}$, satisfying the Courant condition requires time steps on order of 100~yr. We institute a maximum temperature of 8$\times 10^6$~K in gas with densities between 10$^{-30}$~g~cm$^{-3}$ and 10$^{-26}$~g~cm$^{-3}$. These densities were somewhat arbitrarily chosen, but ensure that this threshold does not impact very low density gas in the halo of our dwarf galaxy or higher density gas in supernova injection regions. This threshold decreases the required run-time by factors of a few. The value of the temperature threshold was chosen to ensure the affected hot gas remained just above the high-temperature minimum of our cooling curve (see Appendix~\ref{appendix:cooling}.)

\subsection{Chemistry and Cooling Physics}
\label{sec:chemistry}

We use the chemistry and cooling library \texttt{Grackle}\footnote{https://grackle.readthedocs.io/en/grackle-3.0/} v 3.0 to follow a nine species non-equilibrium chemistry network (H, H$^+$, He, He$^+$, He$^{++}$, e$^{-}$, H$_2$, H$^{-}$, and H$_{2}$) which includes radiative heating and cooling from these species and metals.\footnote{We use a slightly modified version of the the main \texttt{Grackle} repository, available at https://bitbucket.org/aemerick/grackle-emerick at changeset c2c0faf.} \texttt{Grackle} is a freely available, open source, multi-code library, designed to interface with a wide variety of astrophysical codes. We outline specific model choices made in our simulations and refer the reader to \citet{GrackleMethod} for a detailed discussion of the code. We apply the \citet{Glover2008} three-body rate for H$_{2}$ formation and include a model for H$_2$ formation on dust, dust heating, and dust cooling following the methods in \citet{2000ApJ...534..809O} and \citet{2005ApJ...626..627O} as included in \texttt{Grackle}. However, we update the default dust to gas ratio scaling in \texttt{Grackle} to account for the steeper scaling in low metallicity regimes ($Z \lesssim 0.1 Z_{\odot}$), using the broken power law scalings from \citet{Remy-Ruyer2014}. For metallicities above $\sim 0.1 Z_{\odot}$, this is equivalent to the default behavior of \texttt{Grackle}, where dust content scales linearly with metallicity.

As part of the \texttt{Grackle} package, metal line cooling is modeled using pre-computed \texttt{Cloudy} \citep{Cloudy2013} \footnote{http://www.nublado.org/} tables interpolated as a function of density, temperature, and redshift, using the \citet{HM2012} UV metagalactic background. As discussed in more detail in Section~\ref{sec:diffusive heating}, we account for approximate self-shielding of H and He against this UV background. Using this prescription with metal line cooling tables computed under an optically thin assumption can lead to an order of magnitude overestimation of the cooling rate at certain densities, as discussed in \citet{Hu2017} and  Appendix~\ref{appendix:cooling}. To address this issue, we use re-computed metal line tables consistent with the self-shielding approximation.  We have made these new tables public in the main \texttt{Grackle} repository. These are discussed in greater detail in Appendix ~\ref{appendix:cooling}. Finally, we ignore the effect the stellar radiation field (see Section~\ref{sec:diffusive heating}) may have on the interpolated metal line cooling rates. 

\subsection{Star Formation Algorithm}
\label{sec:star formation}
In order to resolve individual star formation events on galactic scales, we implement a stochastic star formation algorithm adopted from \citet{Goldbaum2015,Goldbaum2016}. Each cell at the maximum refinement level is capable of forming stars if it contains gas that meets the following local criteria on number density $n$, temperature $T$, cell mass $M$, and velocity $\vec{v}$: 1) $n > n_{\rm thresh}$, 2) $T < T_{\rm thresh}$, 3) $M > M_{\rm Jeans}$, and 4) $\vec{\nabla} \cdot \vec{v} < 0$, where $n_{\rm thresh}$ is a resolution dependent density threshold, $T_{\rm thresh}$ is a temperature threshold, and $M_{\rm Jeans}$ is the local thermal Jeans mass. Our fiducial values are $n_{\rm thresh} =  200 \mbox{ cm}^{-3}$ and $T_{\rm thresh}= 200$~K. We limit the fraction of a cell's gas mass that is converted into stars by requiring $M > f_{\rm thresh} M_{\rm max,*}$, where $f_{\rm thresh} = 2.0 $ and the maximum star mass $M_{\rm max,*} = 100 \mbox{ M}_{\odot}$. No star formation occurs when $M < f_{\rm thresh} M_{\rm max,*}$, ensuring that a star formation episode can not produce negative densities.

We make the common ansatz that star formation occurs by converting gas into stars in a free fall time $\tau_{\rm ff}$ with a star formation efficiency, $\epsilon_{\rm f} \simeq 0.02$. At high resolution, the choice of $\epsilon_{\rm f}$ should be irrelevant \citep{Orr2018, FIRE2}, as star formation is ultimately self-regulated by feedback.

The stellar mass formed during a timestep $\Delta t$ from a region with a total gas mass $M_{\rm gas}$ is
\begin{equation}
         M_* = \epsilon_{\rm f} M_{\rm gas} \frac{\Delta t}{\tau_{\rm ff}}
\end{equation}
In practice, $\Delta t/\tau_{\rm ff} \ll 1$, and $M_*$ is smaller than the minimum star particle mass at parsec scale resolution. We therefore allow star formation to proceed stochastically, following the methods in \citet{Goldbaum2015, Goldbaum2016}, modified for variable stellar masses. In each cell that could potentially form stars, we compute the probability that 100 M$_{\odot}$ of gas will be converted into stars in that time step, and use a random number draw to determine whether or not star formation actually occurs. If it does, we randomly sample from
the adopted IMF until approximately 100 M$_{\odot}$ of stars form, keeping the last sampled particle when the total stellar mass formed exceeds 100 M$_{\odot}$. The total mass of formed star particles is subtracted from the gas mass in the star forming region to ensure mass conservation. We assume a Salpeter IMF \citep{Salpeter1955} with $\alpha = 2.35$, sampling over the range between a minimum stellar mass of 1 M$_{\odot}$ and an arbitrarily chosen maximum stellar mass of 100 M$_{\odot}$. Our lower limit on stellar masses ensures that we are able to both directly track all particles that contribute in some way to feedback and metal enrichment, and follow longer lived star particles, while reducing the computational expense of following a large number of low mass stars that have no dynamical impact on the galaxy evolution. By ignoring the formation of stars below 1~M$_{\odot}$, our model in effect spreads this mass into higher mass stars, changing the normalization of the IMF slightly from what would be expected for an IMF that extends below  1~M$_{\odot}$.

Formed stars are deposited with random positions within the star forming cell and assigned velocities equal to the cell bulk velocity with a 1~km~s$^{-1}$ velocity dispersion. This dispersion captures some of the unresolved gas motions below the resolution limit that are smoothed out by numerical diffusion; it is comparable to, but less than, the velocity dispersion of the coldest gas in our simulations. Stars are assigned metallicities corresponding to the metallicity of the star forming zone, and are chemically tagged with the 17 individual species abundances (H, He, and the 15 metals) that we follow in our simulations. 

Stars evolve during the simulation, by losing mass from stellar winds and SNe as described below, and by changing types, but persist throughout the entire simulation. For example, low mass stars are tagged as white dwarfs (WDs) at the end of their life, which may eventually explode as a Type Ia SN (discussed below), after which they persist as massless tracer particle remnants. Finally, each star is marked as a ``must refine'' particle, requiring that it be surrounded by a four-cell region at the highest level of refinement. This ensures that both stellar winds and SNe feedback are maximally resolved, and that any ejected yields are deposited over a consistent physical scale throughout the simulation.

\subsection{Stellar Properties}
\label{sec:properties}
Given each star's birth mass and metallicity, we interpolate over the PARSEC grid of stellar evolution tracks \citep{Bressan2012} to assign a lifetime and AGB phase start time (if any) to it, as well as the effective temperature $T_{\rm eff}$ and surface gravity $g$ used in computing radiation properties (see Section \ref{sec:ionizing radiation}). We use the largest subset of the PARSEC models that are regularly sampled in our mass/metallicity space of interest, with 26 mass bins over M$_{*} \in \left[0.95, 120.0 \right] {\rm M_{\odot}}$ and 11 metallicity bins over $Z \in \left[10^{-4}, 0.017 \right]$. Although $T_{\rm eff}$ and $g$ evolve over time for stars, modifying stellar radiative properties, following a stellar evolution track for each of our stars is beyond the scope of this work. We instead fix these properties at their zero age main sequence values.

\subsection{Stellar Feedback and Chemical Yields}

\subsubsection{Stellar Yields}
\label{sec:yields}
For the first time in galaxy-scale simulations, we track galactic chemodynamical evolution using stellar yields ejected from star particles that represent individual stars. We adopt the NuGrid\footnote{http://www.nugridstars.org} collaboration's set of stellar yields given on a uniformly sampled grid in stellar mass and metallicity with 12 mass bins over $M_{*} \in \left[1.0, 25.0\right]$ M$_{\odot}$ and five metallicity bins at metal fractions of $Z =$ 0.02, 0.01, 0.006, 0.001, and 10$^{-4}$ \citep{Pignatari2016, Ritter2018}. This grid includes yields from the AGB phase of stars from 1--7 M$_{\odot}$, as well as yields from both stellar winds and core collapse SNe of massive stars from 12--25 M$_{\odot}$. We complement these tables with tables from Slemer et al.\ (in prep), based on the PARSEC stellar evolution tracks \citep{Bressan2012, Tang2014}, to track stellar winds for stars more massive than 25 M$_{\odot}$. We ignore SN yields from these stars (see next paragraph). We combine all stable isotope yields for a given element into a single elemental abundance for all stable elements from hydrogen to bismuth. Although we can in principle follow an arbitrary number of metal species, practical considerations of memory use prevent this in any given simulation. We refer the reader to previous uses of the NuGrid yields in one-zone galactic chemical enrichment models \citep{Cote2016,  Cote2016_feb,Cote2017a} for a detailed discussion of how various model uncertainties can influence galactic chemical evolution.

%
% KVJ: Possibly include a discussion on uncertainties in stellar yields
%

Above some mass $M_{\rm trans}$ within the unsampled range of 7--12 M$_{\odot}$, stars no longer undergo AGB wind phases but end their lives instead as core collapse SNe. Where this transition occurs is uncertain, but is commonly taken to be at a mass $M_{\rm trans} \sim 8$--10~M$_{\odot}$; we take $M_{\rm trans} = 8$~M$_{\odot}$. In our model, stars below this mass eject their wind yields in an AGB phase only at the end of their lives, typically over a period comparable to or less than a few time steps ($\lesssim 10$ kyr). Stars above this mass are assumed to eject their stellar yields via line-driven stellar winds at a constant mass loss rate throughout their lifetime (neglecting Wolf-Rayet and luminous blue variable phases), ending their lives as a core-collapse SN (see Sect.~\ref{sec:stellar winds} for details on the wind energetics). Varying $M_{\rm trans}$ changes both the time at which yields for stars around this mass are ejected (for reference, the lifetime of an 8 M$_{\odot}$ star is about 35--40~Myr), and the energy injection from these winds. \citet{Cote2017a} explores how the choice of $M_{\rm trans}$ affects galaxy abundances in a one-zone model. We neglect the effects of binary star evolution on stellar feedback, and discuss the significance of this in Sect.~\ref{sec:binary stars}.

There are large uncertainties in stellar yields for stars more massive than 25 M$_{\odot}$ \citep[see ][and references therein]{Cote2016}. Indeed, even the exact fate of these stars is uncertain \citep[e.g.][]{Woosley2002,Zhang2008,Ugliano2012}, particularly as a function of metallicity \citep{Fryer2012} with potentially multiple stable and unstable regimes as a function of mass \citep{Heger2003}. Due to this uncertainty, and to avoid erroneously extrapolating from our yield tables, we adopt the simplest model and assume all stars above 25 M$_{\odot}$ end their life through direct collapse to a compact object with no further mass, metal, or energy ejection.

Type Ia SNe are an important additional source of galactic chemical enrichment. These iron group rich events are responsible for the $\sim$1 Gyr timescale turnover, or ``knee", in $[\alpha/\rm{Fe}]$ vs $[\rm{Fe}/\rm{H}]$ diagrams. We use the Type Ia SN yields given in \citet{Thielemann1986}, adopting a Type Ia SN model as discussed in Sect.~\ref{sec:Type Ia}. We emphasize that we only track Type Ia SNe occurring within the population of stars formed in this model, neglecting SNe from any possible pre-existing population, substantially limiting the number of Type Ia SNe occurring during the initial gigayear in our models.

\subsubsection{Stellar Winds}
\label{sec:stellar winds}
Stellar winds are important sources of enrichment and feedback in galaxies at both early times from massive stars and late times from AGB stars. Although the energy injected via winds over the lifetime of a cluster of stars is much less than that from SNe and radiation \citep{Shull1995}, stellar winds are potentially important sources of pre-SN feedback. We assume complete thermalization of the wind kinetic energy, taking the total injected energy injected in timestep $\Delta t$ as $E_w = \frac{1}{2}\dot{M}  v^2_w\Delta t  + E_{\rm th}$, where $E_{\rm th}$ is the thermal energy of the ejected gas mass $M_w = \dot{M}\Delta t$ given the star's interpolated effective temperature $T_{\rm eff}$. This mass and energy is injected evenly over a three-cell radius spherical volume centered on the star particle. The edges of this spherical region will only partially overlap grid cells. We use a Monte Carlo sampling method to properly compute the volume of this overlap to scale the injection in these cells appropriately. We assume constant mass loss rates for all winds as set by the yield tables over either the lifetime of the star (for massive stars) or the length of the AGB phase (for low mass stars). 

Massive stellar winds have typical velocities of order 10$^{3}$ km s$^{-1}$ \citep{Leitherer1992}. Satisfying the Courant time step becomes prohibitively expensive following this gas, with time steps dropping to $\Delta t \sim$~100~yr. For this reason, we adopt the common simplification of reducing the wind velocity \citep[e.g][]{Offner2015}. In our case, we fix massive stellar wind velocities to $v_w = 20$~km~s$^{-1}$ for stars above 8 M$_{\odot}$. Our initial tests show that turning off energy injection from stellar winds like this does not significantly affect the global star formation rate of our galaxies. Due to the substantial additional computational expense of following stellar winds for gigayear timescales, we reserve examining the detailed importance of winds to future work. These points are discussed in more detail in Sect.~\ref{sec:stellar winds discussion}.

Stars that only undergo an AGB phase deposit their feedback at the end of their lives, as determined by the PARSEC evolution tracks. AGB wind velocities vary dramatically over their relatively short lifetimes, but are typically on the order of 10 km s$^{-1}$. For simplicity, we adopt a fixed wind velocity of 20 km s$^{-1}$ for all AGB stars as well.

\subsubsection{Core Collapse SNe}
\label{sec: core collapse}
Stars between  $M_{\rm trans} = 8$ M$_{\odot}$ and 25 M$_{\odot}$ end their lives as core collapse SNe, ejecting mass and metals as determined by the NuGrid stellar yield tables, along with 10$^{51}$ erg of thermal energy. Due to the high resolution of our simulations (1.8~pc), we generally resolve the Sedov phase of each SN explosion well (see Appendix~\ref{appendix:SN}). We inject thermal energy alone in a three-cell radius spherical region around the star particle, which we find to be sufficient to resolve the SN explosions. We use the same Monte Carlo sampling method as used in our stellar winds to map the spherical injection region to the grid. We continue to track any remaining stellar mass after the SNe occurs as a massive remnant tracer particle. In future work these particles can be used to self-consistently account for more exotic sources of feedback and chemical enrichment such as X-ray binaries and neutron-star binary merger events which, while rare, could be important in long term galaxy evolution \citep[e.g.][]{Artale2015}.

\subsubsection{Type Ia SNe}
\label{sec:Type Ia}
We continue to track low mass stars ($M < 8$ M$_{\odot}$) after their death as WD particles, marking a subset as Type Ia SN progenitors. This is the most self-consistent model for Type Ia SNe in galaxy-scale simulations. We note however that for the low SFRs in our isolated dwarf galaxy simulation, the first Type Ia SN only appears after a few hundred megayears of simulation time. By the end of the simulation presented here (500~Myr), only 18 have gone off. At the end of their life, we assign a new mass to these particles following the initial-to-final-mass relation of \citet{Salaris2009}. We follow the common assumption that progenitor stars with initial masses between 3 and 8 M$_{\odot}$ form WDs that are Type Ia progenitors \citep[see][ and references therein]{Cote2017a}. 

We compute the probability that a given Type Ia progenitor will explode as a function of time using an observationally motivated delay time distribution model. The Type Ia SN rate is taken to be a power law in time, $\Psi (t) \propto t^{-\beta}$, whose slope $\beta$ and normalization $N_{\rm Ia}/M_{\rm SF}$ are observables. The latter represents the number of Type Ia SNe per mass of star formation.  By assuming an IMF  $dN/dm$, one can write down the fraction $\eta$ of stars capable of forming a Type Ia SN progenitor that \textit{will} explode within a Hubble time. This is given as
\begin{equation}
\eta = \frac{N_{\rm Ia}}{M_{\rm SF}} \frac{\int_{M_{\rm min}}^{M_{\rm max}} m (dN/dm) dm }{\int_{M_{1}}^{M_{2}} (dN/dm) dm},
\end{equation}
where $M_{\rm min}$ and $M_{\rm max}$ are the lower and upper bounds of the IMF, and $M_{1}$ and $M_{2}$ are the lower and upper bounds of the range of stars that can form Type Ia candidates. The distribution slope $\beta$ is of order unity, with typical values between 1.0 and 1.2 \citep[see][for a recent review]{Maoz2014}. $N_{\rm Ia}/M_{\rm SF}$ can be derived by taking observed values of the Type Ia SN rate and integrating over a Hubble time. Typical values for this are on order of 10$^{-3}$ M$_{\odot}^{-1}$ \citep{Maoz2014}. For our fiducial values, we adopt $\beta = 1.2$ \citep{Maoz2010} and $N_{\rm Ia}/M_{\rm SF} = 0.8\times 10^{-3}$ M$_{\odot}^{-1}$ \citep{GraurMaoz2013}. Given our choice of IMF, and with $M_{\rm min} = 1$~M$_{\odot}$, $M_{\rm max}=100$~M$_{\odot}$, $M_{1}=3$~M$_{\odot}$, and $M_{2}=8$~M$_{\odot}$, this gives $\eta = 0.041$.

Finally, we can normalize $\Psi(t)$ to give the probability per unit time $\dot{P}(t)$ that a Type Ia candidate will explode at a time $t$ after the formation of its main sequence progenitor. Integrating this gives the total probability at any given time as
\begin{equation}
P(t) = \int \dot{P}(t)dt = \frac{\eta}{{ \int_{t_{\rm o}}^{t_{\rm H} + t_{\rm o}} \tau^{-\beta} d\tau}} \int t^{-\beta} dt,
\end{equation}
where $t_{\rm o}$ is the formation time of the WD and the leading term on the right hand side properly normalizes the total probability over a Hubble time to $\eta$. This naturally accounts for both a prompt and delayed Type Ia SN population in our simulations.
In practice, rather than drawing a random number for each candidate every timestep, we make a single random number draw, $u$, at the formation time of the white dwarf particle. For $u \in [0,1]$, we interpolate its position on a pre-tabulated and inverted cumulative probability distribution function to assign a single time at which the WD particle will explode as a Type 1a supernova. We institute a minimum delay time by defining $P(t)$ only for $t > t_o$, such that a particle cannot be assigned an explosion time before its formation time.

\subsubsection{Ionizing Radiation from Discrete Sources}
\label{sec:ionizing radiation}
Radiation feedback, including ionization, ionization heating, and radiation pressure, is an important source of feedback in galaxies. \ion{H}{2} regions carved out by stellar radiation change the ISM structure in regions where SNe eventually explode, generally increasing their dynamical importance. However, accounting for angular effects, radiation can also allow energy from SNe to dissipate more readily by escaping out of channels carved through dense clouds. Radiation feedback effects have been included with various approximations in a wide range of simulations \citep[e.g.][]{OppenheimerDave2006, Krumholz2007, HopkinsQuataertMurray2012, Agertz2013, Renaud2013, Stinson2013, Roskar2014, Ceverino2014, FIRE, AgertzKravtsov2015, Forbes2016, Hu2016, Hu2017, FIRE2}, with a smaller subset using full radiation hydrodynamics \citep{WiseAbel2012,Wise2012a,Wise2014,Kim2013a, Kim2013b,Pawlik2013,Rosdahl2015,Aubert2015,Ocvirk2016,BaczynskiGloverKlessen2015,Pawlik2017} due to the additional computational expense of direct ray tracing. As we seek a complete accounting of stellar feedback physics, we follow HI and HeI ionizing radiation from our stars through the ray tracing methods described below. 

Enzo includes an adaptive ray tracing implementation, \textsc{Enzo+Moray} \citep{WiseAbel2011}, to solve the equations of radiative transfer coupled to the hydrodynamics of the simulation. We follow HI and HeI ionizing photons which are coupled to the \texttt{Grackle} primordial chemistry and heating and cooling routines to track photoionization and heating, as well as radiation pressure on hydrogen. 

We determine the HI and HeI ionizing photon rates for each star using the OSTAR2002 \citep{Lanz2003} grid of O-type stellar models, appropriate for $M_{*} \gtrsim 15$~M$_{\odot}$ at solar metallicity\footnote{The exact stellar mass range on the OSTAR2002 grid is model dependent and a function of metallicity}. We use linear interpolation in stellar effective temperature, surface gravity, and metallicity to compute the ionizing photon fluxes and rates for each star. Stars less massive than about 15 M$_{\odot}$ and very massive stars with sub-solar metallicity are generally not well sampled by the OSTAR2002 grid. In this case, we integrate a black body spectrum at $T_{\rm eff}$ to obtain the ionizing photon fluxes, but normalize the result to be continuous with the OSTAR2002 grid (see Appendix \ref{appendix:radiation}).

Instead of assigning a fixed ionizing photon energy across all sources, we integrate over each star's blackbody curve to compute the average ionizing photon energy individually for each source (see Appendix~\ref{appendix:radiation}). The average energy for HI and HeI ionizing photons changes significantly over the OSTAR2002 temperature range $\log(T_{4,\rm{eff}} [K]) \in \left[2.75,5.5\right]$, ranging from 15.72 eV to 20.07 eV and 26.52 eV to 31.97 eV respectively.

We also include the effects of radiation pressure on HI. This has been shown to be important in suppressing the star formation rates of dwarf galaxies by influencing turbulence and the dense gas content of the ISM \citep{WiseAbel2012,Ceverino2014}. We ignore the absorption of ionizing radiation by dust and re-radiation in the infrared. This is included in other models \citep[e.g.][]{Rosdahl2015,FIRE,FIRE2} as this may increase by a factor of a few to several the effective radiation pressure \citep{ZhangDavis2017}. However, the importance of multiple scattering is still unclear. Other works have shown the effect to only increase the radiation pressure by a factor of order unity \citep{Krumholz2012,Krumholz2013a,Krumholz2018,Reissl2018,Wibking2018}. Due to these uncertainties, and given that our dwarf galaxy has a low dust content, and therefore a low infrared opacity, we ignore this effect. 

\subsubsection{Diffuse  Heating}
\label{sec:diffusive heating}
We include two forms of diffuse heating in our simulations, each tied directly to the non-equilibrium primordial chemistry network in \texttt{Grackle}: 1) the optically thin, uniform metagalactic UV background \citep{HM2012}, and 2) localized photoelectric heating from the FUV (6 eV $<h\nu< 13.6$ eV) radiation from each of our star particles. The FUV flux for each star is again obtained from the OSTAR2002 grid by directly integrating over the spectral energy distributions for each gridded star. Like the ionizing radiation, we again use an adjusted black body spectrum to compute the flux for stars off of the grid (see Sect.~\ref{sec:ionizing radiation} and Appendix~\ref{appendix:radiation}). Photoelectric heating can be a dominant heating mechanism in the ISM of the Milky Way \citep{Parravano2003}, and could be significant in regulating star formation in dwarf galaxies \citep{Forbes2016}. However, this conclusion warrants further research as its exact importance in dwarf galaxies relative to other feedback mechanisms is contentious \citep{Hu2016,Hu2017}. Generally, models for photoelectric heating and Lyman-Werner radiation in hydrodynamic simulations of galaxies adopt a constant value or a static, radial profile. Only recently has the localization and time variation of these processes been considered.

Self-shielding of gas against the metagalactic UV background is important in high-resolution simulations, particularly for low-mass, low-metallicity dwarf galaxies where the UV background is capable of gradually photoevaporating unshielded gas from the galaxy \citep{Simpson2013}. We have implemented the \citet{Rahmati2013} approximate self-shielding method in \texttt{Grackle} to account for HI self-shielding against the UV background \citep[see][ for more details of this implementation]{GrackleMethod}. We assume HeI ionization generally follows HI. This allows us to approximate HeI self-shielding using the same form (A. Rahmati, private communication). We ignore HeII photoionization from the UV background entirely. For consistency, we additionally reduce the reaction rates for direct H$_2$ ionization (15.4 eV) and H$_2^+$ destruction (30 eV) by the same shielding factors computed for HI and HeI shielding.\footnote{Ignoring this effect leads to unrealistically high electron fractions in self-shielding gas from direct $H_2$ ionization, which drives significant production of H$_2$ via gas-phase reactions.} Accounting for self-shielding in this manner leads to an inconsistency in using tabulated, optically-thin metal line cooling rates from \textsc{Cloudy} (see Section 4.1.1 of \citet{Hu2017}). As mentioned previously, we have re-computed metal line cooling tables using \textsc{Cloudy} models of optically thick clouds to be consistent with our self-shielding prescription. This is described in more detail in Appendix ~\ref{appendix:cooling}. 

%Ignoring this effect leads to an overestimation of the cooling rates when in the self-shielding regime. %We refer to \citet{2006agna.book.....O,2010MNRAS.408.1945M,2012MNRAS.421.2232F,Rahmati2013} for a more detailed description of this approximation and \citet{Gracklemethod} for its specific implementation, as used here.

We assume the galaxy is mostly optically thin to stellar FUV and use only local approximations for shielding.  We calculate the stellar FUV flux in each cell as summed over the contributions from each star to parameterize the local photoelectric heating rate as \citep{BakesTielens1994,Wolfire2003,Bergin2004}
\begin{equation}
\label{eq:PE}
\Gamma_{\rm pe} = (1.3 \times 10^{-24} \rm{erg~s^{-1}~cm^{-3}})\, \epsilon n_{\rm H} G_{\rm eff} D 
\end{equation}
where $\epsilon$ is an efficiency factor that depends on $G_{\rm{o}} T^{1/2} /n_{\rm{e}}$, the attenuated local FUV flux \begin{equation} G_{\rm eff} = G_{\rm o}~\exp(-1.33\times10^{-21}~D~N_{\rm H}), \end{equation} $D$ is the dust-to-gas ratio, normalized to the solar value, and $G_{\rm o}$ is the local FUV flux normalized to the solar neighborhood \citep{Habing1968}. Aside from a different treatment of $D$ and the attenuation, both discussed below, this is equivalent to the method used in \citet{Hu2016,Hu2017}.

The value of $D$ is computed consistently with our \texttt{Grackle} dust model, using the broken power law fit from \citet{Remy-Ruyer2014}, as described in Section~\ref{sec:chemistry}. The extremely low dust-to-gas ratio in our modeled galaxies leads to a reduction in the photoelectric heating rate by approximately two orders of magnitude, as compared to a model that assumes a ratio $D$ that scales linearly with metallicity at very low metallicity. At these low metallicities, the FUV field only becomes optically thick at length scales of $\sim$ 100 pc for densities of $n \sim 10^2$~\ccunit. Given that the ambient density of the ISM is generally 1--10~cm$^{-3}$, we can safely assume the FUV field to be optically thin. However, we do include a localized attenuation prescription that may influence high-density or metal-enriched regions of the galaxy. We approximate $N_{\rm H}$ given in the equation above locally, as $n_{\rm H}\Delta x$, where $\Delta x$ is the cell width; this approximation is necessarily resolution dependent, but substantially more computationally efficient than direct ray tracing.

Properly computing $\epsilon$ in Eq.~\ref{eq:PE} requires an accurate account of the electron number density $n_e$. This is non-trivial in dense, neutral regions where $n_e$ is dominated by contributions from carbon, dust, and PAH ionizations. Our chemical network only includes contributions from H, He, and H$_2$ to n$_e$. Instead, we use a power-law fit of $\epsilon$ as a function of $n_{\rm H}$ from the \citet{Wolfire2003} model of $\Gamma_{\rm{Pe}}$ in the solar neighborhood (see Figure 10b of that work); we adopt $\epsilon = 0.0148n_{\rm{H}}^{0.235}$. %This is a compromise between adoption of a constant efficiency, as in \citet{Forbes2016}, and a more detailed model with more detailed non-equilibrium chemistry \citet[e.x.][]{Hu2016}.

\subsubsection{Lyman-Werner Radiation}
\label{sec:LW}
In addition to the Lyman-Werner radiation from the UV background, we account for localized Lyman-Werner flux from each of our stars to compute the total, local H$_2$ dissociation rate. We compute the stellar Lyman-Werner flux again from the OSTAR2002 grid by integrating the spectral energy distributions over photon energies from 11.2~eV to 13.6~eV (see Appendix~\ref{appendix:radiation}). Given the local Lyman-Werner flux, the H$_2$ dissociation rate is taken as $k_{\rm diss} = \sigma_{\rm H2} F_{\rm LW}$, where $\sigma_{\rm H2}$ is the H$_2$ dissociation cross section. We account for approximate H$_2$ self-shielding against these sources of Lyman-Werner flux by implementing the Sobolev-like approximation from \citet{Wolcott-Green2011} in \texttt{Grackle}. 

\section{Galaxy Initial Conditions}
\label{sec:IC}
We apply these methods to a first test case of the evolution of an isolated, low mass dwarf galaxy. The galaxy is constructed to have initial properties similar to those observed for the Local Group dwarf galaxy Leo P \citep{Giovanelli2013,McQuinn2013,McQuinn2015a,McQuinn2015}, although it is not intended to be a matched model to this galaxy. Leo P is gas rich, with a neutral hydrogen mass $M_{\rm HI} = 8.1\times 10^{5}$~M$_{\odot}$ and stellar mass $M_{*} = 5.6^{+0.4}_{-1.9} \times 10^{5}$~M$_{\odot}$ \citep{McQuinn2015a} extending to an observed neutral hydrogen radius $r_{\rm HI} = 500$~pc. LeoP has a low metallicity, with an oxygen to hydrogen abundance ratio (O/H) of $12 + \rm{log(O/H)} = 7.17 \pm 0.04$ \citep{Skillman2013}, or a metallicity of $Z \sim 5.4\times10^{-4}$ ($Z/Z_{\odot} = 0.03$, adopting $Z_{\odot} = 0.018$ from \citet{Asplund2009}). Our dwarf galaxy model is constructed without an initial background stellar population, with a total gas mass of $1.8 \times 10^{6}$~M$_{\odot}$, of which $M_{\rm HI} = 1.35 \times 10^{6}$~M$_{\odot}$, and $Z = 4.3\times 10^{-4}$, comparable to the average redshift $z = 0$ metallicity from the stellar models computed in \citet{McQuinn2015}.

The galaxy initially consists of a smooth, exponential gas disk in hydrostatic equilibrium with a static, background dark matter potential. The gas profile follows \citet{Tonnesen2009} and \citet{Salem2015}, with
\begin{equation}
\rho_{\rm gas} (R,z) = \frac{M_{\rm o}}{2\pi a^2_{\rm gas}b_{\rm gas}} 0.5^2{\rm sech}\left(\frac{R}{a_{\rm gas}}\right){\rm sech}\left(\frac{z}{b_{\rm gas}}\right)
\end{equation}
where $a_{\rm gas}$ and $b_{\rm gas}$ are the radial and vertical gas disk scale heights, and $M_{\rm o}$ is approximately 70\% of the total gas mass. We set $a_{\rm gas} = 250$~pc, $b_{\rm gas} = 100$~pc, and $M_o = 1.26\times 10^6$~M$_{\odot}$. We adopt a \citet{Burkert1995} dark matter potential with virial mass and radius $M_{\rm vir} = 2.48\times 10^{9}~M_{\odot}$ and $R_{\rm vir}~=~27.4$~kpc as defined in \citep{BryanNorman1998} and scale radius r$_{\rm s} = 990$~pc. This gives a maximum circular velocity $V_{\rm max} = 30.1$~km~s$^{-1}$ at $R_{\rm vmax}~=~3.2$~kpc. These parameters were adopted specifically to match the observed dynamical mass of Leo P interior to 500 pc of $M_{\rm dyn} (r < 500~\rm{pc}) = 2.7\times 10^{7}$ M$_{\odot}$, and represent virial properties within the halo mass expected for galaxies of this size \citep{Ferrero2012,Read2017}.

Following the initialization procedure of \citet{Hu2017}, we use artificial SN driving to generate realistic initial densities and turbulent properties in the galaxy ISM. This prevents an otherwise uniform collapse of the gas disk at the beginning of the simulation. During this period, SNe explode at a fixed rate of $0.4$~Myr$^{-1}$, corresponding to the SFR obtained given the central HI surface density and the relation presented in \citep{Roychowdhury2009}. We stop the artificial driving 25~Myr after the first star particle forms. These artificial SNe do not drive chemical evolution of the galaxy; their metal yields correspond to the mean ISM abundances. We note this initial driving is ad hoc in that we do not include other effects from the stellar population that would have caused these SNe.

We emphasize here that our model, with no initial stellar distribution, is not intended to identically reproduce the evolution of Leo P, which has formed stars continuously over cosmological timescales. In addition, the mass fractions for the individual metals we track are set to zero so that we follow only the evolution of metals self-consistently produced in the simulations. Otherwise, the galaxy chemical properties would be dominated by the somewhat arbitrary choice of initial abundances. In some ways this is similar to the first pollution of pristine gas in the early Universe, but we note that we cannot directly make this comparison as the environmental conditions and UVB properties were different at high redshift, and we explicitly ignore Pop III and Pop II stellar evolution. Instead, this work is intended as a numerical experiment to investigate how metal enrichment from ongoing star formation proceeds in a gas / dark matter environment similar to low mass halos at both low redshift and the early Universe. The subsequent metal enrichment from the stars in our simulation can be thought of as tracking a change in abundances from arbitrary initial conditions. We will discuss how to make proper comparisons to observed stellar and gas abundance properties of dwarf galaxies in future work, where we will investigate the abundance evolution of our simulations in more detail.

\section{Results}
\label{sec:results}
We present our initial results here, providing an overview of the morphological (Section~\ref{sec:structure}), star formation (Section~\ref{sec:sfr}), ISM (Section~\ref{sec:phase}), radiation field (Section~\ref{sec:ISRF}), outflow (Section~\ref{sec:outflows}), and chemical (Section~\ref{sec:chemical evolution}) properties of our dwarf galaxy during the 500~Myr after the first new star forms. Unless otherwise noted, $t = 0$ is defined as the time at which that first star particle forms, which is 43~Myr after the actual beginning of the simulation run. The galaxy disk is defined as the fixed physical region within a cylindrical radius of 600~pc and vertical height $|z| < 200$~pc relative to the center of the galaxy. ISM properties are calculated considering only the gas contained within the disk of the galaxy. We include a resolution study in Appendix \ref{appendix:resolution_study}.

Our analysis makes extensive use of the open-source \textsc{yt} toolkit \citep{yt}. All analysis scripts used in this work can be found at https://github.com/aemerick/galaxy-analysis at changeset dd76ad10.

\subsection{Morphological Structure and Evolution}
\label{sec:structure}

\begin{figure*}
\centering
\includegraphics[width=0.975\linewidth]{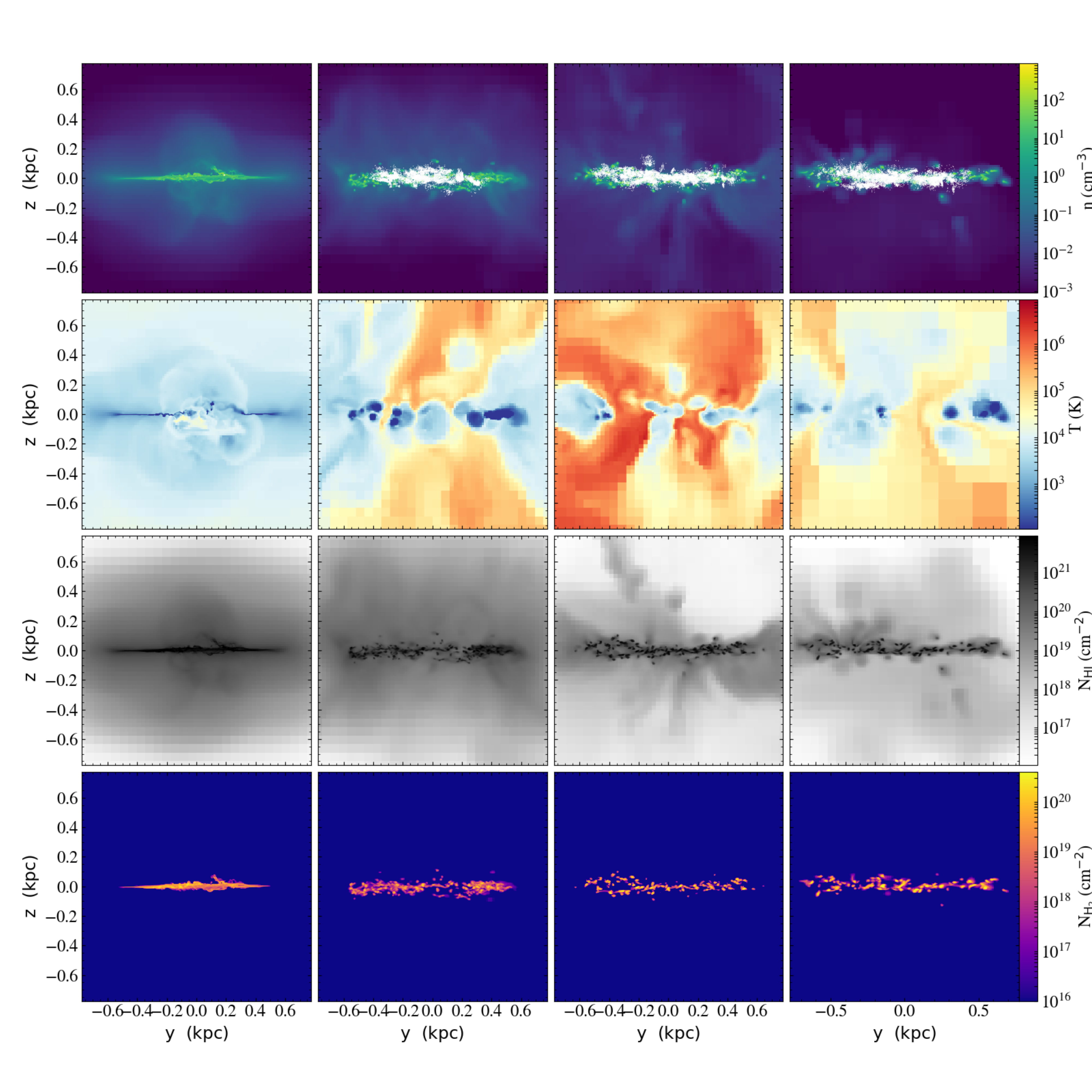}
\caption{Edge-on views of our dwarf galaxy at four different times in its evolution, 0, 150, 300, and 500 Myr after the beginning of star formation. Shown are the density weighted projection of number density (top row), temperature slices (second row), HI column density (third row), and H$_2$ column density (fourth row). Each individual main sequence star particle is shown in the number density projections as a single white dot.}
\label{fig:panel_x}
\end{figure*}

\begin{figure*}
\centering
\includegraphics[width=0.975\linewidth]{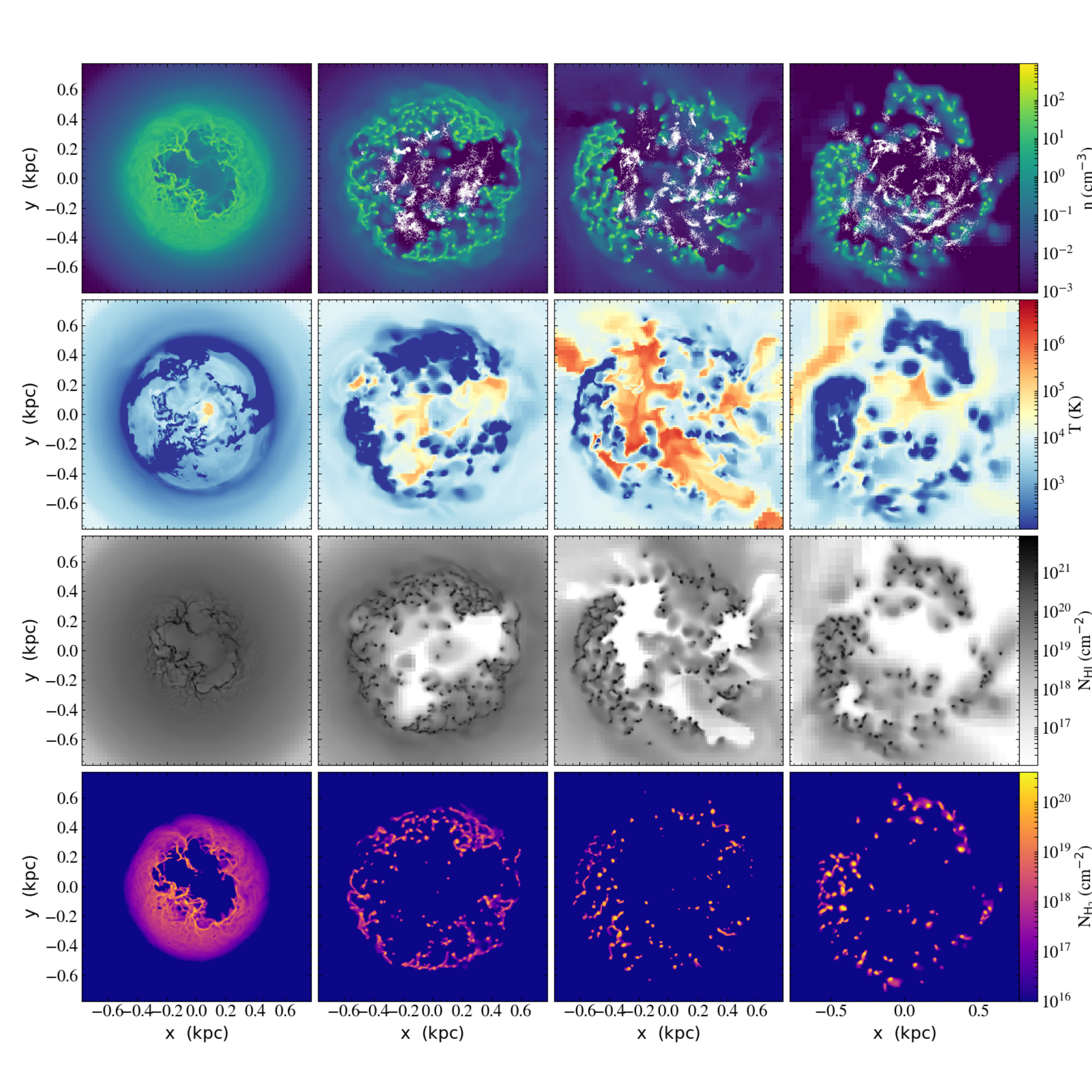}
\caption{Same as Fig.~\ref{fig:panel_x}, but showing face-on views.}
\label{fig:panel_z}
\end{figure*}

We begin by characterizing the morphological properties of our dwarf galaxy, as demonstrated in a series of face-on and edge-on images, presented in Fig.~\ref{fig:panel_x} and Fig.~\ref{fig:panel_z}. These figures show inside-out star formation, as star formation propagates from the inner regions outward during the galaxy's evolution. This is clear in the face-on panels, which demonstrate the growth of the stellar population from the center outward, and the declining gas densities inside-out as a result of stellar feedback driven winds. This central region quickly fills with warm and hot gas generated from radiation feedback and SNe respectively. Both the ISM and the halo gas are multi-phase, containing gas at cold, warm, and hot temperatures with a range of densities, as evident in the temperature slices in both panels. The ISM properties are quantified further in Section~\ref{sec:phase}.

\begin{figure}
\centering
\includegraphics[width=0.975\linewidth]{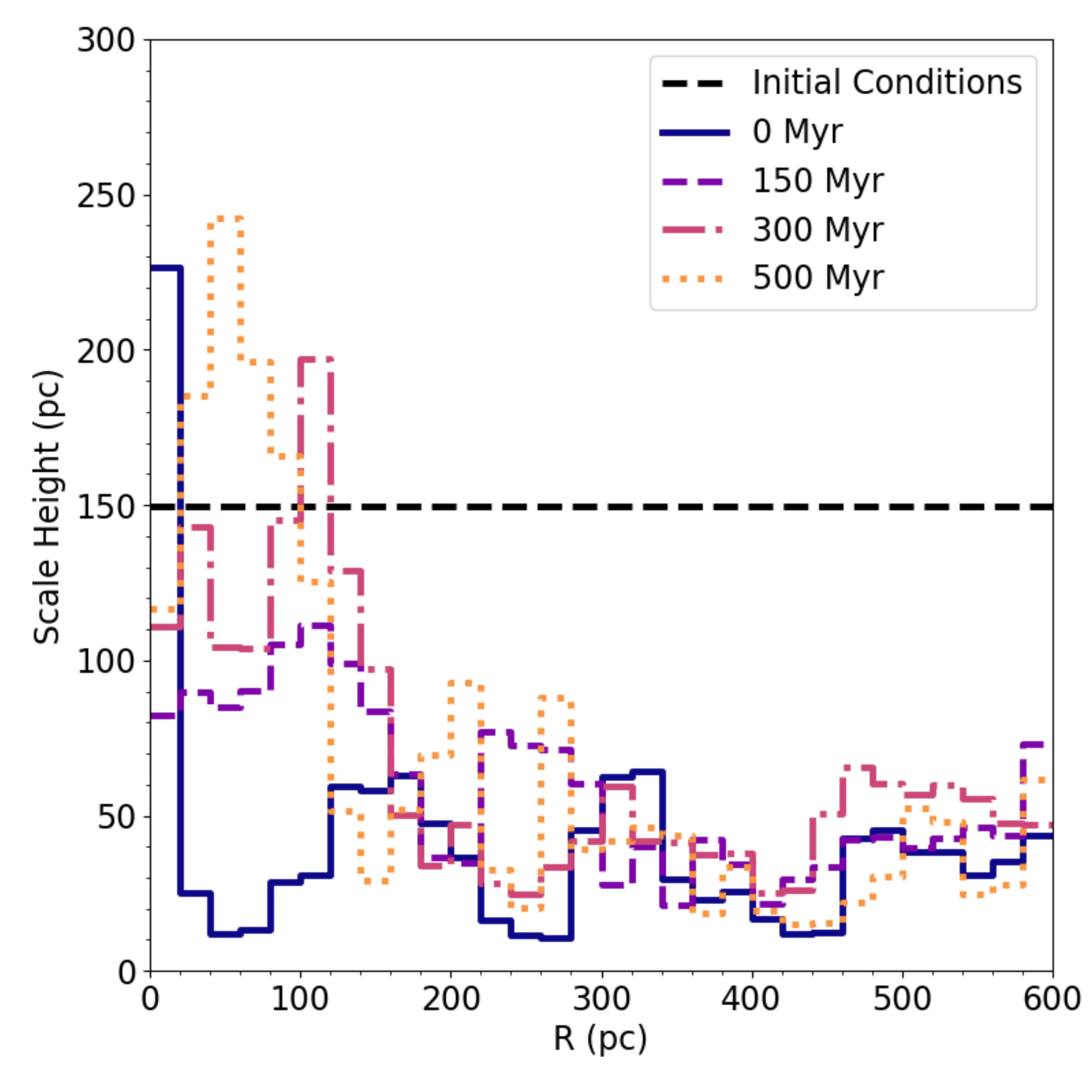}
\caption{The total gas scale height at various times throughout the simulation. These times match the images in Figs.~\ref{fig:panel_x} and ~\ref{fig:panel_z}.}
\label{fig:scale_height}
\end{figure}

The initially puffy gas distribution collapses to a thin disk, with scale heights between 10--30~pc, as shown by the blue line in Fig~\ref{fig:scale_height}. This figure shows the scale height of all gas in the disk, averaged over 20~Myr periods centered on each given time. Stellar feedback acts to heat up this initially thin disk substantially, creating typical scale heights of 50--120~pc. Towards the end of the simulation time, the half-light radius is $391 \pm 19$ pc, where the uncertainty represents the 1$\sigma$ variation in this quantity during the final 20 Myr. Although the disk remains thin beyond the half-light radius, with a scale height of around 50~pc, it is fully resolved at all radii. By the end of the simulation, the majority of the disk has a scale height of $\sim$100~pc. 

Constraining the gas scale height in ultra-faint dwarf galaxies observationally is challenging. For Leo P, located at 1.7~Mpc, HI observations that are capable of detecting the diffuse HI throughout the galaxy have a resolution of 100--200 pc, with higher resolution observations identifying only the densest HI clumps in the galaxy \citep[e.g.][]{Bernstein-Cooper2014}. In the final column of Fig.~\ref{fig:panel_z}, the peak HI column density reaches $N_{\rm HI} = 9.4 \times 10^{21}$~cm$^{-2}$, but is located in dense regions with sizes $<$ 10$\sim$pc. With a resolution of 100~pc, the peak column density in an edge-on view is $N_{\rm HI} = 4.3 \times 10^{20}$~cm$^{-2}$, and $N_{\rm HI} = 2.8 \times 10^{20}$~cm$^{-2}$ in a face-on view. At 500 pc resolution, this corresponds to $N_{\rm HI} = 7.5 \times 10^{19}$~cm$^{-2}$, and $N_{\rm HI} = 6.2 \times 10^{19}$~cm$^{-2}$. These column densities are consistent with the resolution-dependent peak column densities found in the low mass dwarf galaxy sample of \cite{Teich2016}, and consistent with the observed peak column density of Leo P, $N_{\rm HI} = 6.5 \times 10^{20}$~cm$^{-2}$, observed with a spatial resolution of about 33~pc.

\subsection{Star Formation Rate and Mass Evolution}
\label{sec:sfr}

\begin{figure*}
\centering
\includegraphics[width=0.475\linewidth]{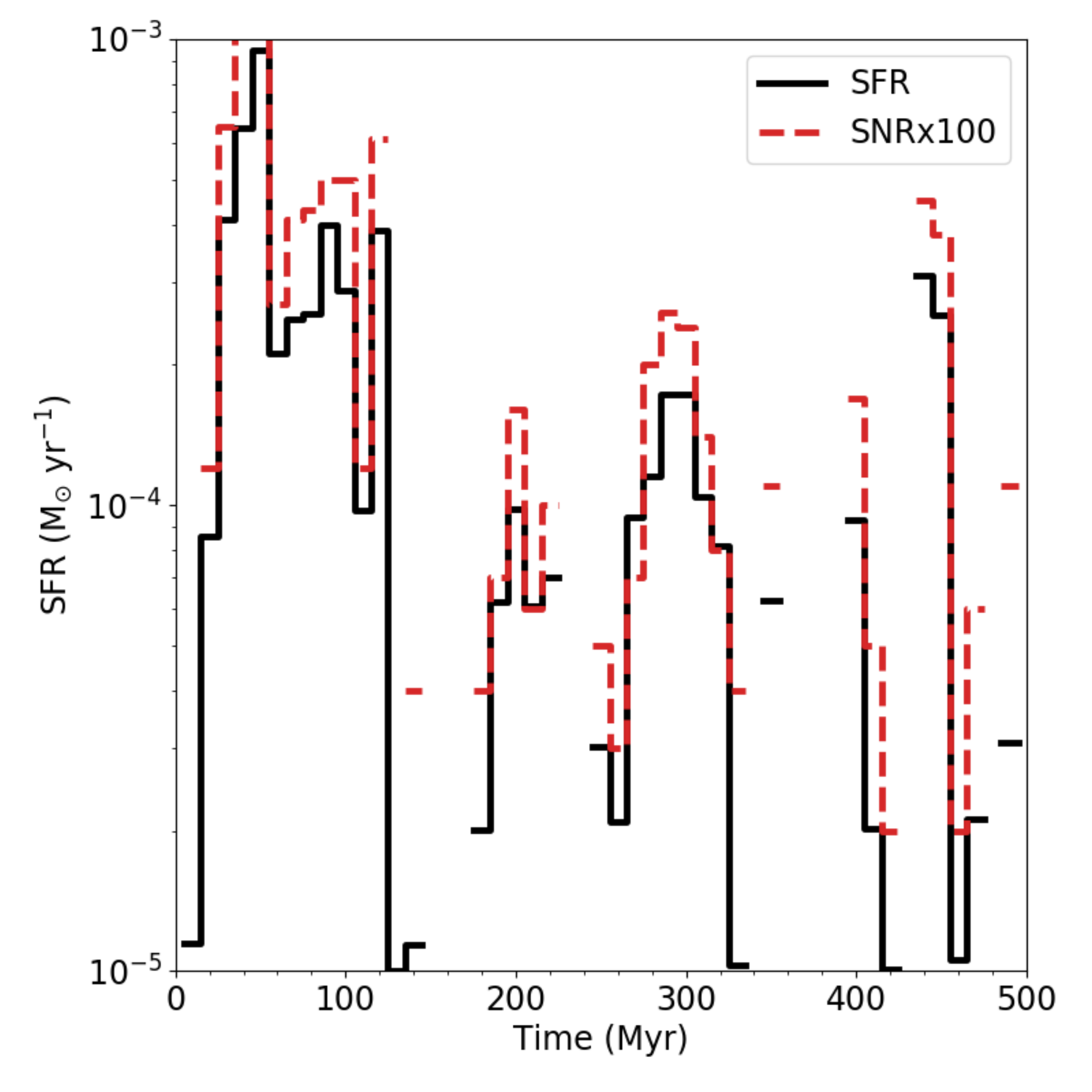}
\includegraphics[width=0.475\linewidth]{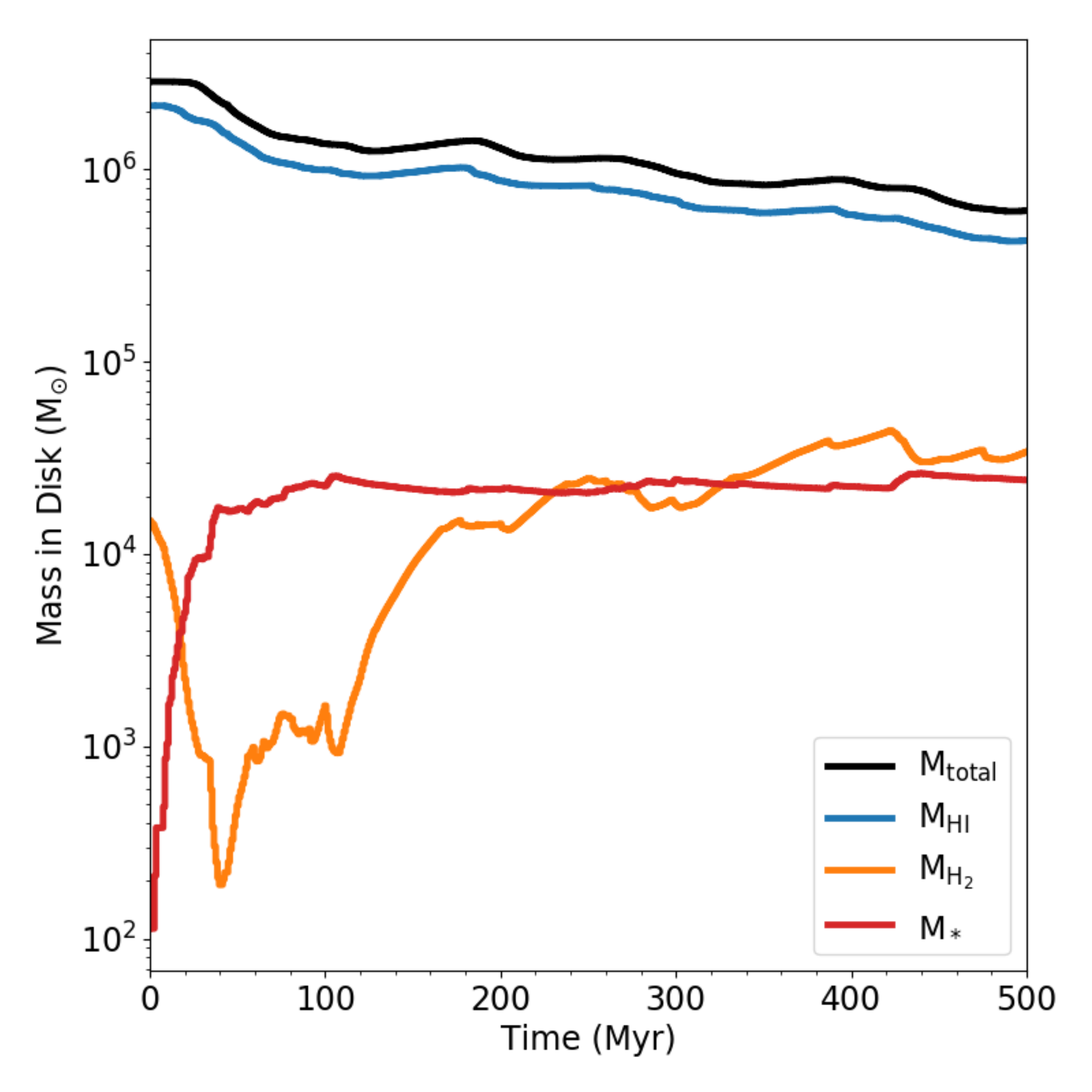}
\caption{Left: The SFR and core collapse SN rate in our dwarf galaxy in 10 Myr bins. Broken portions of this histogram are time periods with no star formation or supernovae.  Note that the SN rate has been scaled by a factor of 100 to fit on the same vertical axis as the SFR. Right: The evolution of the total gas mass (black), HI mass (blue), H$_2$ mass (orange), and stellar mass (red) in the disk of our galaxy over time.}
\label{fig:sfr_mass_evolution}
\end{figure*}

We present the star formation rate (SFR) and core collapse SN rate (SNR) evolution of our dwarf galaxy as measured in 10 Myr bins in the left panel of Fig.~\ref{fig:sfr_mass_evolution}.\footnote{We do not show the Type Ia rate as there have only been 16 by the end of the simulation.} Within the first 50 Myr of evolution the SFR rises quickly to nearly $10^{-3}$ M$_{\odot}$ yr$^{-1}$, declining to $\sim 3 \times 10^{-4}$~M$_{\odot}$~yr$^{-1}$ until a significant drop off at about 130 Myr. The remainder of the evolution is characterized by periods of little to no star formation interspersed with periods of continual, but low star formation around $10^{4}$~M$_{\odot}$~yr$^{-1}$. The SNR tracks the SFR with a time delay, with roughly one core collapse SNe per 100 M$_{\odot}$ of star formation. Averaging over the entire simulation time, we obtain  $<\rm{SFR}> = 1.19 \times 10^{-4}$ M$_{\odot}$ yr$^{-1}$. We discuss how the SFR of this galaxy compares to observed galaxies in Section~\ref{sec:observation}.

We note that the granularity in our star formation algorithm creates a lower limit to the SFR that depends on the period $\Delta t$ over which the SFR is measured. Since we produce stars in $\sim 100$~M$_{\odot}$ sets, the smallest value for our measured SFR is $\sim 100/ \Delta t$. For $\Delta t = 10$ Myr this is 10$^{-5}$ yr$^{-1}$. Removing the granularity requires a fundamental change in our star formation algorithm, likely at the cost of increased complexity and computational expense. Sink particles, which track pre-main sequence stellar mass accumulation, would be the most viable way to do this \citep[see for example ][]{Krumholz2004,Federrath2010,GongOstriker2013,BleulerTeyssier2014,Sormani2017}.

At initialization, all H and He of our dwarf galaxy is neutral, with no molecular hydrogen component. By the time of first star formation ($t=0$ in Fig.~\ref{fig:sfr_mass_evolution}), HI still dominates the mass of the galaxy, with a molecular hydrogen mass fraction of only $\sim$~0.3~\%. The molecular component declines rapidly as this gas is both converted into stars and is destroyed by stellar radiation feedback. For the remainder of the simulation, the H$_2$ mass generally increases,  with small fluctuations during periods of star formation, reaching a peak mass fraction of 5\% at 500~Myr. The growth of the molecular fraction is due in part to a decline in the total gas content of our galaxy from feedback-driven galactic winds. During these outflows, the densest gas, the molecular gas, is preferentially retained over the more diffuse ISM. Examining the molecular properties of the ISM in low mass dwarf galaxies in more detail is a vital avenue of future research, as there are significant observational uncertainties in deriving H$_2$ content of galaxies in this low metallicity regime \citep{Leroy2008,McQuinn2012,Amorin2016}. The molecular properties of our galaxy are discussed further in context with other works in Section~\ref{sec:observation}.

\subsection{ISM Properties}
\label{sec:phase}

Our simulations include sufficient resolution and microphysics to capture a multi-phase medium within the ISM and halo of our simulated galaxy. We define five different gas phases following those defined in \citep{Draine2011}: molecular gas, cold neutral medium (CNM), warm neutral medium (WNM), warm ionized medium (WIM), and hot ionized medium (HIM). We emphasize that the molecular ISM phase is defined as all cells with M$_{\rm H_2}$/M$_{\rm gas} = $~f$_{\rm H_2} > 0.5$, and is thus somewhat different than simply considering the total H$_2$ content. By this definition, although our galaxy certainly contains molecular hydrogen, molecular gas as a phase does not exist; the peak f$_{\rm H_2}$ in any single cell remains below 30\%. See Appendix \ref{appendix:phases} for a quantitative definition of these phases. The properties of these phases are regulated by the complex interplay between cooling, turbulence, self-gravity, and radiative and shock heating from stellar feedback throughout the galaxy's evolution. Here we discuss the thermodynamic properties of the gas within the inner halo of our dwarf galaxy.

\begin{figure}
\includegraphics[width=0.95\linewidth]{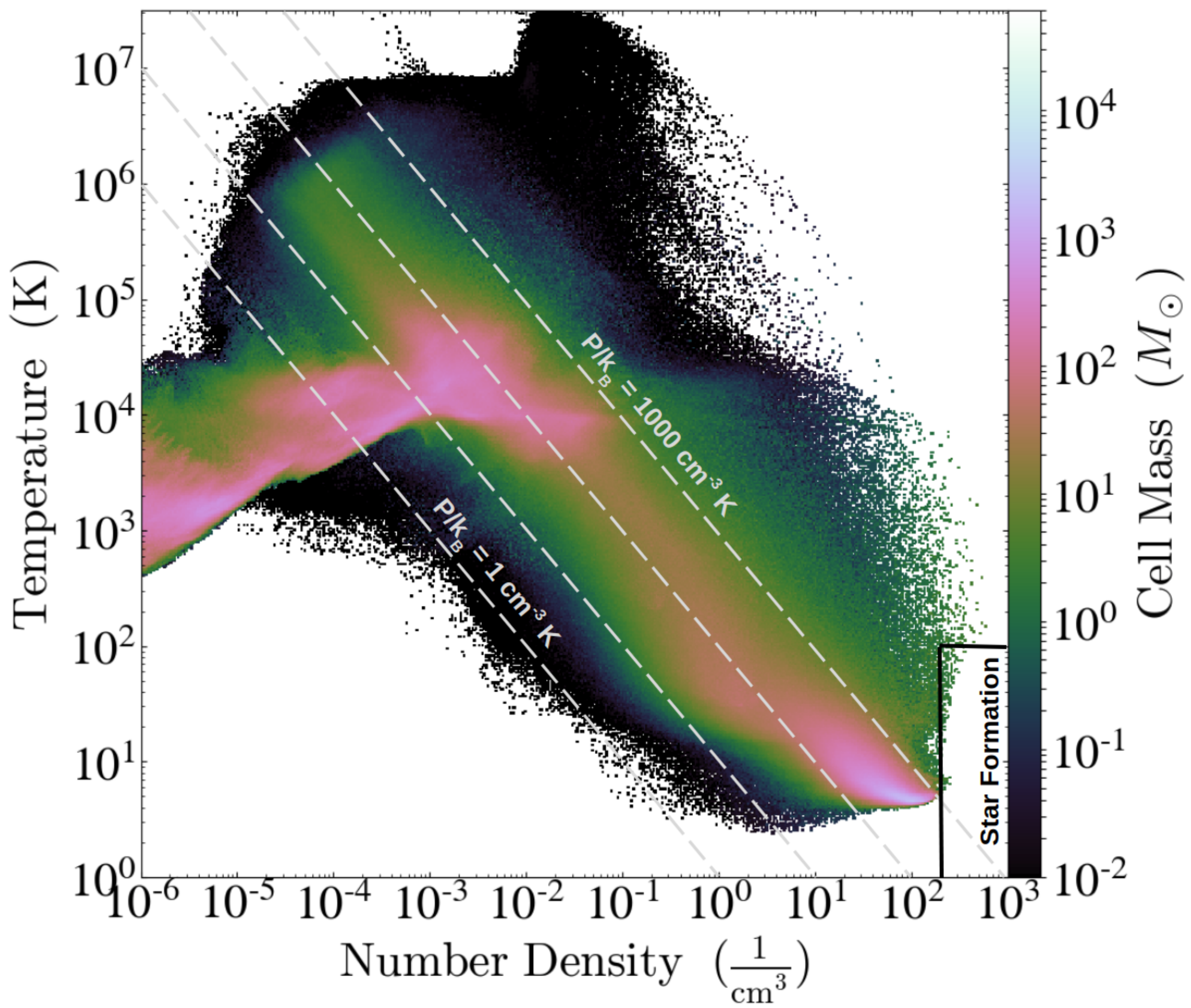}
\caption{The temperature vs.\ number density phase diagram of our dwarf galaxy simulation showing all gas interior to 0.25~$R_{\rm vir}$, averaged over a 50~Myr period from t~=~300~Myr to t~=~350~Myr. The dashed lines are lines of constant pressure, separated by factors of 10. The region in the lower right corner indicates our density and temperature thresholds for star formation.}
\label{fig:phase}
\end{figure}

\begin{figure*}
%\centering
\includegraphics[width=0.45\linewidth]{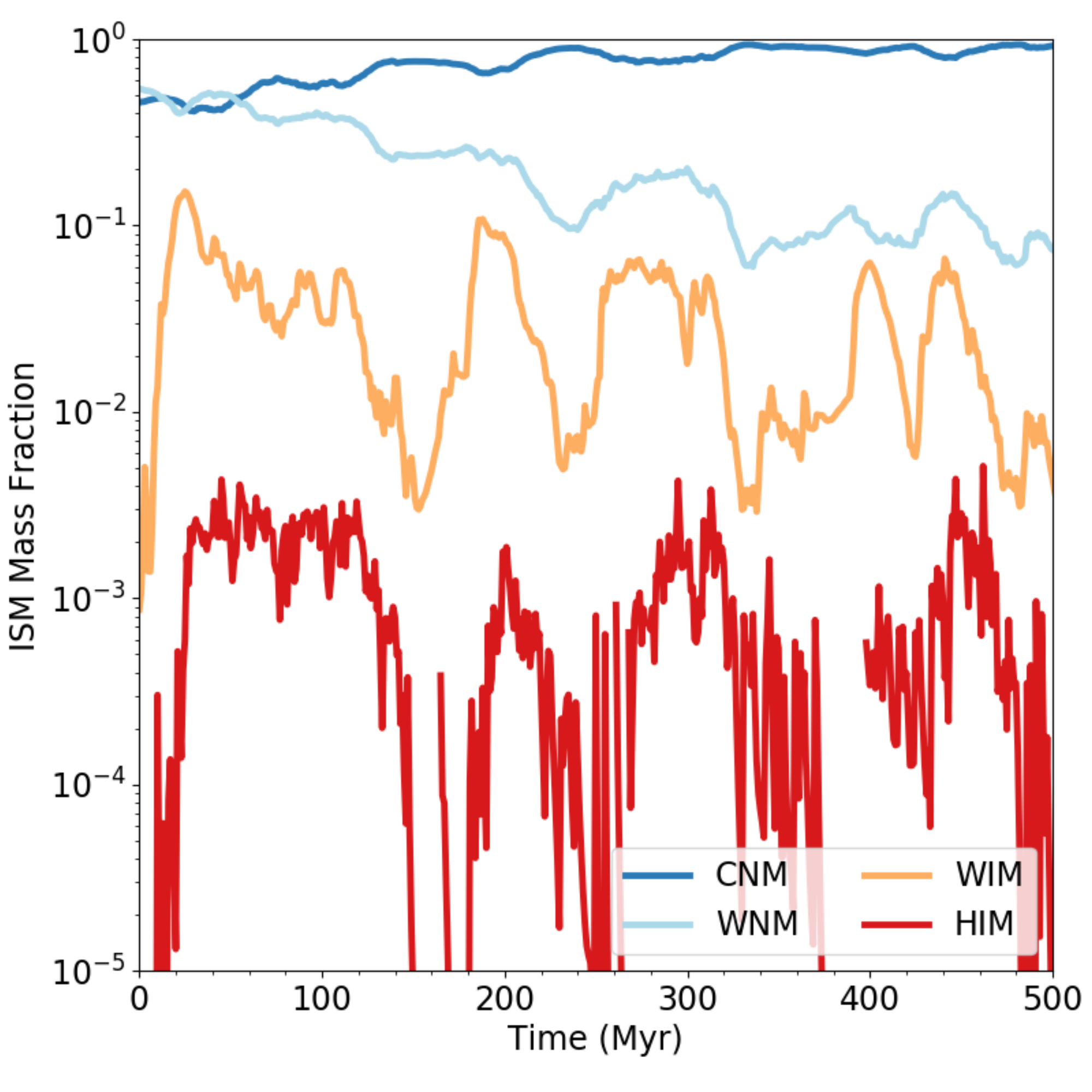}
\includegraphics[width=0.45\linewidth]{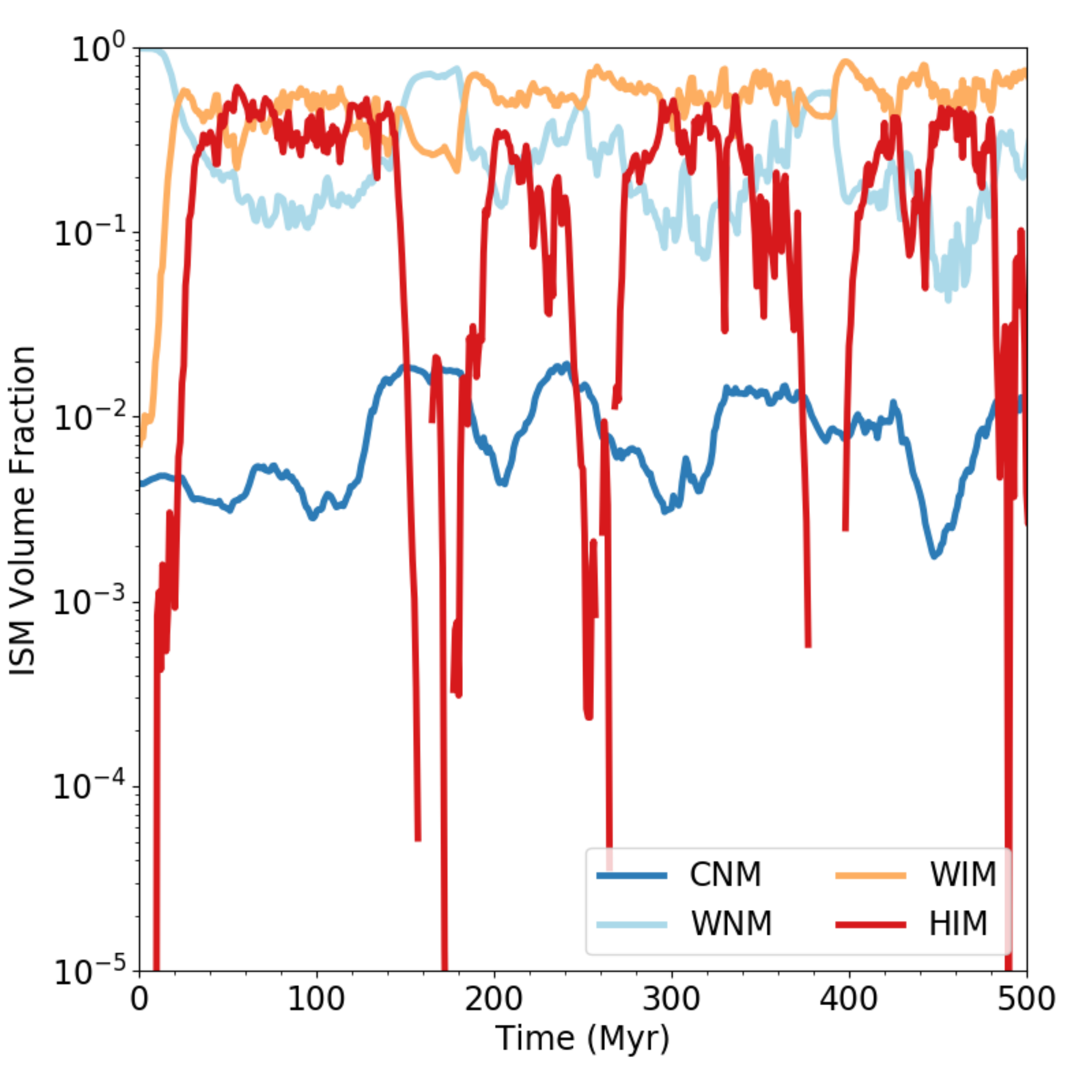}
\caption{The evolution of the mass and volume fractions for each phase of the model galaxy's ISM. See Appendix~\ref{appendix:phases} for definitions of each phase.}
\label{fig:ISM_evolution}
\end{figure*}

Fig.~\ref{fig:phase} shows the temperature-density distribution of all gas within 0.25 R$_{\rm vir}$ of the center of the galaxy, averaged over the time period 300--350~Myr. One can readily identify the two regimes containing most of the mass in the simulation: low density, warm gas produced through ionization and SN heating, and cold, high density gas that makes up most of the mass in the galaxy's disk (see Fig.~\ref{fig:ISM_evolution}). Several notable features of the distribution include: broad ranges of temperature even in quite dense gas, perhaps produced by photoionization and photoelectric heating, a substantial amount of extremely cold gas below 10~K, and the lack of well-defined thermal phases due to the complexity of both the heating and cooling in a turbulent medium. We note that we are likely missing important physics, such as cosmic ray heating and ionization, that would prevent the formation of the coldest gas in this diagram (below about 10~K), but we do not expect this to significantly alter our results. Our artificial temperature ceiling in diffuse gas (see Section~\ref{sec:hydro}) is seen clearly by the horizontal feature in the top left. The boxed regime in the lower right corner shows our star formation density and temperature threshold. Gas in this regime is rapidly consumed by star formation and subsequent feedback. Given the small size of our dwarf galaxy, the total amount of mass in this regime at any given instant can be small, but does appear in this time average. 

The mass of the ISM in our dwarf galaxy is dominated by the CNM for the entirety of the simulation, as shown in the left panel of Fig.~\ref{fig:ISM_evolution}. The mass fraction of the remaining phases are ordered by temperature, with the WNM as the next most-significant component. The WNM is initially comparable to the CNM, but comprises a mass fraction of about 0.1 by the end of the run. The WIM and HIM fluctuate significantly, corresponding to fluctuations in the SFR and associated feedback, but are subdominant throughout the simulation. During periods of peak stellar feedback, however, the WIM can reach a mass fraction above 0.1. Although the CNM dominates the mass fraction, it is a negligible component of the ISM volume, which is WIM dominated. However, the large, anti-correlated fluctuations in the WNM and HIM make these three phases often comparable. Together, these figures better quantify the general properties observed in the panel plots in Fig.~\ref{fig:panel_x} and Fig.~\ref{fig:panel_z}. 

These results are in contrast with those found for the more massive dwarf galaxy modeled by \citet{Hu2016,Hu2017}. They find the mass and volume fraction of the ISM are nearly entirely dominated by warm gas (defined in those works as gas with 100~K$<$T$< 3\times 10^{4}$), with cold gas having between 1 and 10\% of the mass, and occupying negligible volume. Hot gas (defined as gas with $T > 3\times 10^{4}$~K) occupies 10\% of the volume, with negligible mass, in their galaxy, while our WIM alone occupies $>$ 50\% of the volume. Our lower mass, lower metallicity galaxy contains more cold gas (by mass fraction) and hot gas (by volume fraction) that seen in the more massive dwarf galaxy in these works. The driver of these differences, which are likely somewhat related to differences in the dark matter halo potential, will be investigated in future work. We have compared our cooling curves to those used in \citep{Hu2017} and found them to be comparable; though this could contribute to the differences, it is likely not the dominant source.

\subsection{Interstellar Radiation Field}
\label{sec:ISRF}

\begin{figure}
\includegraphics[width=0.95\linewidth]{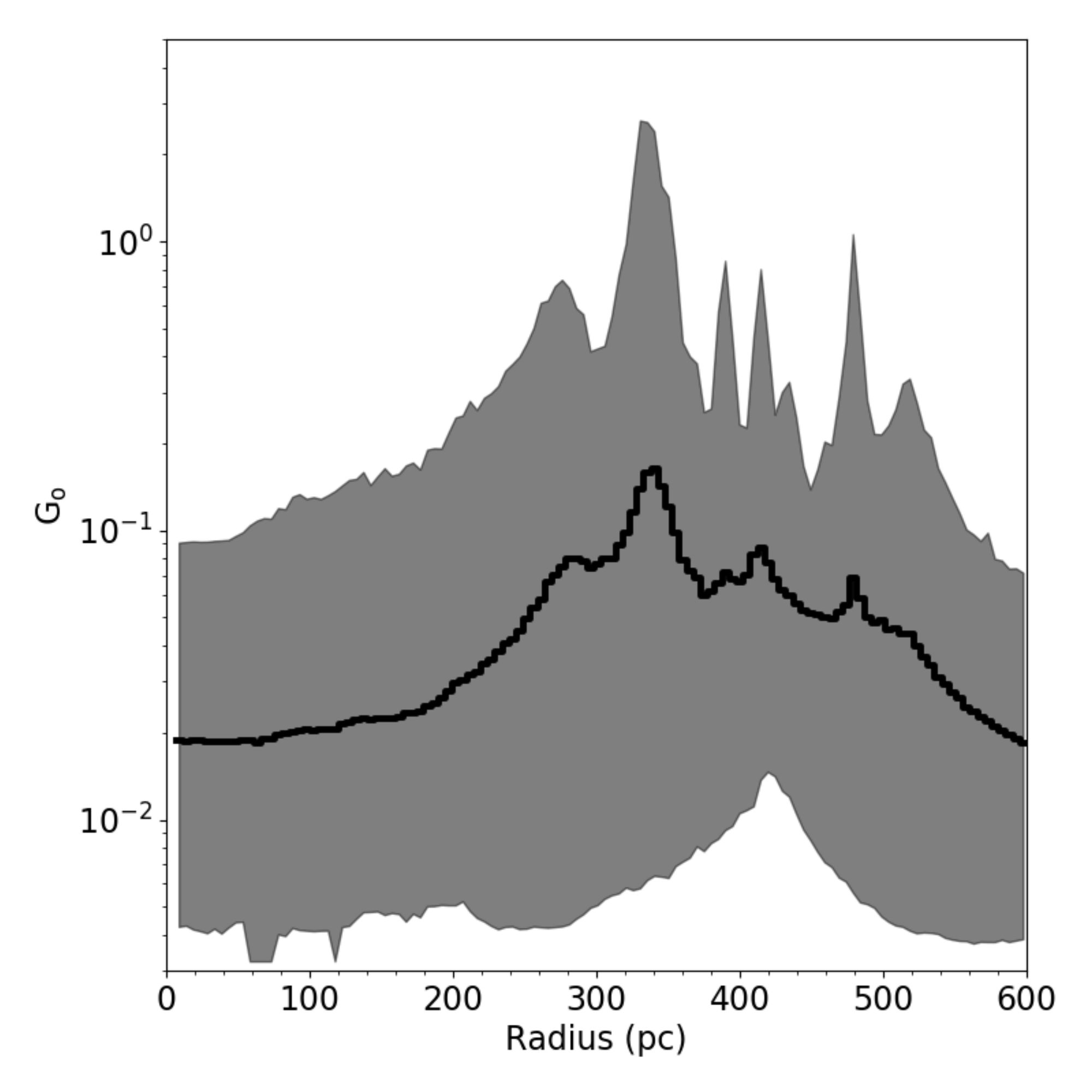} \\
\includegraphics[width=0.95\linewidth]{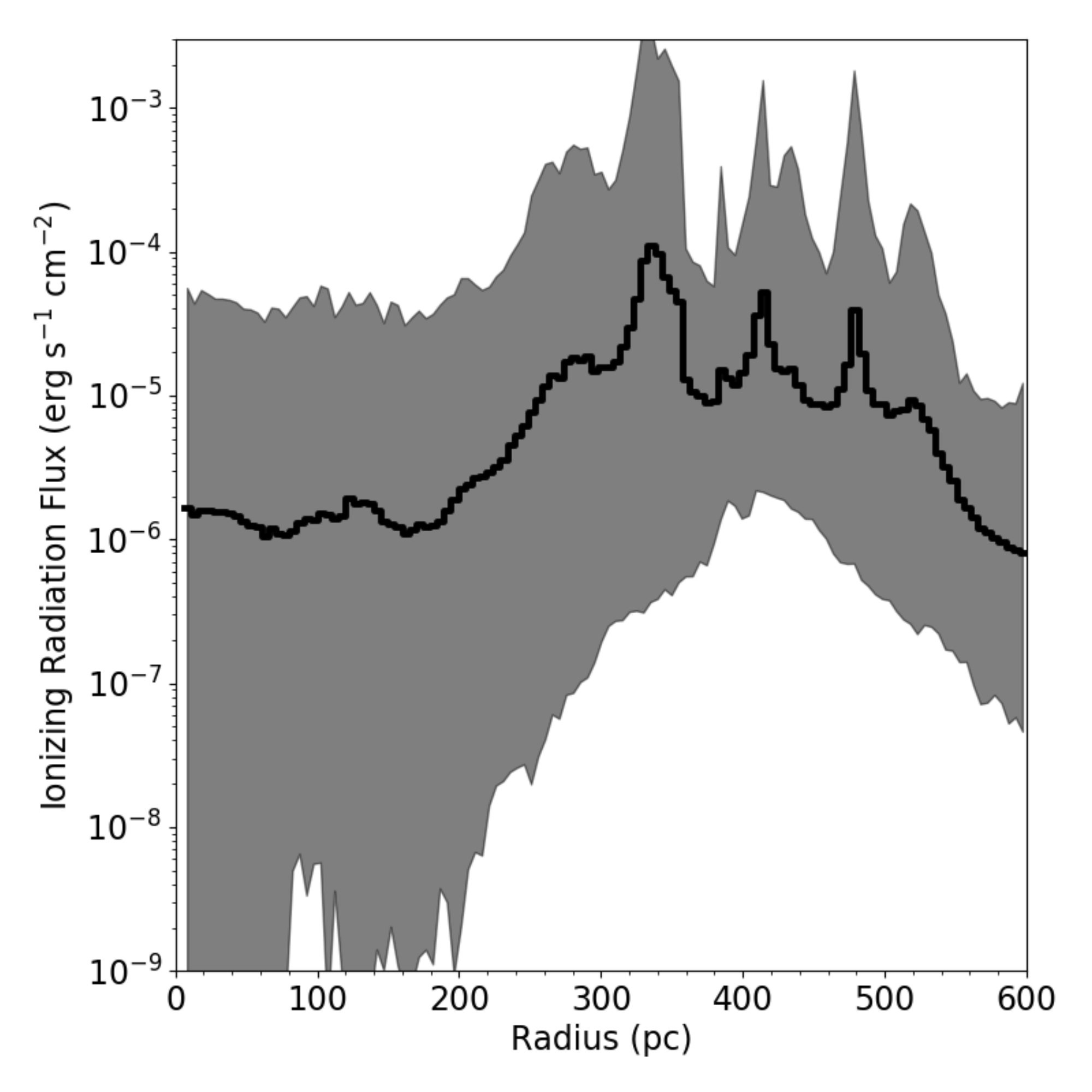}
\caption{
Azimuthally averaged radial profiles of the ISRF in the mid-plane of our galaxy in two different bands, time-averaged over 50 Myr from 300 -- 350~Myr. Here we define the midplane as within 2.5~$dx$ of z = 0, or 4.5~pc. The top panel gives $G_{\rm o}$, the flux of radiation between 6--13.6 eV normalized to the value in the solar neighborhood, shaded between minimum and maximum values at each position, with the average shown as a black line. The bottom panel gives the HI ionizing stellar radiation flux. Since this radiation is tracked directly through radiative transfer, the minimum value at all radii is 0 at some point. For this reason we only shade between the first quartile and maximum values. HeI ionizing radiation is very similar to HI ionizing radiation, with a small vertical offset, and is not shown for clarity. In the top panel, the minimum of the vertical axis is the UVB value of $G_o$.}
\label{fig:ISRF}
\end{figure}

The interstellar radiation field (ISRF) of our galaxy varies dramatically in both space and time, as has been seen previously in works modeling varying radiation fields both as expected from stellar motions in our own galaxy \citep{Parravano2003}, and in models including radiation \citep[e.g.][]{Hu2017}. This is not particularly surprising in our low SFR regime, where there can be large fluctuations over time as individual massive stars form, move about, and evolve. In Fig.~\ref{fig:ISRF} we present azimuthally averaged radial profiles of the ISRF in various bands, time averaged over 100 Myr during the period of star formation from roughly 250--350~Myr. The top panel shows $G_{\rm o}$, the ISRF flux between 6--13.6~eV normalized to the value in the solar neighborhood of the Milky Way (see Sec.~\ref{sec:diffusive heating}). The averaged profile varies between values of 0.02 and 0.1, with peaks located at radii of the few active star formation regions. At any given radius, there is over a two order magnitude variation in the ISRF during this period of time.

The bottom panel gives the HI ionizing photon flux from stellar radiation. The ionizing radiation profile follows a similar trend, yet with significantly more variation, anywhere from two to four orders of magnitude. As this radiation is followed through radiative transfer, the profile encodes information about local attenuation by dense, neutral gas. This is the main driver of the differences between the two panels. The total fluctuation in both panels is due in part to the low-level, stochastic star formation in our galaxy. A higher star formation rate would produce a more regular population of massive stars and more uniform (in time) ISRF.

\begin{figure*}
%\centering
\includegraphics[width=0.45\linewidth]{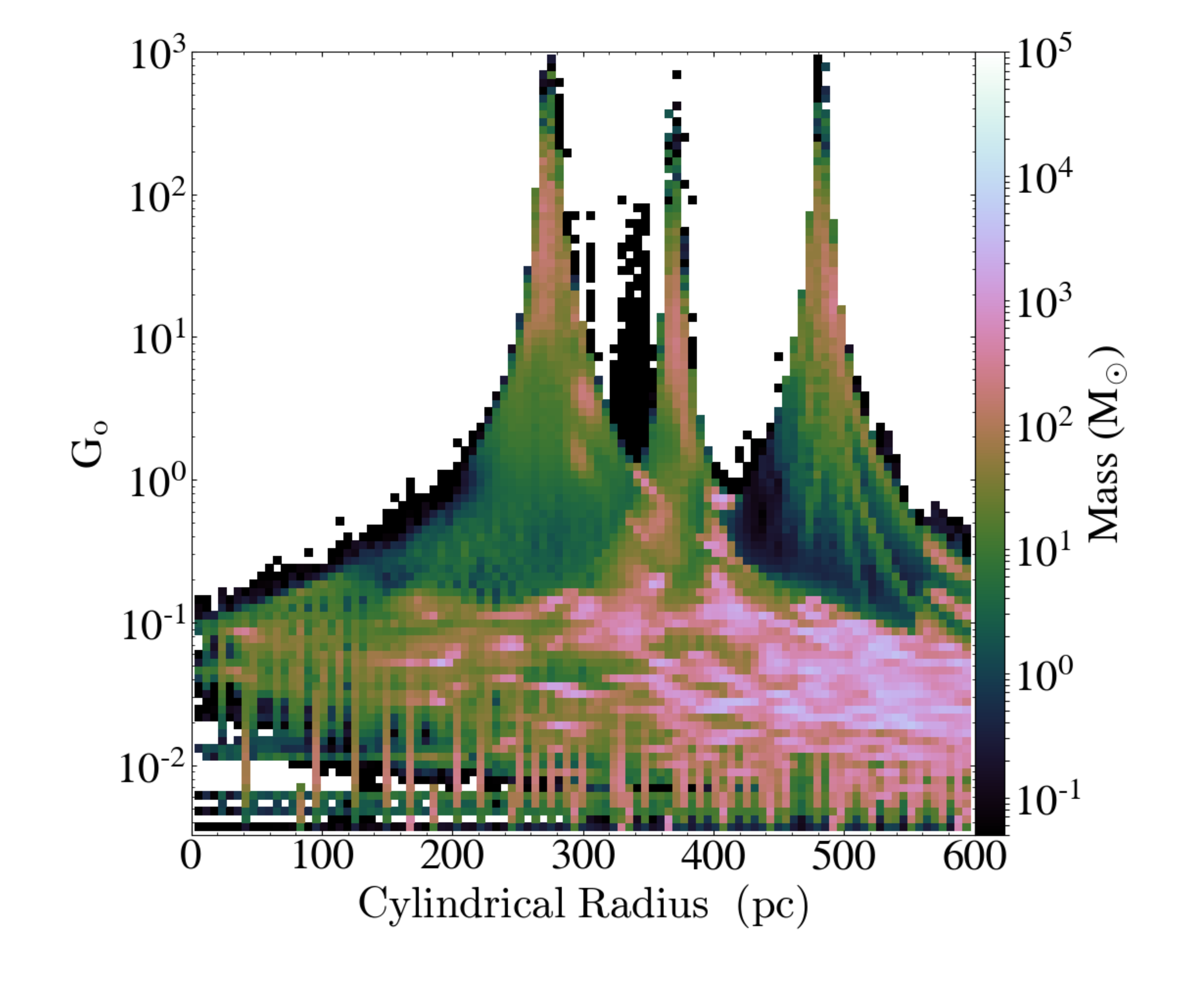}
\includegraphics[width=0.45\linewidth]{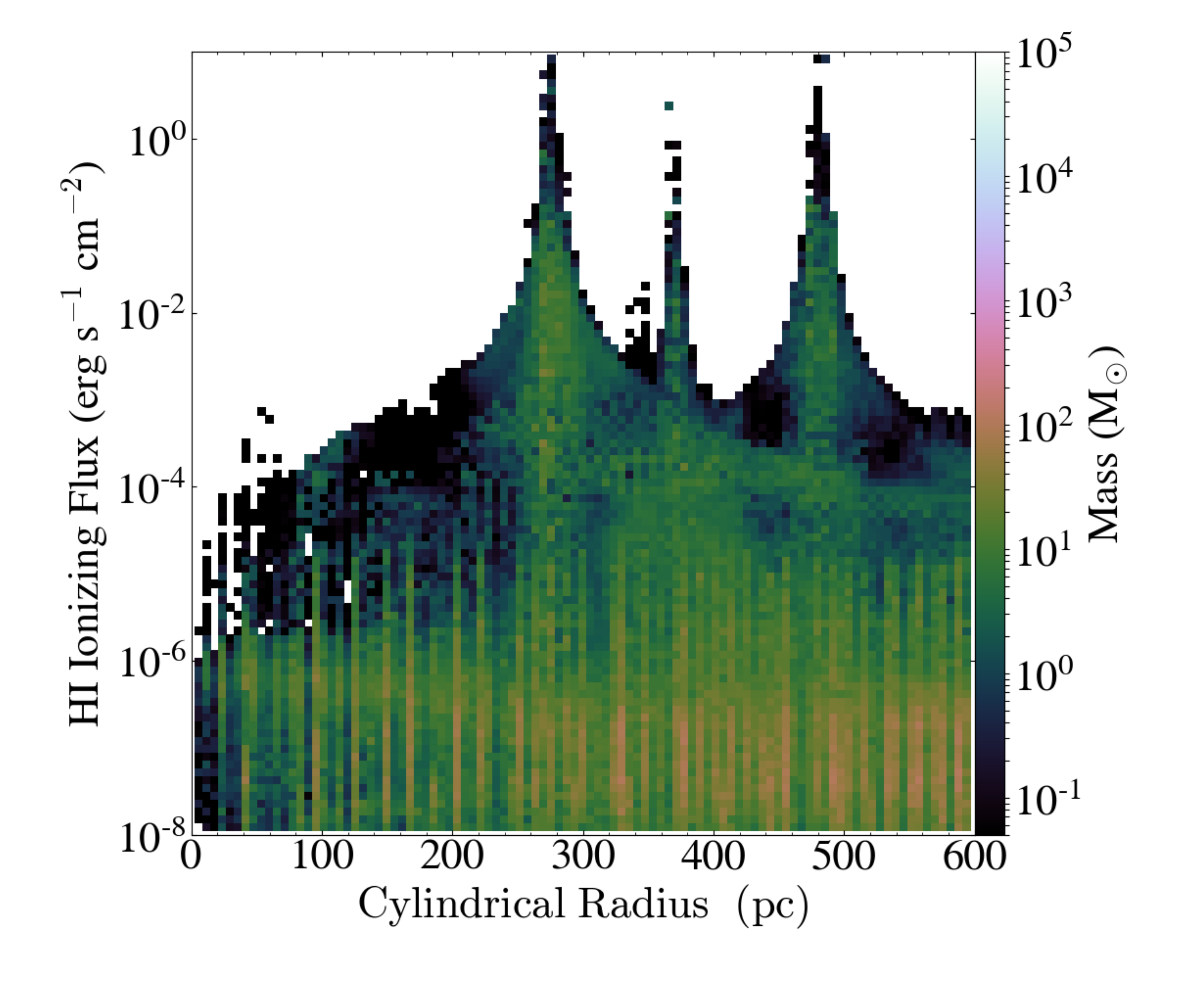}
\caption{Single-snapshot 2D radial profile plots at 300~Myr of the ISRF in two flux bands, $G_{\rm o}$ and HI ionizing radiation, illustrating the full dynamic range of radiation flux at a given radius in the galaxy. Here, we include all gas within the mid-plane of our dwarf galaxy. Since a majority of the mass of the galaxy is in the cold phase (see Fig.~\ref{fig:phase}), and is therefore optically thick to HI ionizing radiation, it does not show up in the HI ionizing radiation diagram. This gas readily appears in the $G_{\rm o}$ diagram since we assume it to be optically thin, though we do apply a localized shielding approximation.}
\label{fig:ISRF_2D}
\end{figure*}

To further quantify the local variations in these radiation fields, we present the full distribution of $G_{\rm o}$ and the HI ionizing flux in Fig.~\ref{fig:ISRF_2D} at a single snapshot at 300~Myr. This diagram shows how dramatic the increase in ISRF near young, massive stars is (the spikes in both diagrams), while much of the mid-plane sees an ISRF orders of magnitude lower. The striking contrast between the two diagrams is due to the shielding of the HI ionizing flux in the most massive (cold and dense) regions of the galaxy through the radiative transfer calculations; shielding of the FUV radiation is approximate and in general weaker, making these regions more prominent in the left hand figure (the pink/white clumps). From both of these diagrams, it is clear that the ISRF of a low mass dwarf galaxy varies greatly over time and space in a way that cannot be appropriately captured by an analytic profile. Although one could adopt an averaged radial profile to provide a realistic, global source of energy for thermal pressure support of the gas against collapse, it is unclear how sufficient this would be in suppressing star formation. In particular, the large increases around sites of recent star formation could be important sources of feedback to destroy molecular clouds and reduce their effective star formation efficiency. It remains to be seen which of these two modes of feedback is more important in regulating star formation.

\subsection{Outflow Properties}
\label{sec:outflows}

\begin{figure*}
\includegraphics[width=0.45\linewidth]{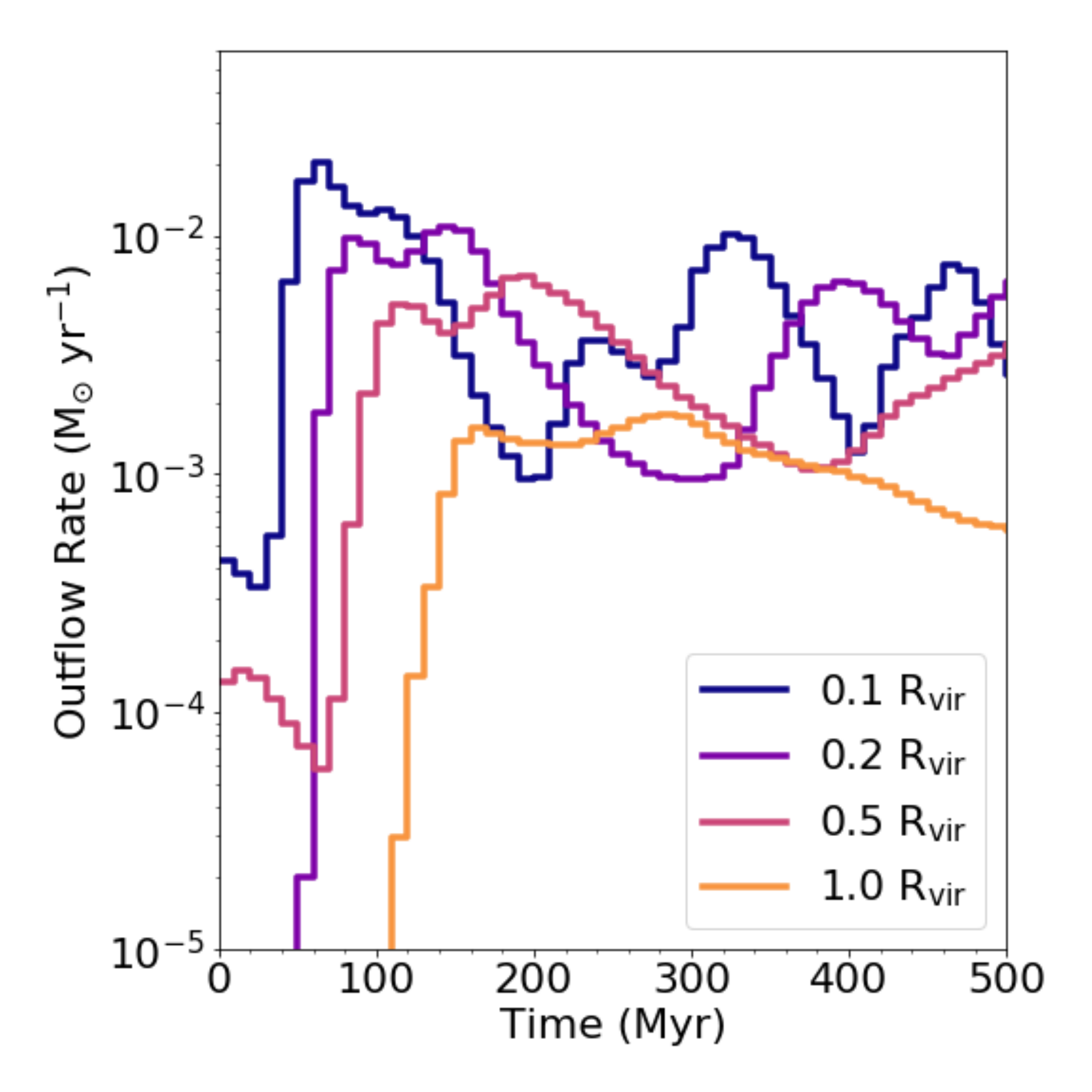}
\includegraphics[width=0.45\linewidth]{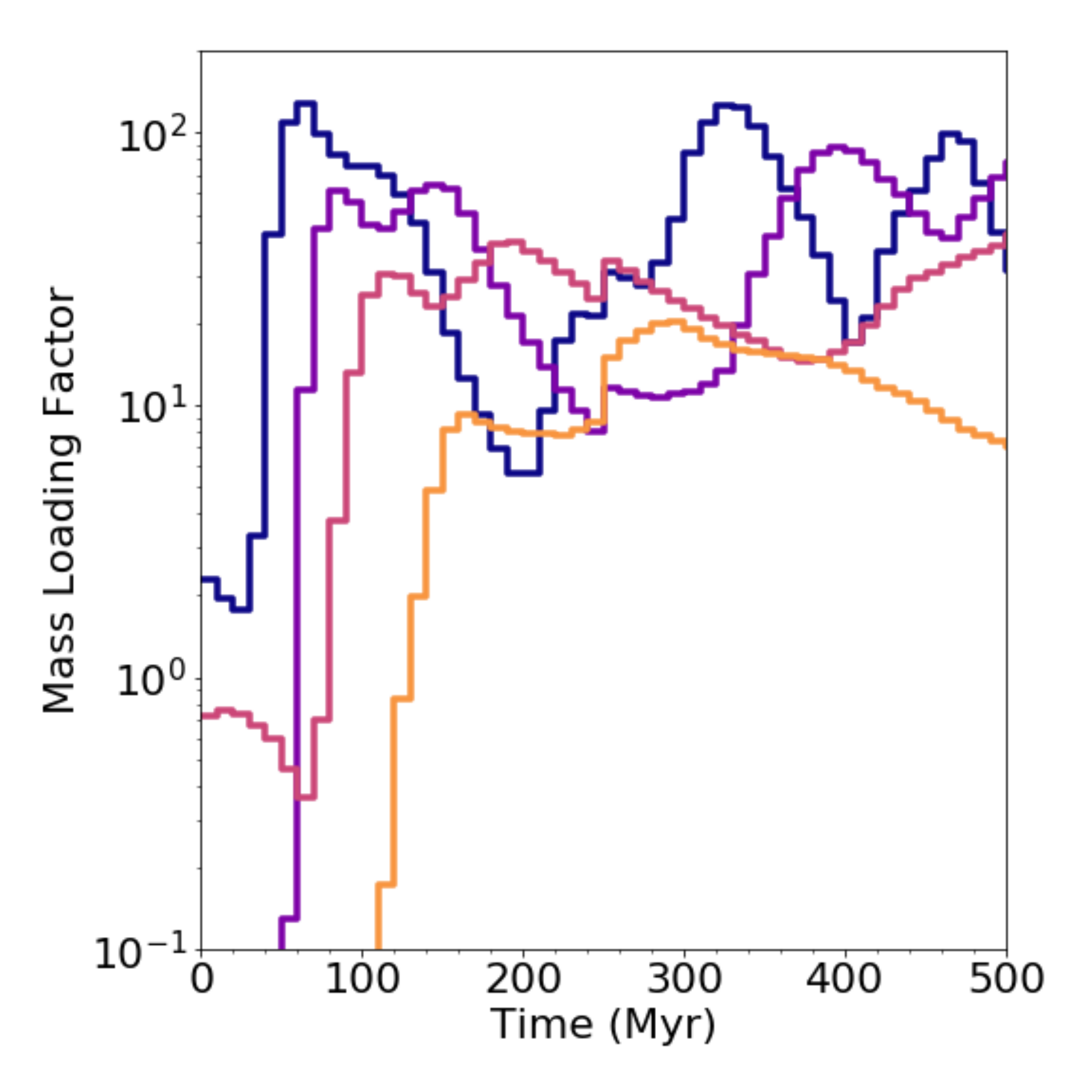}
\caption{Spherical mass outflow rates (Eq.~[\ref{eq:dotM}]) and mass loading rates over time at 4 different radii from the galaxy.}
\label{fig:mass_outflow}
\end{figure*}

The recent FIRE cosmological simulations of dwarf galaxies over a range of dark matter halo masses find that they exhibit large outflows, with mass loading factors ($\eta = \dot{M}_{\rm out}/<\rm{SFR}>$) on order of 100--1000 \citep{Muratov2015}. However, comparable models of idealized dwarf galaxies with detailed feedback and physics treatments find more modest mass loading factors \citep{Hu2016,Hu2017}. In Fig.~\ref{fig:mass_outflow} we present the mass outflow and mass loading rates for our dwarf galaxy as a function of time, computed at five different positions from the galaxy. We follow \citet{Muratov2015} in defining the mass outflow rate at any given radius to be the sum of the outflow rate in all cells in a spherical annulus of width $dL$ centered at that radius,
\begin{equation} \label{eq:dotM}
\dot{M}_{\rm out} = \sum M_{\rm gas} \times v_{\rm r} / dL.
\end{equation} 
We choose $dL = 0.1~R_{\rm vir}$, or 2.74 kpc. 

The total mass outflow rates and mass loading factors at 0.1, 0.25, 0.5, and 1.0 $R_{\rm vir}$ are shown in Fig.~\ref{fig:mass_outflow}. Generally, other works use gigayear timescale measurements of the SFR to compute the mass loading factor. For consistency with those works, we use the 500~Myr average SFR for computing the mass loading factor. The outflow rate at 0.1 $R_{\rm vir}$ is high, corresponding to mass loading factors between 20--100 throughout the simulation time. This declines towards larger radii, however, with substantially less outflow past the virial radius. \citet{Muratov2015} finds typical mass loading factors at 0.25 R$_{\rm vir}$ on order of 20--40 for galaxies with $v_{c} = 30$ km s$^{-1}$ at low redshift, consistent with our results. The fluctuations in both of these panels are directly correlated with the SFR, with increased outflow during periods of star formation, and decreased outflow during periods of quiescence.

Interestingly, the v$_{\rm c}\sim$ 30 km s$^{-1}$ halos examined in \citet{Muratov2015} are more massive than the M$_{\rm vir} = 2.5\times 10^9$ M$_{\odot}$ halo examined here by a factor of a few. Using a fit provided in \citet{Muratov2015} to extrapolate and compare $\eta$ at fixed halo mass, one would expect mass loading factors on order of 100 at 0.25 R$_{\rm vir}$ for our dwarf galaxy, a factor of a few higher than what we find. These differences could be attributed to our lack of cosmological evolution in these isolated simulations, but ultimately requires a larger set of dwarf galaxy simulations to make a more robust comparison. We note, however, that our results are closer to the \citet{Muratov2015} results than those in \citet{Hu2016,Hu2017}, which find lower mass loading factors even closer to the disk, at 0.05 $R_{\rm vir}$, between 1 and 10 for a dwarf galaxy with M$_{\rm vir} = 10^{10}$ M$_{\odot}$; certainly this implies even smaller mass loading factors at 0.25 $R_{\rm vir}$.

\begin{figure}
\includegraphics[width=0.95\linewidth]{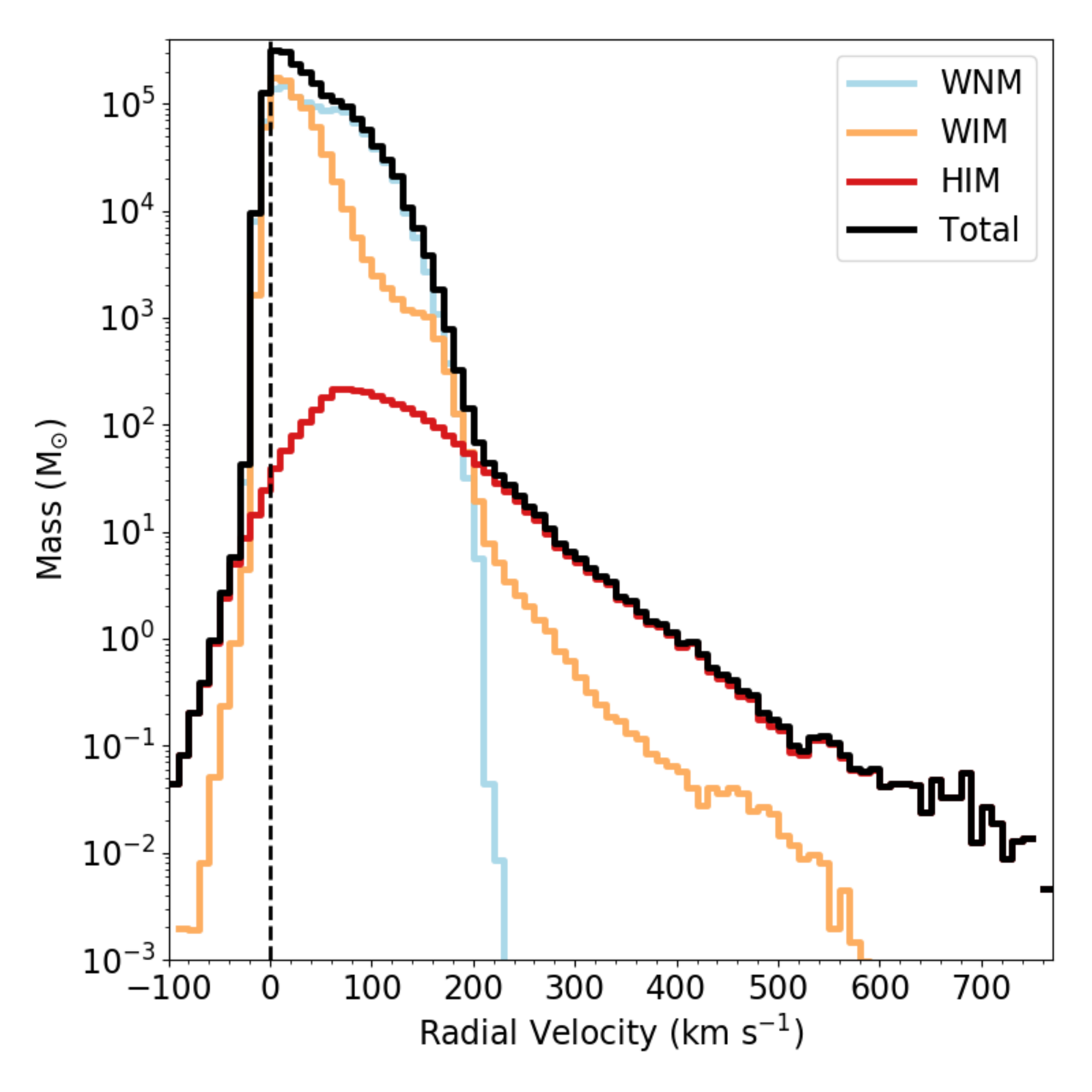}
\caption{The time averaged radial velocity distribution of outflowing material external to the disk and within the virial radius of our dwarf galaxy. This is averaged over the same time interval as Fig.~\ref{fig:ISRF}. The outflowing material is multiphase, broken into WNM, WIM, and HIM. See Section~\ref{sec:phase} for definitions of these regimes. We note that WNM is often labeled simply as ``cold'', or gas below $T = 10^{4}$~K. There is little to no outflowing mass in the CNM.}
\label{fig:outflow_velocity}
\end{figure}

Detailed outflow properties, beyond outflow rates and mass loading factors, can help discriminate between the model dependent feedback physics included in galaxy simulations. In Fig.~\ref{fig:outflow_velocity} we present radial velocity distributions of all material outside our dwarf galaxy's disk, and within the halo, broken into three gas phases. Gas with a negative velocity is moving towards the center of the halo. Roughly 25\% of this mass is inflowing, mostly with modest negative velocities, and corresponds to previously ejected gas mixing and recycling throughout the halo. Half of the outflowing gas (positive velocities) is moving at velocities below 30 km s$^{-1}$, 75\% at velocities below 70 km s$^{-1}$, and 95\% at velocities below 100 km s$^{-1}$. Although the mass contained in the tails of these distributions is a sub-dominant fraction of the total, there is still a non-negligible amount of gas moving at velocities of a few hundreds of km/s, with a peak velocity of over 700 km s$^{-1}$. The WNM and WIM together dominate the mass of both the inflowing and outflowing gas, with the WIM and HIM dominating at velocities above 200 km s$^{-1}$. The dominant launching mechanism in this simulation is SN feedback, which generates a rapidly moving and volume-filling WIM and HIM, consistent with the results in \citet{Hu2016,Hu2017}. However, as shown, the HIM, which is mostly the SN ejecta itself, comprises very little of the outflow by mass. Most of the outflowing gas (by mass) comes from the warm phase, pushed out by the high pressure, fast moving HIM. Some of this warm gas certainly originates from adiabatically and radiatively cooled HIM, however. The amount of transfer between phases in the halo of our galaxy will be investigated in future work.

\subsection{Metal Evolution} 
\label{sec:chemical evolution}

\subsubsection{Metal Enriched Outflows}
\begin{figure}
\includegraphics[width=0.9\linewidth]{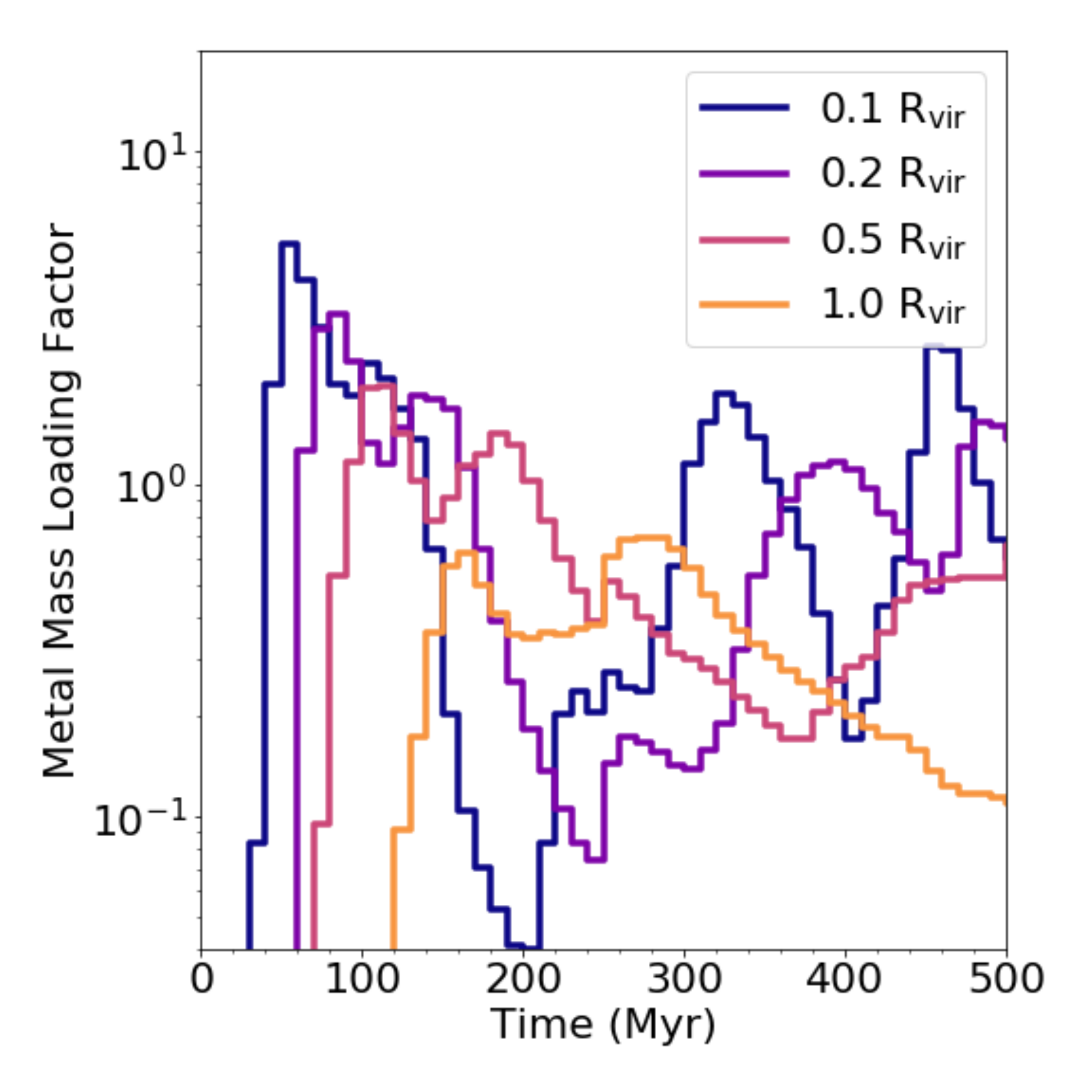} \\
\includegraphics[width=0.9\linewidth]{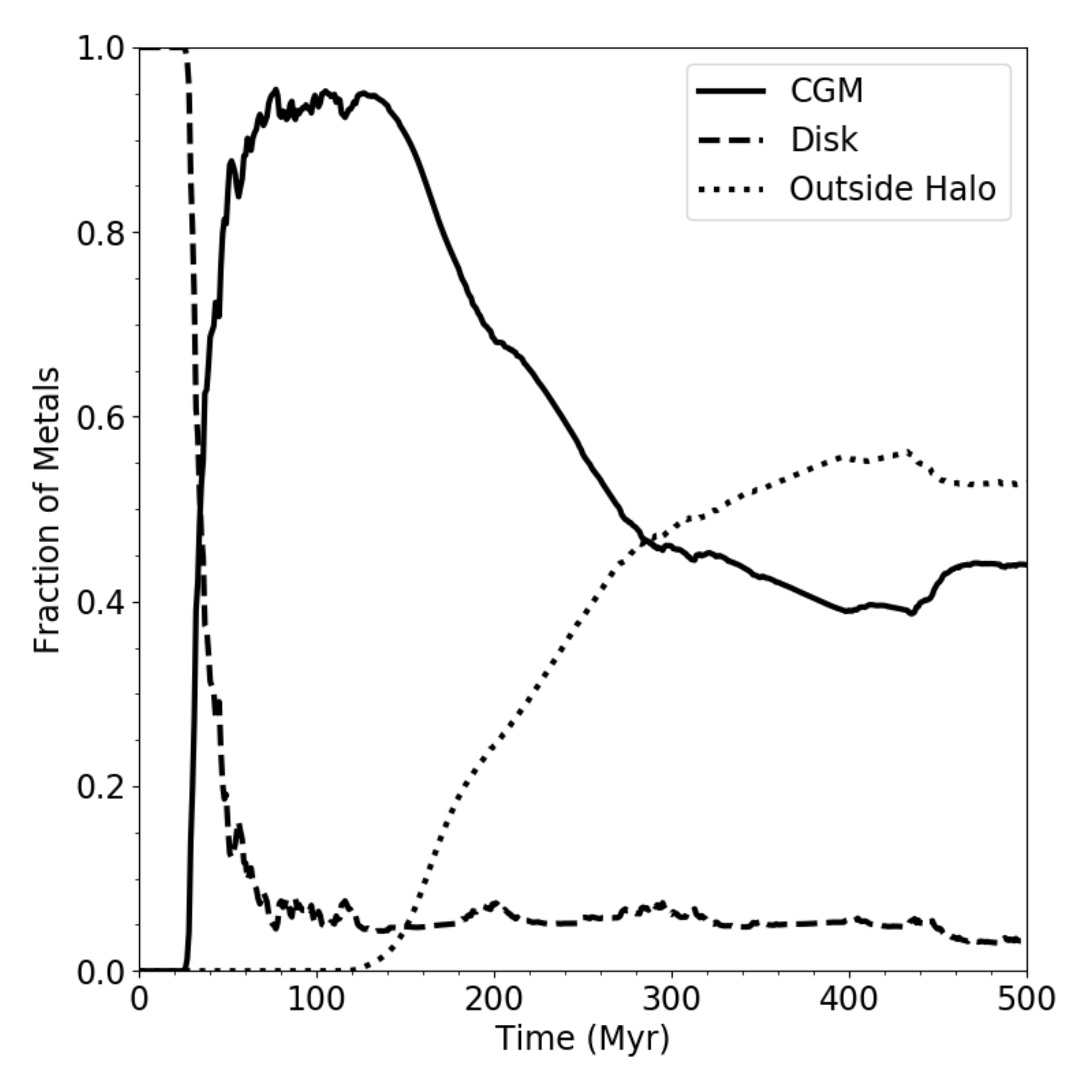}
\caption{{\bf Top}: Metal mass loading factor (see Eq.[~\ref{eq:eta-metal}]) at the same radii as in Fig.~\ref{fig:mass_outflow}. This is the ratio between the metal outflow rate and the metal production rate. {\bf Bottom}: The fraction of metals contained in the disk, CGM, and outside the halo of our dwarf galaxy over time. In both panels, we consider the total mass of all individually tracked metal species, which is zero at initialization, not the aggregate total metallicity field, which is non-zero at initialization.}
\label{fig:metal_evolution}
\end{figure}

Dwarf galaxies efficiently, and preferentially, eject metals released in stellar feedback from their shallow potential wells \citep{MacLowFerrara1999,FerraraTolstoy2000}. This has been better quantified recently both observationally \citep[e.g.][]{Kirby2011-metals,Zahid2012,Peeples2014,McQuinn2015} and with more detailed cosmological simulations \citep{Simpson2013,Angles-Alcazar2017,Muratov2017}. In the top panel of Fig.~\ref{fig:metal_evolution}, we give the metal mass loading factor for our galaxy over time, at the same radii as in Fig.~\ref{fig:mass_outflow}. The parameter used to quantify metal outflow efficiencies varies among works. Here, we define the metal mass loading factor as the metal outflow rate divided by the metal production rate, or
\begin{equation} \label{eq:eta-metal}
\eta_{\rm metal} = \frac{\dot{M}_{\rm metal}}{\rm{SFR} \times (M_{\rm metal}/M_{*})},
\end{equation}
where $\dot{M}_{\rm metal}$ is the metal mass outflow rate, $M_{\rm metal}$ is the total mass in metals produced, and $M_{*}$ is the total mass in stars. These metal loading factors fluctuate significantly with the SFR, just as was shown in Fig.~\ref{fig:mass_outflow}, reaching a minimum of about 0.05, but peaking at around 5. On average, over the simulation time, $\eta_{\rm metal}$ is below unity (around 0.5). Recent simulations of outflows from a Milky Way type disk indicate typical $\eta_{\rm metal}$ comparable to our results, usually between 0.5 and 1 \citep{Li2017,Fielding2017}. \cite{Muratov2017} computes a slightly different quantity for their galaxies, the normalized metal outflow rate $\eta_{Z} = \dot{M}_{\rm metal}/{\rm SFR}$, finding values of about 0.02 at 0.25~$R_{\rm vir}$ regardless of galaxy circular velocity. Our galaxy is consistent with this value, with an average $\eta_Z = 0.015$, fluctuating between 0.007 and 0.02.

These large metal mass loading factors indicate that a majority of the metals produced in our dwarf galaxy are ejected from the disk. This is quantified in the lower panel of Fig.~\ref{fig:metal_evolution}, where we show the mass fraction of metals in the disk, circumgalactic medium (CGM), and outside the virial radius of our galaxy over time. After the first 20 Myr, SN-driven winds rapidly drive out large quantities of metals from the disk and into the galaxy's halo. This continues throughout the simulation, with only $\sim$4\% of produced metals residing in the disk of the galaxy. It only takes 150 Myr for some metals to reach the virial radius of the halo, with a steadily increasing fraction continually leaving the virial radius until about 350 Myr where the fraction levels off to just above 50\%. Likewise, the CGM metal content continually decreases until the end of the simulation from loss through the virial radius of the halo. See Section~\ref{sec:obs_metals} for further discussion.

\subsubsection{Differential Evolution of Elements Within the ISM}
It is important to understand how metals from each source of stellar yields enrich the ISM. Observations of more massive dwarf galaxies than those simulated here indicate fairly uniform radial gas-phase metallicity profiles, even beyond the stellar radius \citep[e.g.][]{Werk2011,Belfiore2017}. This requires that metal mixing and transport occur on hundred megayear timescales, much more rapidly than the gigayear timescale expected from assuming transport at the cold gas sound speed. Therefore, either metals are transported first through a hot phase with high sound speed, or through efficient turbulent mixing within the ISM \citep[e.g.][]{Avillez2002,Tassis2008,YangKrumholz2012}. It remains uncertain how metal abundances vary in detail within these galaxies, beyond one-dimensional radial profiles, and whether or not abundance distributions depend on the metal species. It is even more unclear how metals are transported and distributed within low-mass dwarf galaxies, which generally host too few \ion{H}{2} regions for a detailed examination.

\begin{figure*}
\includegraphics[width=0.98\linewidth]{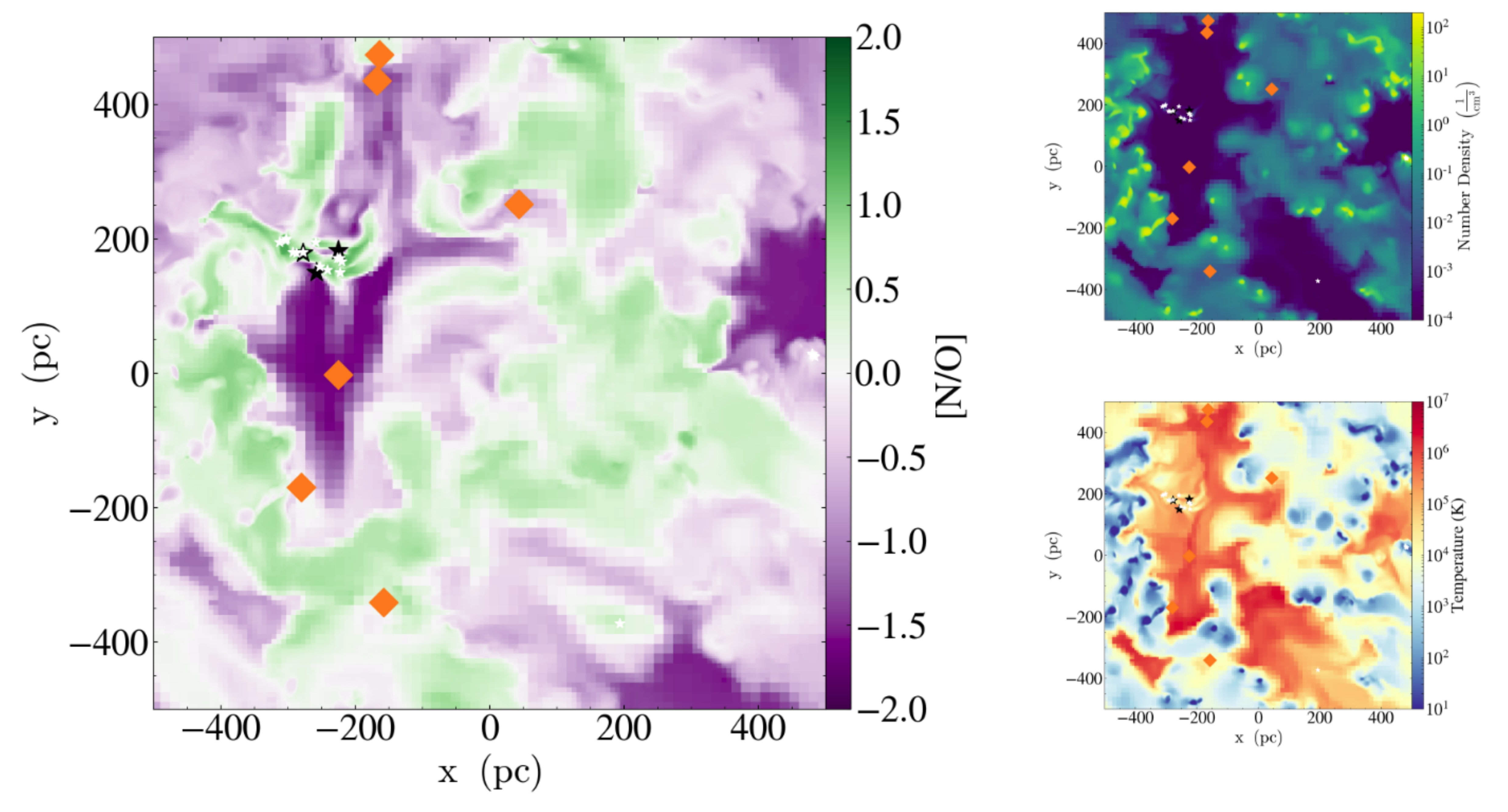}
\caption{Three slices in the mid-plane of our dwarf galaxy at 300~Myr after the start of star formation showing the variation in gas phase metal abundances. The left slice gives the ratio of the abundance between N and O, normalized to the solar abundance, while the number density and temperature are shown on the right. In each, we mark massive stars with active stellar winds as white points and SNe and AGB-phase enrichment events that occurred in the preceding 5 Myr as black stars and orange diamonds respectively.}
\label{fig:metal_slices}
\end{figure*}

We demonstrate the power of our simulations, which capture a realistic ISM at high resolution with multiple feedback sources, by addressing these questions in Fig~\ref{fig:metal_slices}. The left panel gives the abundance ratio of N to O throughout the ISM. The right 
two panels give the slices of number density (top right) and temperature (bottom right) in the mid-plane of a portion of our dwarf galaxy. These show regions with dense, cold gas clouds ($n \sim 100~\rm{cm}^{-2}$, $T \lesssim 100~$K) connected by cold filaments, warm, diffuse gas ($n\sim 0.1$~cm$^{-3}$, $T\sim 10^{4}$~K), and hot gas from a recent SN explosion ($T\sim10^{6}$~K). 

As shown in the left panel, [N/O] varies significantly over this section of the ISM with notable differences across the various phases and ISM structures. The hottest gas, dominated by recent supernova explosions is overabundant in oxygen, relative to solar (purple). However, the relative abundance of nitrogen increases in the WIM and WNM, being overabundant relative to solar (green). 

At our adopted metallicity, nitrogen is predominantly produced in AGB star winds, with very little production in core collapse SNe and winds from more massive stars. Therefore, nitrogen is injected into the ISM with significantly less energy ($v \sim 10$~km~s$^{-1}$) than elements produced in SNe, like oxygen, ($v\sim 10^3$~km~s$^{-1}$). Given the variations in Fig.~\ref{fig:metal_slices}, the energetic differences between injection sources can drive abundance variations within the ISM of our dwarf galaxy. The regions most rich in N are sites of recent AGB winds that have yet to mix with the rest of the ISM. This suggests that metal mixing within the ISM (and also metal ejection from the ISM) is species dependent. A more detailed analysis is beyond the scope of this work, but we investigate this in detail in \cite{Emerick2018c}.

\section{Discussion}
\label{sec:discussion}

\subsection{Comparison to Observed Low Mass Dwarf Galaxies}
\label{sec:observation}

As noted in Section~\ref{sec:IC}, our galaxy model is not intended to directly reproduce the observed properties of Local Group ultra-faint dwarfs. Notably, our initial conditions neglect a pre-existing stellar population and are only followed for 500~Myr, a fraction of the age of $z = 0$ dwarf galaxies. However, we can still place our model in context with observations using simple comparisons to the star formation rate (Section~\ref{sec:gas_sf}), molecular gas (Section~\ref{sec:molecular gas content}), and metal retention fraction (Section~\ref{sec:obs_metals}) properties of observed dwarf galaxies. We show that these properties are broadly consistent with observations. 

\subsubsection{Gas and Star Formation}
\label{sec:gas_sf}

\begin{figure*}
\centering
\includegraphics[width=0.475\linewidth]{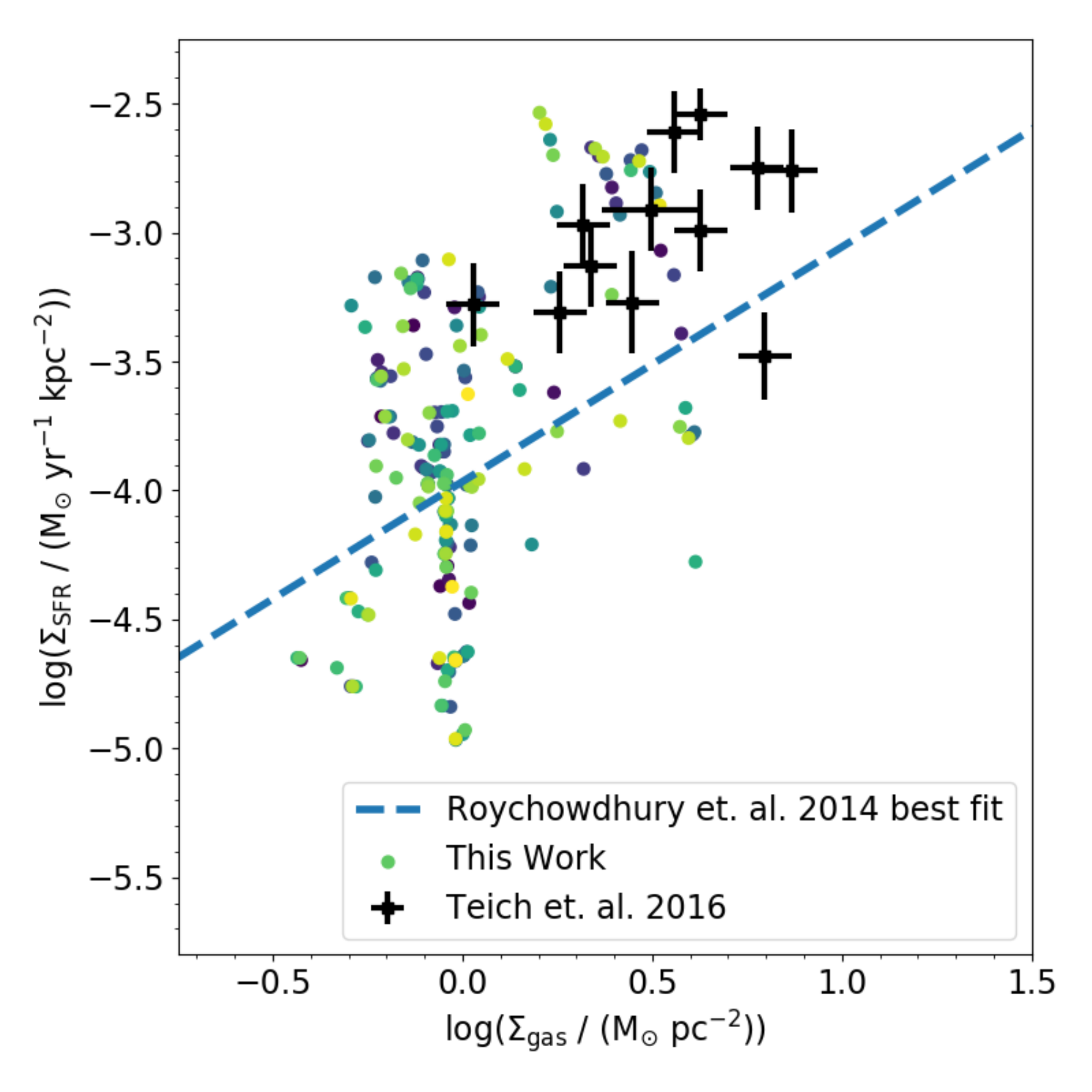}
\includegraphics[width=0.475\linewidth]{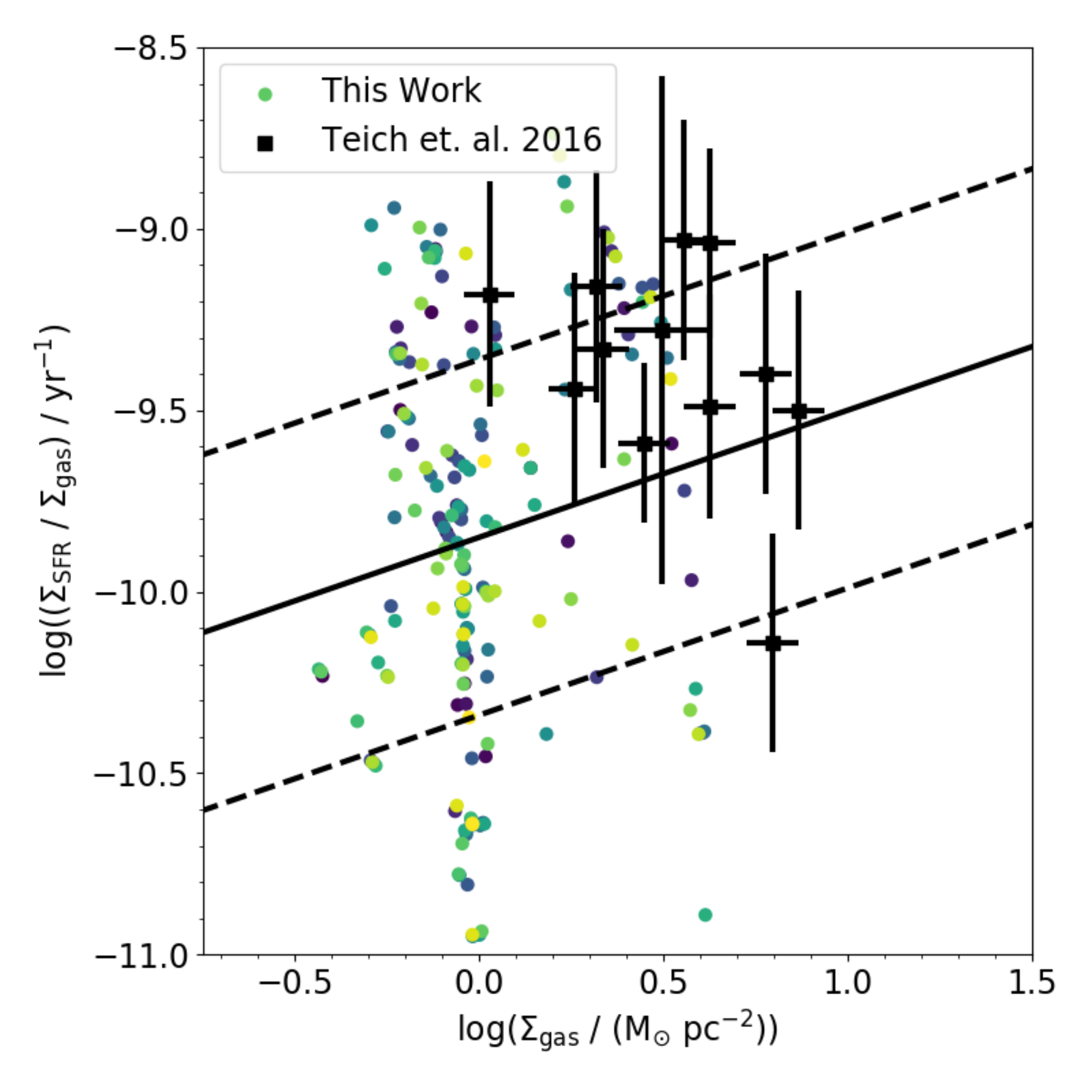}
\caption{The Kennicutt-Schmidt law (left) and extended Schmidt law (right) relationships for our galaxy as measured every megayear, plotted as points colored by time, with dark / purple early and light / green late. See text for details of the calculation. Recent observations from the SHIELD sample \protect\citep{Teich2016} are plotted as black points with error bars. On the left, we also give the best fit line to galaxies from the FIGGS sample from \protect\cite{Roychowdhury2014}, and on the right we also show the best fit line and 1 $\sigma$ errors from \protect\cite{Shi2011}. There is no clear correlation with time in this diagram.}
\label{fig:KS}
\end{figure*}

The observational sample of isolated, gaseous, low mass dwarf galaxies is limited compared to more massive galaxies, but has improved substantially with recent blind and targeted HI surveys \citep[e.g.][]{Giovanelli2005, Geha2006, Geha2012, Walter2008, Cannon2011, Haynes2011, Hunter2012, Bradford2015, James2015, Tollerud2015, Sand2015, Wang2017}. However, the sample of isolated, gaseous dwarf galaxies with $M_{*} < 10^{7}$~M$_{\odot}$ remains small. In Figure ~\ref{fig:KS} we show where our galaxy lies relative to the observed Kennicutt-Schmidt relation and extended Schmidt law for low mass galaxies. In both diagrams, our simulations are given by the colored points, sampled every megayear throughout the entire simulation. 

Although simple to measure in simulations, these quantities are challenging to directly compare to observations. We have attempted to make a reasonable analog to how $\Sigma_{\rm sfr}$ and $\Sigma_{\rm gas}$ are measured observationally for low mass dwarfs \citet[see ][]{Roychowdhury2014}. We define $\Sigma_{\rm sfr} = \dot{M}_{*,10} / A_{*,10}$, where $\dot{M}_{*,10}$ is the SFR measured over the preceding 10~Myr, and $A_{*,10}$ is the area of the disk within the radius of the outermost star formed within the previous 10 Myr. Likewise, $\Sigma_{\rm gas} = M_{\rm gas,10} / A_{*,10}$, where M$_{\rm gas,10}$ is the total gas mass within this defined disk. However, the total gas content cannot be determined observationally. To match this limitation, we follow \cite{Roychowdhury2014} and take $\Sigma_{\rm gas, obs} = 1.34 \times \Sigma_{\rm HI}$, where the factor 1.34 attempts to account for He. We note that there is generally no correction made for any possible H$_{\rm 2}$ or HII content. As shown in Section~\ref{sec:molecular gas content}, these components may be significant.

We include recent observational constraints on these relationships in Figure ~\ref{fig:KS}. Our galaxy fluctuates significantly about both relationships with no clear trends in time. However, in both cases, it is consistent with the available observational sample. At times our galaxy exhibits gas surface densities below the observational constraints. The trend is still consistent with higher densities at this point, but with a larger scatter towards lower star formation rate densities and efficiencies.

In constructing our galaxy model, we employed \textit{no} tuning of the underlying physics, adopting only canonical values for any available free parameters. It is thus non-trivial that our galaxy should oscillate about the median relationships in Fig.~\ref{fig:KS}, and signifies a proper accounting of the relevant physics governing gas and star forming properties in our galaxy. This result is consistent with galactic evolution simulations run at high resolution with a detailed accounting of stellar feedback physics \citep[see ][ and references therein]{NaabOstriker2017} and with the demonstration by \citet{Li2005} that the Kennicutt-Schmidt law can be reproduced by gravitation acting on an isothermal disk. The net result of the complex interactions of heating, cooling , chemistry, and feedback physics on star formation is to offset to a level not too dissimilar to more simple simulations considering gravity alone.

\subsubsection{Molecular Gas Content}
\label{sec:molecular gas content}
The molecular gas content of low-mass dwarf galaxies is generally assumed
to be small, but is not well constrained by either theory or observations. Assuming the relationship in \citet{Leroy2013} and \citet{Momose2013}, \citet{Roychowdhury2014} finds typical molecular gas mass fractions can be anywhere from $f_{\rm H_2} = 0.05$ to $f_{\rm H_2} = 0.5$; a significant range in values. As discussed in Section~\ref{sec:properties}, our galaxy has $f_{H_2}$ of 0.001 -- 0.05, just overlapping with this range. H$_2$ formation in our model is possible through formation on dust, the three body interaction, or the gas-phase reaction H$^-$~+~H~$\rightarrow$~H$_2$~+~e$^{-}$. The gas-phase reaction dominates in our low-metallicity galaxy over the other two channels by several orders of magnitude. In our model, H$^{-}$ is produced solely through the reaction H~+~e$^{-} \rightarrow$ H$^{-}$~+~$\gamma$. Thus the presence of some ionizing background is required to generate the molecular fractions we find in our simulations, as confirmed in separate test simulations. 

In contrast, \citet{Hu2016} and \citet{Hu2017} find low molecular fractions ($f_{\rm H_2} \sim 10^{-4}$) even in their simulations without any feedback ($f_{\rm H_2} \sim 2 \times 10^{-3}$). However, although these works do contain a non-equilibrium chemical model, they do not include either H$^{-}$ or a background radiation field. Our model suggests that both H$^{-}$ and the UV background are critical components in H$_2$ formation in small dwarf galaxies. Indeed it is not surprising that the gas phase reactions dominate over grain catalysis at low metallicity \citep{Glover2003}, though it is worth noting that the rate coefficients associated with gas-phase H$_2$ formation are still uncertain by an order of magnitude \citep{Glover2006,Glover2007}. Our model does lack additional chemistry that may be important to the formation and destruction of H$_2$, including HD chemistry, C and O chemistry, a detailed dust model, and cosmic rays.

In addition, our model does not account for the stellar component of the radiation field that leads to H$^{-}$ photodetachment ($E_\gamma > 0.76$~eV), though we do account for the contribution from the UVB. We find that the H$_2$ fraction is more strongly dependent upon the Lyman-Werner radiation field, which is followed for each star, than the H$^{-}$ photodetachment rate. Our tests suggest that, by ignoring the stellar contribution to H$^{-}$ photodetachment, our results may represent upper limits on the H$_2$ mass. However, including this component will likely only make a substantial difference during periods of no star formation, when there are no massive stars with significant Lyman-Werner luminosities. We do not anticipate this to have a significant dynamical impact on our simulations, as is discussed in more detail in Appendix~\ref{appendix:H minus}.

It is unclear how the combination of all of the above effects will behave, especially considering that, even at this resolution, we are unable to resolve the high density turbulent density perturbations in which H$_2$ forms most efficiently \citep{Glover2007}. These uncertainties certainly warrant further study of molecular gas in low metallicity dwarf galaxies.

\subsubsection{Metal Retention}
\label{sec:obs_metals}

The simulations presented here have not yet been run for the gigayear timescales required to begin to make direct comparisons to the observed stellar and gas phase metallicities of comparable dwarf galaxies at $z = 0$. However, we can compare to a key observable: the retention fraction of metals within stars and the galaxy's ISM compared to what would be expected from closed box stellar evolution models given the galaxy's star formation history. This can be done readily with Milky Way dSph's. Their stars seem to retain very little of the expected metal production: on order of a few percent or less depending on the galaxy and the species \citep{Kirby2011-metals}. However, environmental effects, namely ram pressure and tidal stripping, complicate the understanding of how these metals were removed from the galaxy. 

Leo P, the dwarf galaxy we approximate in our initial conditions, is extremely valuable as a gas-rich, star forming, low-mass dwarf galaxy, with an observable HII region, necessary for determining gas phase abundances, that is close enough to the Milky Way to conduct this experiment. Leo P retains $\sim$ 5\% $\pm$ 2\% of its metals, $\sim$ 1\% in stars and the rest in ISM gas \citep{McQuinn2015}. As discussed in Section~\ref{sec:chemical evolution}, more than 90\% of the tracked metals produced during our simulation no longer reside within the galaxy's disk, agreeing with observations. However, this is an evolving quantity that also depends on how much (if any) subsequent re-accretion of these metals occurs. Although more than half of these metals are expected to eventually re-accrete \citep{Christensen2016,Christensen2018,Angles-Alcazar2017}, it is still possible that most of the metals that have been produced at these early times will remain outside the galaxy disk.

It is interesting to consider whether ejected metals in our model reside in the CGM or have been ejected into the intergalactic medium. While this cannot be observationally determined for dwarf galaxies at this mass, cosmological simulations of an $M_{\rm vir} = 10^{10}$ M$_{\odot}$ galaxy show that, by redshift zero 40\% is ejected from the galaxy's halo \citep{Angles-Alcazar2017}. We find 5.3\% of all metals lie within 1 kpc of the center of our galaxy, 24.5\% within 0.25 $R_{\rm vir}$ (or 6.85 kpc), and 51\% outside $R_{\rm vir}$. This is consistent with previous works, but this is again an evolving quantity. In addition, the amount of gas that escapes the virial radius is certainly sensitive to the details of gas accretion from the IGM on this galaxy, which we cannot capture in this model.

The re-accretion or final ejection of this gas is directly relevant to the chemical evolution of low mass dwarf galaxies. Recycling of metal enriched gas could be a significant driver of long-term chemical evolution in low mass galaxies, particularly if a majority of metals ejected from the disk (itself nearly all the metals produced by the galaxy) return. In addition, the accretion of pristine gas from the intergalactic medium could significantly affect the gas flows around the galaxy, possibly promoting the retention of ejected metals. This effect is not included in our isolated galaxy simulations, and its role is beyond the scope of this work.

\subsection{Missing Physics}
Although we include many detailed physical models in our simulations, there remain additional physical processes that may be relevant, which we now discuss.

\subsubsection{Massive Stellar Wind Energy}
\label{sec:stellar winds discussion}
Our massive stellar wind model drastically reduces the injected wind velocity from $\sim$1000~km~s$^{-1}$ to 20~km~s$^{-1}$. Although our algorithm is entirely capable of generating realistic stellar winds with velocities comparable to those observed, such fast winds place a near constant and severe constraint on the Courant time step that renders $\gtrsim$100~Myr simulations impractical. When considered in isolation, stellar winds are an important source of pre-SN feedback and can dramatically influence dynamical evolution on molecular cloud and galaxy scales \citep{Dale2008,Peters2017,Gatto2017}. However, when considered together with ionizing radiation, stellar winds contain less total energy \citep{Agertz2013} and do not seem to have a significant dynamical influence in either idealized simulations \citep{Geen2015} or  individual giant molecular clouds \citep{Dale2014}, unless densities near the ionizing source are high enough to trap the  HII region in the source cell. In that case, they can clear out a cavity to allow initial establishment of the HII region. 

They are even less relevant in the low-metallicity regime studied here, as stellar winds become weaker with decreasing metallicity \citep{Puls2000, Vink2005}. Although they likely have minimal dynamical importance at resolutions where peak densities are not anyway high enough to trap ionization fronts, a full model of stellar winds may affect detailed ISM properties and metal mixing, warranting closer examination in future work.

\subsubsection{Cosmic Rays} 
\label{sec:CRs}

Recent work has explored the importance of cosmic ray feedback in regulating the ISM and wind properties in galactic disks \citep{Hanasz2013,GirichidisCR,Simpson2016,Farber2018}, isolated galaxies \citep{SalemBryanCorlies,Salem2015,Pakmor2016,Ruszkowski2017}, and galaxies in cosmological context \citep{SalemBryanHummels}. These relativistic charged particles act as a source of non-thermal pressure support in the galaxy's ISM, capable of driving outflows at different velocities and containing different thermal phases than those driven through thermal feedback alone \citep{SalemBryanCorlies}. Modeling cosmic rays is challenging, however, as they encompass a wide range of energies, and there are significant uncertainties in how they propagate through the ISM \citep[e.g.][]{Wiener2017}. Their propagation is often modeled as a diffusive process, but in truth this diffusion should vary depending on cosmic ray energy. In addition, cosmic rays couple effectively to the magnetic fields of galaxies, diffusing preferentially along structured magnetic field lines within the ISM. Modeling cosmic ray feedback completely thus requires both an accurate cosmic ray model and magnetohydrodynamics in order to capture the interplay of these two physical phenomena. Finally, including MHD presents additional difficulties in untangling the effects of each individual feedback mechanism on galaxy chemodynamics.

We do note that an isotropic, two-fluid model for cosmic ray feedback exists in  \textsc{Enzo} \citep{SalemBryan2014,Salem2015} and has been well tested. Mechanically, including this relatively simple treatment of cosmic ray feedback in our model is trivial. However, the cosmic ray population, their diffusion coefficient, and the magnetic field structure of the lowest mass dwarf galaxies each have significant enough uncertainties to warrant reserving their full inclusion into our model to later work.

\subsection{Detailed Stellar Evolution and Binary Stars}
\label{sec:binary stars} 

Roughly half of massive stars live in binary pairs \citep{Sana2013}. Their interactions, primarily through mass transfer, can significantly alter their radiation properties and lifetimes. This can change both how much and how long these stars emit ionizing radiation, an important source of stellar feedback, and where and when these stars explode as SNe. This effect could be significant, but is rarely accounted for in galaxy evolution models, which are commonly based on calculations of single star evolution (e.g. STARBURST99). For example, \citet{Zapartas2017} finds that binarity extends the timescales over which core collapse SNe occur from a given star formation event, from a maximum time of $\sim$ 50~Myr to $\sim$~200~Myr. Although they find only $\sim 15\%$ of core collapse SNe explode after 50~Myr, this could still be an important effect. Properly accounting for the delay times due to variations in individual star lifetimes has already been shown to change the significance of feedback and influence galaxy metallicity properties \citep{Kimm2015}. Extending the lifetimes of these stars combined with accounting for binary effects that change the luminosities of these stars \citep{Gotberg2017,Gotberg2018} could increase the importance of radiation feedback; however this may be less important as these additional photons are more likely to escape the galaxy \citep[e.g.][]{Ma2016-binary}. 

Since we model stars on a star-by-star basis, both of these effects could be reasonably accounted for by stochastically assigning binary star properties to some subset of our individual stars. This is beyond the scope of this project, but will be investigated in future work.

\section{Conclusion}
\label{sec:conclusion}
We have developed a new method for simulating galaxy evolution with detailed feedback and chemical enrichment. For the first time on galaxy scales, we simultaneously model multi-channel stellar feedback in detail, using individual star particles to model core collapse and Type Ia SNe, ionizing radiation followed through radiative transfer, photoelectric heating, Lyman-Werner radiation and pollution from AGB and massive stellar winds. This treatment of feedback, coupled with the detailed chemistry and heating/cooling physics followed with \textsc{Grackle}, allows us to capture realistic galaxy evolution in detail. In this work, we apply these methods to simulate the evolution of an isolated, low-mass, dwarf galaxy modeled after the $z=0$ properties of the Leo P dwarf galaxy. We present an overview of the properties of this simulation in this work.

For our simulated dwarf galaxy, we find:
\begin{enumerate}
\item Multi-channel feedback is effective in regulating star formation to a rate consistent with the Kennicutt-Schmidt relationship and the extended Schmidt law in observed galaxies. (See Figs.~\ref{fig:sfr_mass_evolution} and \ref{fig:KS}).

\item This feedback drives large outflows having mass loading factors of $\eta \sim 50$ at 0.25~R$_{\rm vir}$, falling to $\eta \sim 10$ at R$_{\rm vir}$,  and
metal mass loading factors near unity. By mass, nearly all of this outflow is moving with velocities below 100~km~s$^{-1}$, but there is a significant tail towards velocities up to 1000~km~s$^{-1}$ (See Fig.~\ref{fig:outflow_velocity}). 

\item 
Only $\sim$4\% of metals are retained in the disk of our simulated galaxy, consistent with the observed metal retention fractions of low-mass dwarfs.  By the end of the simulation $\sim$45\% of the remaining metals stay within the virial radius (but outside the galaxy), while $\sim$50\% 
have been ejected beyond the virial radius. (See Figs.~\ref{fig:mass_outflow}, \ref{fig:outflow_velocity}, and \ref{fig:metal_evolution}.)

\item 
Beyond the stellar radius, the gas scale height is thin ($\sim 50$~pc), yet resolved, with larger scale heights ($\sim 100$--200~pc) driven by feedback interior to the stellar radius. This is comparable to the resolution limit of the diffuse HI in observed, gaseous low-mass dwarfs. At a spatial resolution of 100~pc, our galaxy has a peak HI column density $N_{\rm HI} = 2.8-4.3 \times 10^{20}$~cm$^{-2}$, depending on inclination (See Fig.~\ref{fig:scale_height}).

\item The ISRF of our galaxy varies strongly in both space and time by orders of magnitude. It is unclear how important these fluctuations are as a source of feedback, or if the affect can be approximated with a time-averaged radial profile. The importance of radiation feedback in our model is investigated in more detail in \cite{Emerick2018b}. (See Figs.~\ref{fig:ISRF} and \ref{fig:ISRF_2D}.)

\item 
We find H$_2$ fractions below 5\% in our dwarf galaxy, consistent with the poor constraints on molecular gas formation in low metallicity dwarf galaxies. This H$_2$ forms entirely through gas-phase reactions facilitated by H$^{-}$ in self-shielding regions; H$_2$ formation on dust grains and in the three body reaction are both insignificant. Cold, neutral hydrogen dominates the mass of our galaxy. While warm, neutral hydrogen is present, it does not dominate the mass fraction (See Fig.~\ref{fig:sfr_mass_evolution} and Fig.~\ref{fig:ISM_evolution}.)

\item Finally, we present gas-phase oxygen and nitrogen distributions as examples to briefly demonstrate that there are marked differences in how individual metal species are distributed within the ISM of our galaxy. These variations could be tied to differences in elemental yields among different sources (for example, AGB winds vs. SNe), as suggested by \cite{KrumholzTing2018}. This is explored in more detail in \citep{Emerick2018c}.
\end{enumerate}

\section*{Acknowledgments:} 
We would like to thank the following for their advice and valuable discussions, without which this work would not have been possible: B. C\^ot\'e, S. Glover, K. Hawkins, C. Hu, K. Johnston, B. O'Shea, M. Putman, B. Smith, J. Wall, and J. Wise. We would additionally like to thank the referee for their careful report. A.E. is funded by the NSF Graduate Research Fellowship DGE 16-44869. G.L.B. is funded by NSF AST-1312888, NASA NNX15AB20G, and NSF AST-1615955. M.-M.M.L. was partly funded by NASA  grant NNX14AP27G. We gratefully recognize computational resources provided by NSF XSEDE through grant number TGMCA99S024, the NASA High-End Computing Program through the NASA Advanced Supercomputing Division at Ames Research Center, Columbia University, and the Flatiron Institute. This work made significant use of many open source software packages, including \textsc{yt}, \textsc{Enzo}, \textsc{Grackle}, \textsc{Python}, \textsc{IPython}, \textsc{NumPy}, \textsc{SciPy}, \textsc{Matplotlib}, \textsc{HDF5}, \textsc{h5py}, \textsc{Astropy}, \textsc{Cloudy} and \textsc{deepdish}. These are products of collaborative effort by many independent developers from numerous institutions around the world. Their commitment to open science has helped make this work possible. 

\bibliographystyle{mnras}
\bibliography{msbib}

\appendix
\renewcommand\thefigure{\thesection.\arabic{figure}}    
\setcounter{figure}{0}

\section{Gas Phases of the ISM}
\label{appendix:phases}

To aid in comparison to other works, we define our ISM phases here, as adopted and modified from \citet{Draine2011}, Table 1.3, and used consistently throughout this analysis. By construction, these phases are mutually exclusive. We take $f_{\rm H_2}$ to be the molecular hydrogen fraction of the total gas mass.

\begin{enumerate}
\item Hot Ionized Medium (HIM): $T \geq 10^{5.5}$ K
\item Warm Ionized Medium (WIM): $10^{4}~\rm{K} \leq T < 10^{5.5}~\rm{K} $
\item Warm Neutral Medium (WNM): $10^{2}~\rm{K} \leq T < 10^{4}~ \rm{K}$
\item Cold Neutral Medium (CNM): $T < 10^2$ K, $f_{\rm H_2} \leq 0.5$
\item Molecular: $T < 10^2$~K, $f_{\rm H_2} > 0.5$
\end{enumerate}

\section{Stellar Radiation Properties}
\label{appendix:radiation}
\begin{figure*}
\centering
\includegraphics[width=0.4\linewidth]{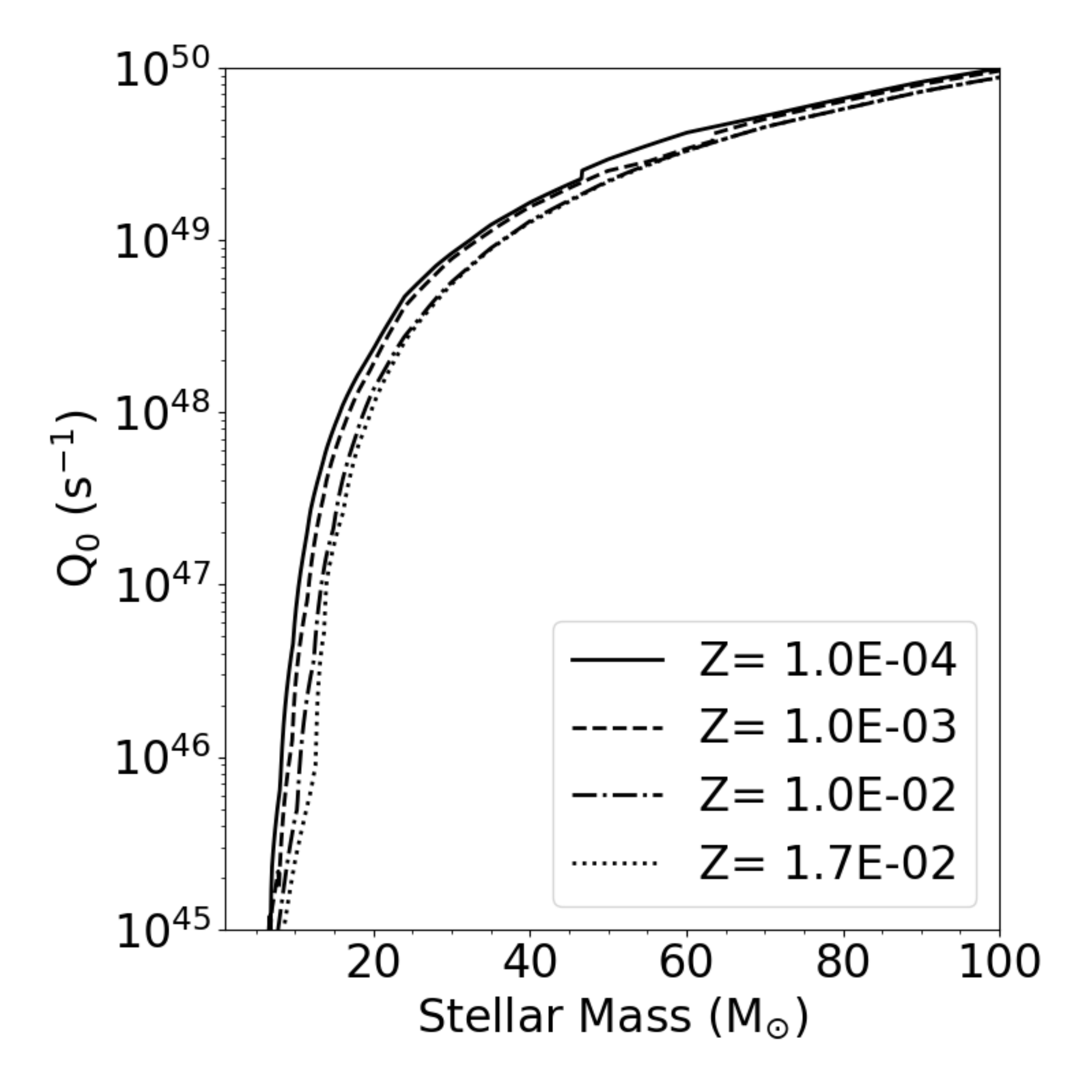}
\includegraphics[width=0.4\linewidth]{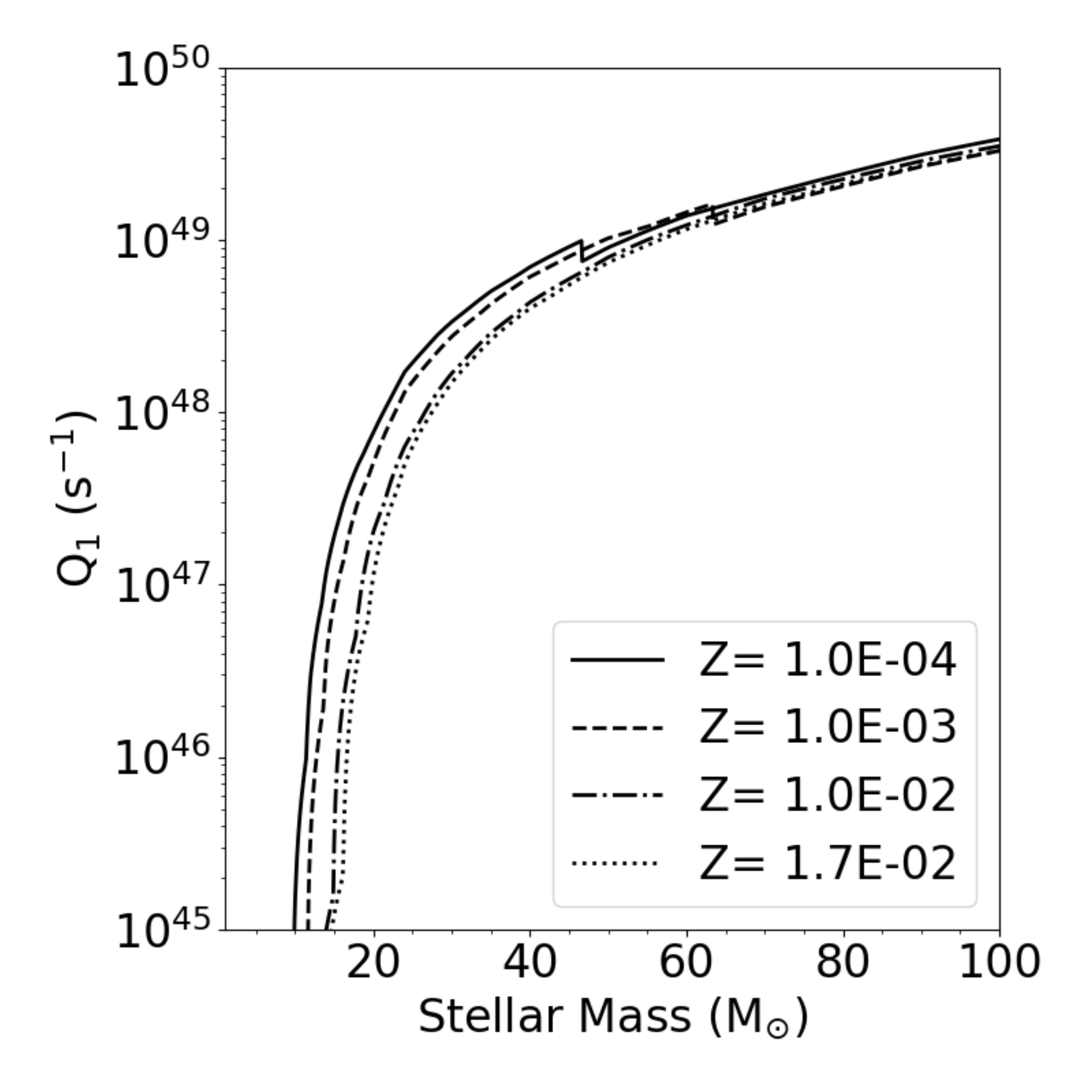}
\vspace{0.01cm}
\includegraphics[width=0.4\linewidth]{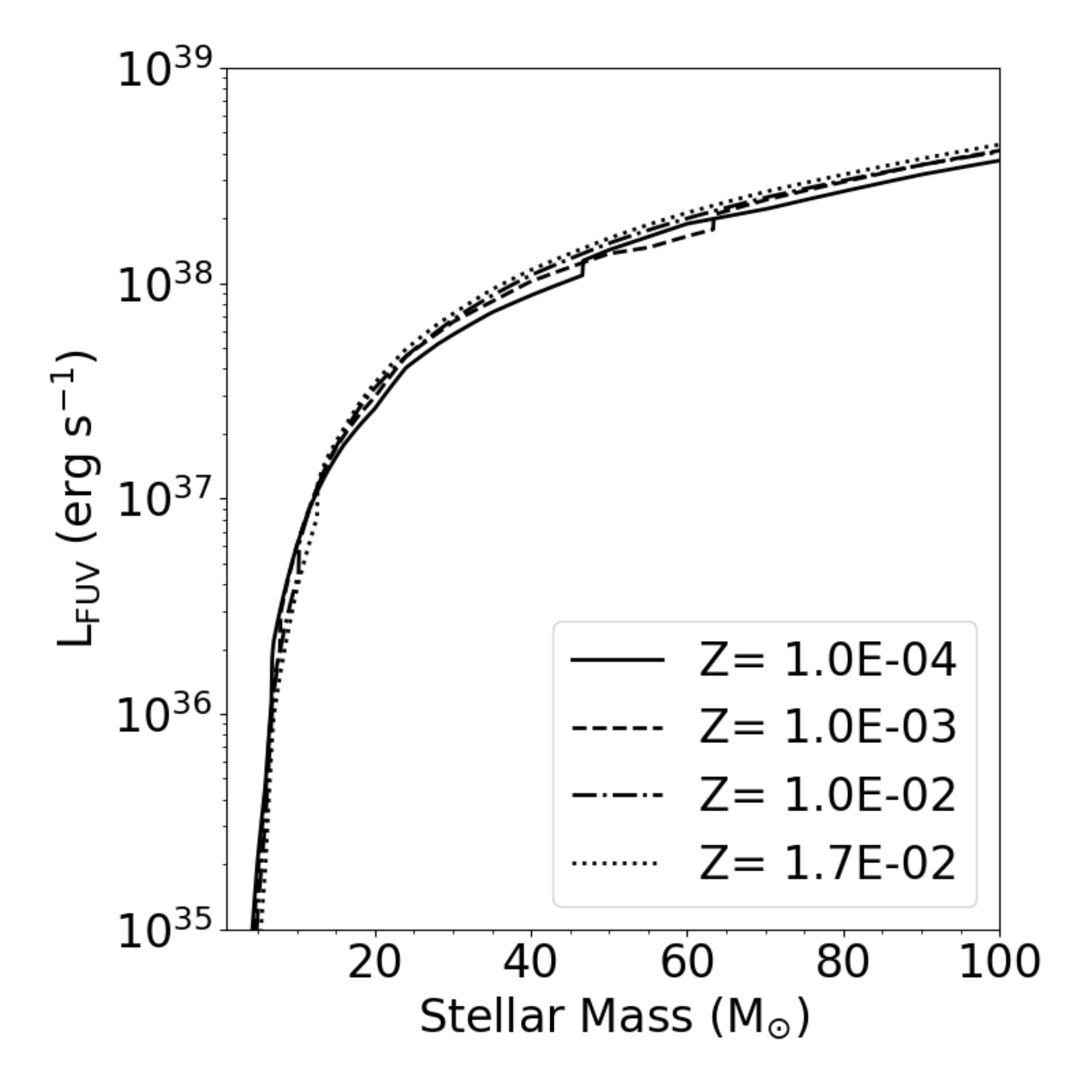}
\includegraphics[width=0.4\linewidth]{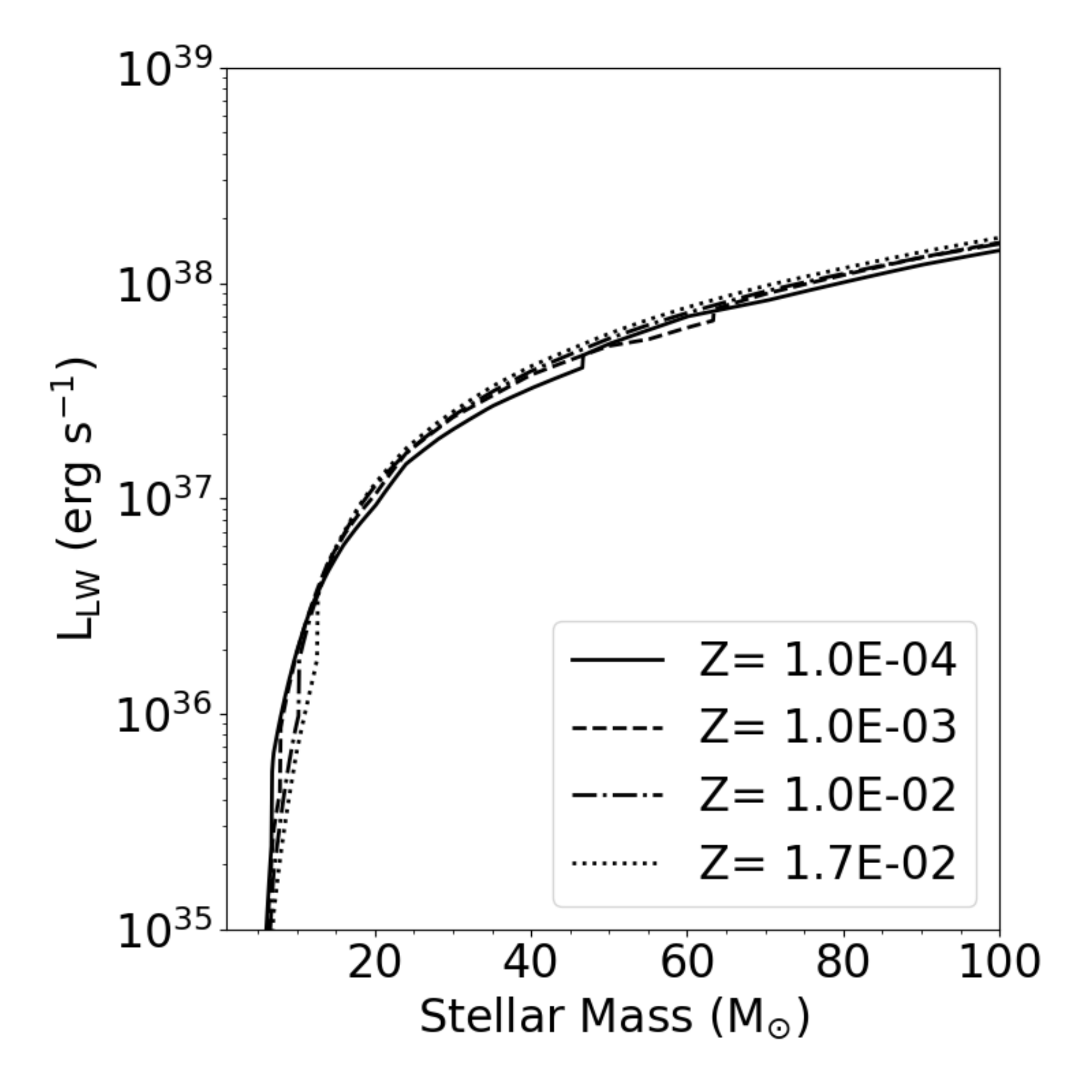}
\vspace{0.01cm}
\includegraphics[width=0.4\linewidth]{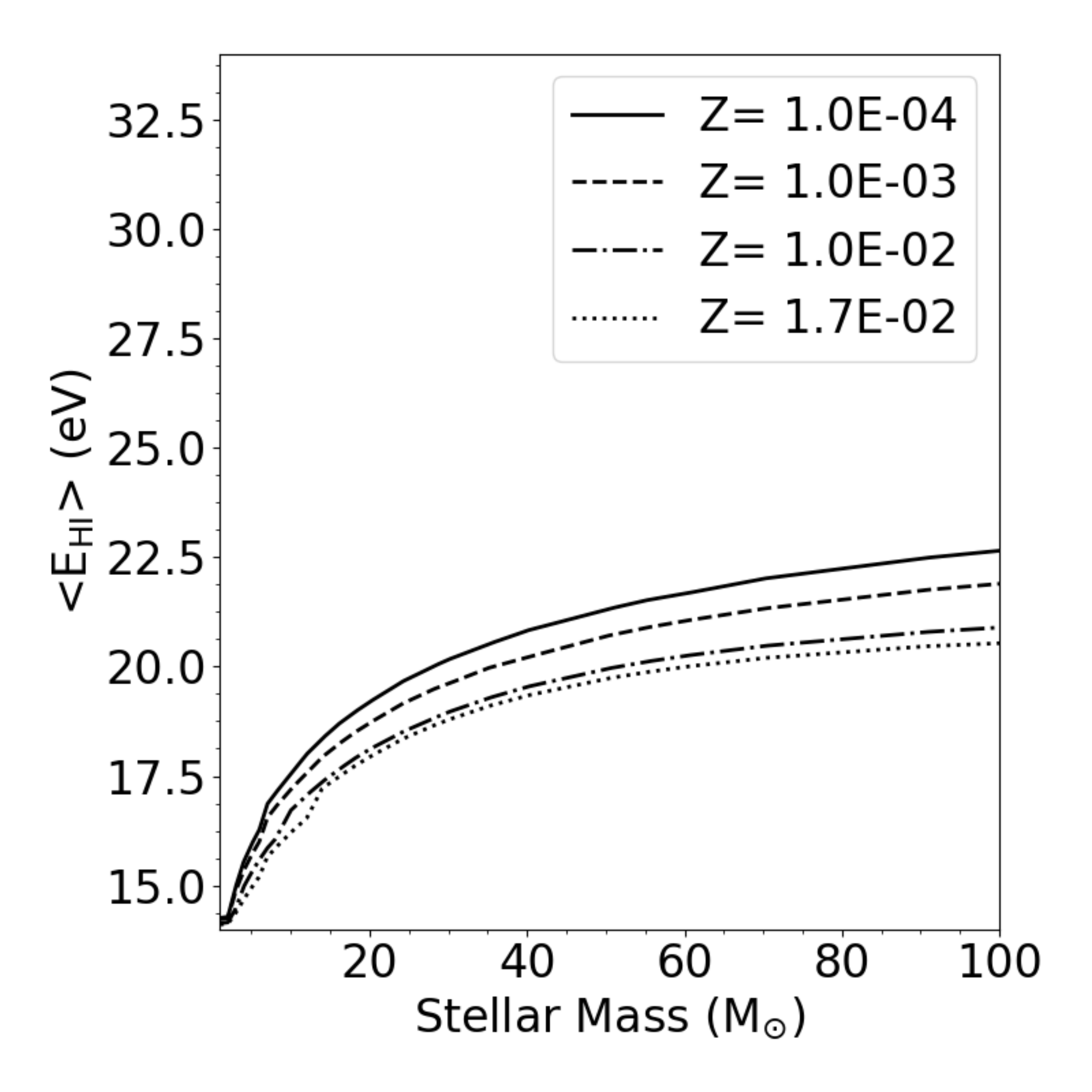}
\includegraphics[width=0.4\linewidth]{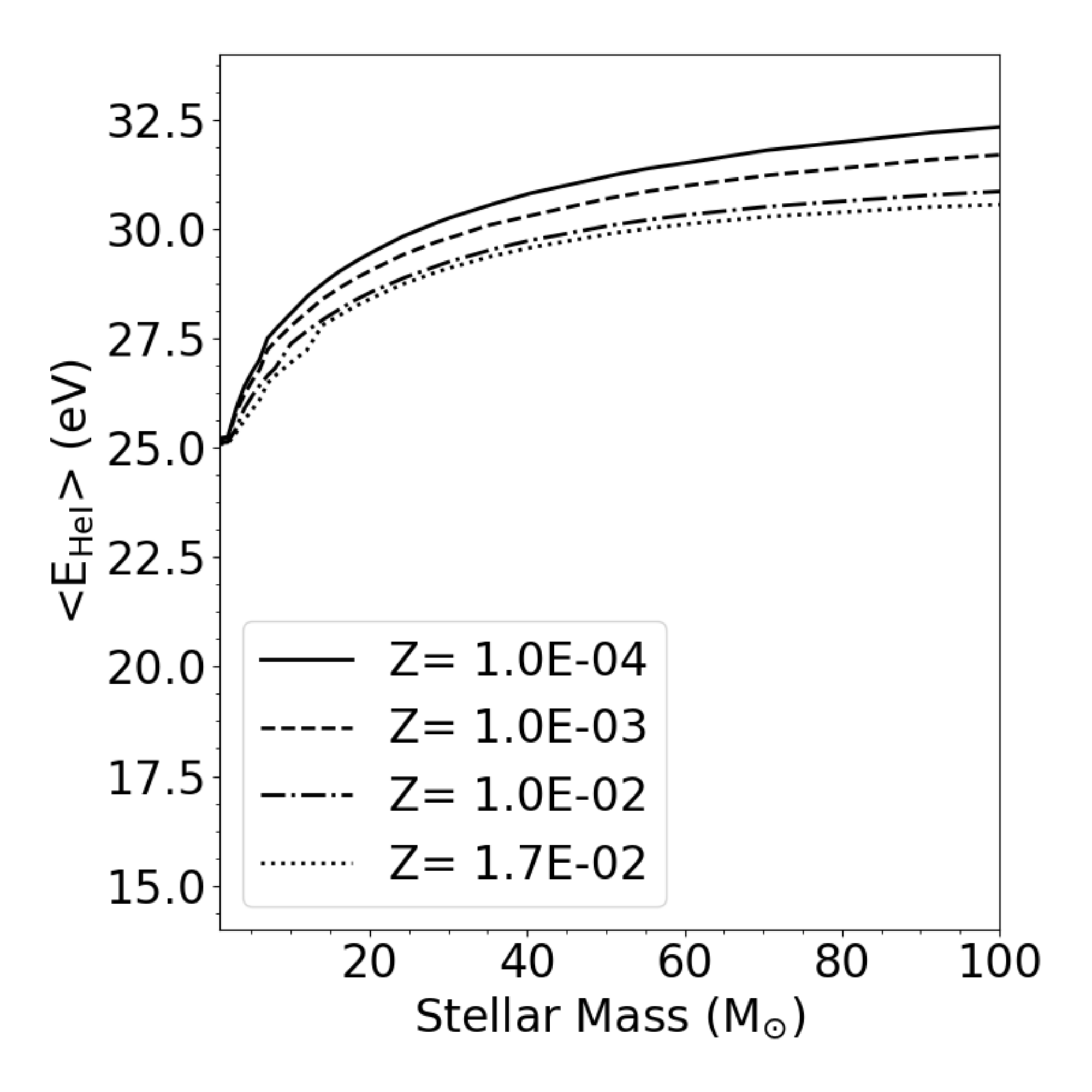}
\caption{Radiation properties for our model stars, showing the ionizing photon luminosities for HI (top left) and HeI (top right) for each star, the FUV (middle left) and Lyman-Werner (middle right) luminosities, and finally the average ionizing photon energy for HI (bottom left) and HeI (bottom right). Note, we only track radiation from stars above 8.0 \msun, which dominate over less massive stars, even when accounting for IMF weighting.}
\label{fig:stellar radiation properties}
\end{figure*}

We determine the radiation properties of our model stars as a function of stellar mass and metallicity from the OSTAR2002 grid \citep{Lanz2003} in the regime that it covers, or integrated from an adjusted black body curve for other stars, given the ZAMS stellar radius obtained from the PARSEC \citep{Bressan2012,Tang2014} stellar evolution data set. In Fig.~\ref{fig:stellar radiation properties} we plot these properties. We use a constant factor across metallicities for stars not sampled on the OSTAR2002 grid to shift the black body radiation fluxes to be roughly continuous with the ionizing photon rates and luminosities as a function of stellar mass. This requires two factors for each radiation type, one for low mass and one for high mass stars. We use the following multiplicative factors to adjust the black body spectrum for HI and HeI ionizing radiation respectively: [0.1, 3.2] and [0.0001, 4.0]. We do not find this adjustment to be necessary for the FUV and Lyman-Werner radiation bands.

The ionizing radiation photon energies are taken as the average ionizing photon energy for a black body of the star's given T$_{\rm eff}$ obtained from the PARSEC grid. A more accurate approach to compute this energy would convolve the full stellar spectrum and the frequency-dependent absorption cross section. We tested this approach using the frequency-dependent photoionization cross sections from \citet{1996ApJ...465..487V}.\footnote{Source code containing the analytic fits given in \citet{1996ApJ...465..487V} was obtained from \url{http://www.pa.uky.edu/~verner/photo.html}} 
The blackbody approximation is accurate to within 5\%, yet substantially easier to compute on the fly, as integrals over the blackbody spectrum can be expressed as infinite series that rapidly converge to high precision. Unlike for the ionizing radiation, we assume constant FUV and Lyman-Werner band energies for each star, at 9.8 eV and 12.8 eV respectively.

\setcounter{figure}{0}
\section{Typical Gas Densities in Supernova Injection Regions}
\label{appendix:SN}

Modeling SN feedback with the injection of thermal energy alone can lead to rapid overcooling of the injected energy, and a significant underestimate of the effects of SN feedback. Often, ad hoc solutions to this problem are used in large simulations with coarse resolution, such as momentarily turning off cooling in feedback affected regions or decoupling affected regions from the hydrodynamics for some time. However, physically consistent solutions have been developed \citep[e.g][]{Simpson2016} that inject a mixture of kinetic and thermal energy with a ratio that depends on resolution and local gas density. Overcooling becomes less important with higher resolution and lower ISM densities, until eventually pure thermal energy injection is sufficient to resolve each SN. We take advantage of our high resolution and employ a simple thermal energy injection model for our SN feedback. We demonstrate how well this resolves our SNe in the comparison simulation presented in this work.

\begin{figure*}
\centering
\includegraphics[width=0.4\linewidth]{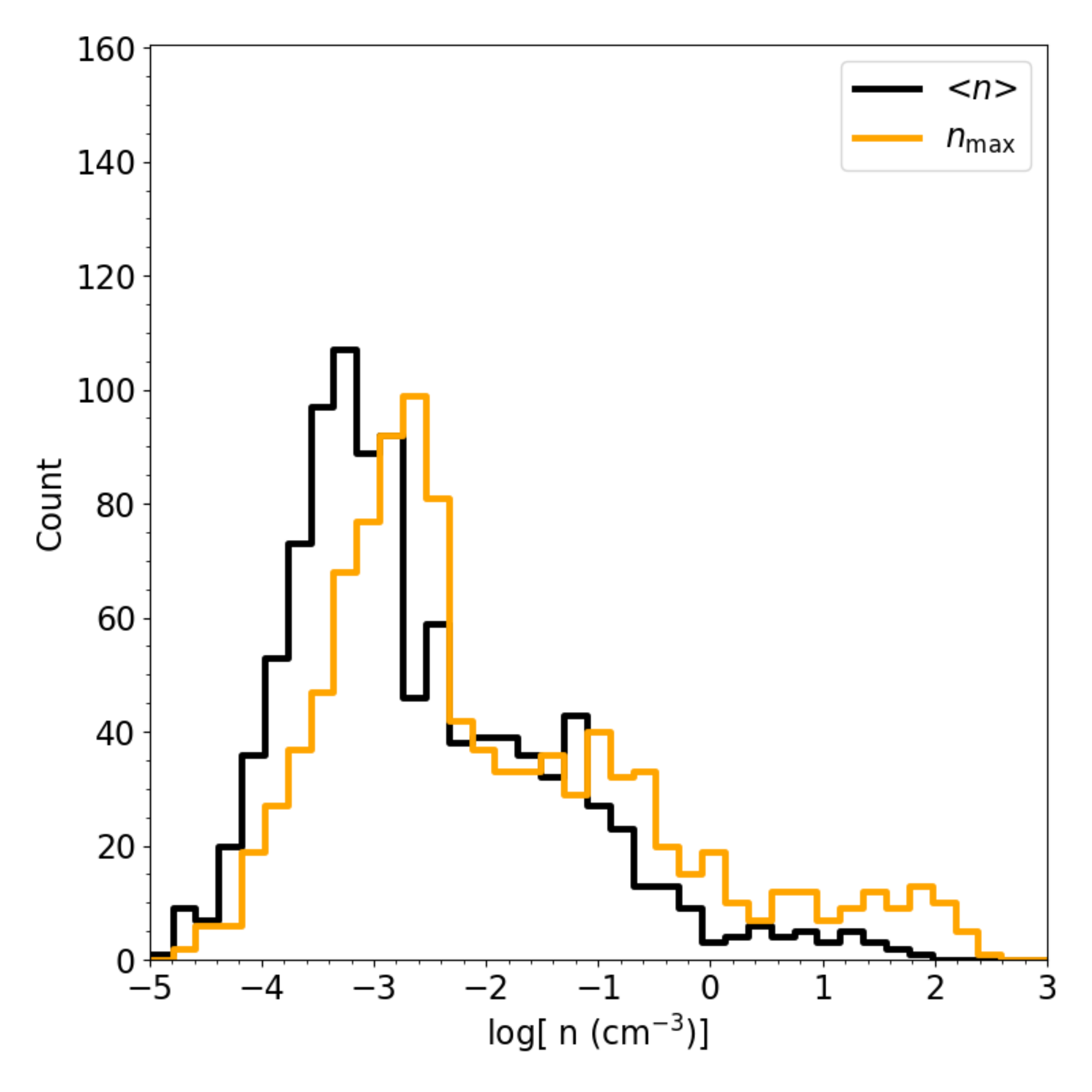}
\includegraphics[width=0.4\linewidth]{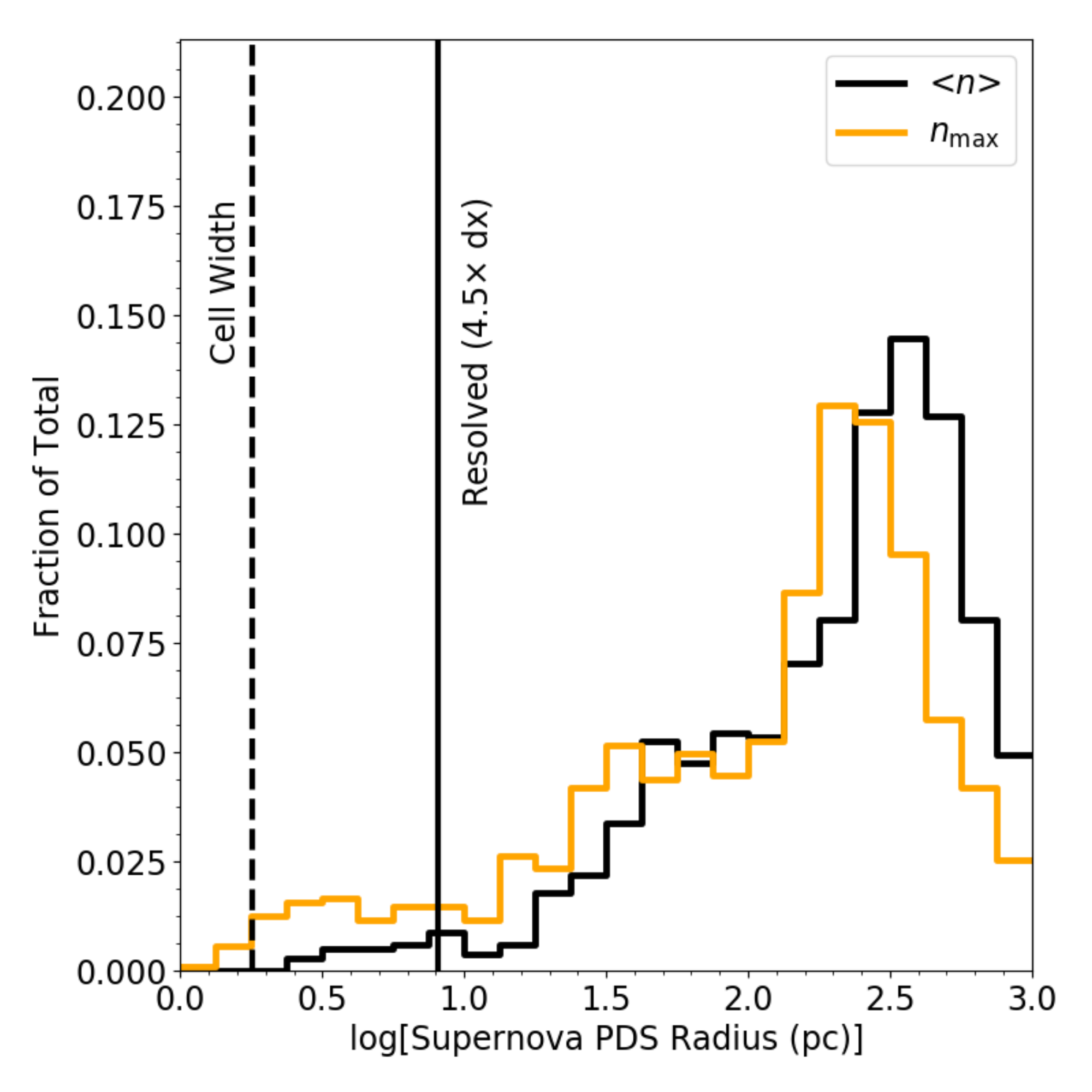}
\caption{{\em Left:} Distribution of peak (orange) and average (black) gas densities in the injection region of each of the SNe from the simulation. {\em Right:} Distribution of the radius of the pressure-driven snowplow phase for each SN assuming a uniform medium at either the peak or average density in the injection region. The vertical lines show one and 4.5 cell widths. This shows that a fraction of these SNe are certainly unresolved, with R$_{\rm PDS}$ less than a single cell size, and somewhat more don't satisfy the 4.5 cell criterion, but the majority are well resolved.}
\label{fig:SN histogram}
\end{figure*}

The left panel of Fig.~\ref{fig:SN histogram} gives the distributions of the peak and average ISM number densities in the SN injection regions for each SN in our fiducial simulation. As shown, a majority explode in regions at substantially lower densities than the star formation threshold of 200~cm$^{-3}$. For most SNe, $n_{\rm max} \le 1.0$~cm$^{-3}$, which is due to the substantial pre-SN feedback included in our simulations. The right panel of Figure ~\ref{fig:SN histogram} gives the fractional distribution of the calculated radius of the pressure-driven snowplow ($R_{\rm PDS}$) phase of the Sedov-Taylor expansion, adopting the definitions from \citet{Simpson2016}
\begin{equation}
R_{\rm PDS} = 
\begin{cases}
49.3E^{1/4}_{51}n^{-1/2}_{o} & if  Z \le 0.01 \\
18.5E^{2/7}_{51}n^{-3/7}_{o}Z^{-1/7} & if  Z \ge 0.01 \\
\end{cases}
\end{equation}
where $E_{51}$ is the injection energy in units of 10$^{51}$~erg, $n_{o}$ is the number density of the medium, and $Z$ is the metallicity in units of $Z_{o}$. \citet{Simpson2016} found that $R_{\rm PDS} > 4.5 \Delta x$ is needed to resolve the SN explosion with thermal energy injection alone, assuming uniform density in the injection region. In practice, the injection region is never uniform. We give $R_{\rm PDS}$ assuming uniform density at both the average and maximum number density in the injection region of each SN. As shown, a majority of SNe are resolved (to the right of the 4.5$\Delta x$ line), but up to 8.3\% are not well resolved and 2.0\% are completely unresolved (below a single cell width). In general, these are SN explosions from the most massive (i.e. most prompt) stars. We do not expect that resolving these SNe will dramatically alter our results. However, we note that using our feedback model at resolutions much less than a few parsecs will require implementing an different injection mechanism.

\setcounter{figure}{0}
\section{Cooling and Heating Rates}
\label{appendix:cooling}

\begin{figure*}
\centering
\includegraphics[width=0.95\linewidth]{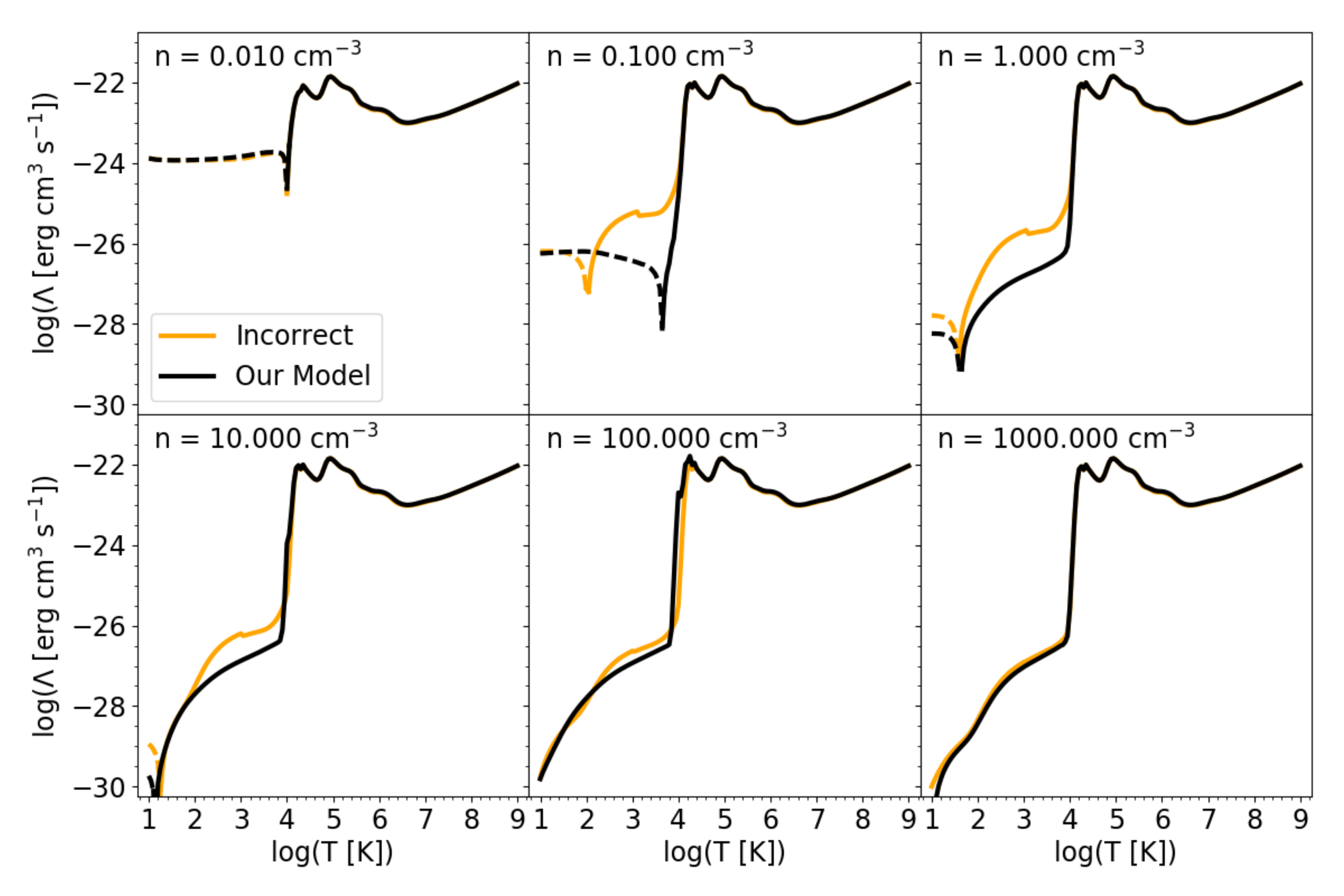}
\caption{The absolute value of the net cooling or heating rate in two models that account for self-shielding of primordial gas against a metagalactic UV background. Our model (black) uses self-consistent metal line cooling rates, and is compared to an incorrect model (orange) that adopts the uncorrected (i.e. optically thin) metal line cooling rates. The regimes where heating dominates are plotted with dashed lines for clarity.}
\label{fig:cooling comparison}
\end{figure*}

\begin{figure*}
\centering
\includegraphics[width=0.95\linewidth]{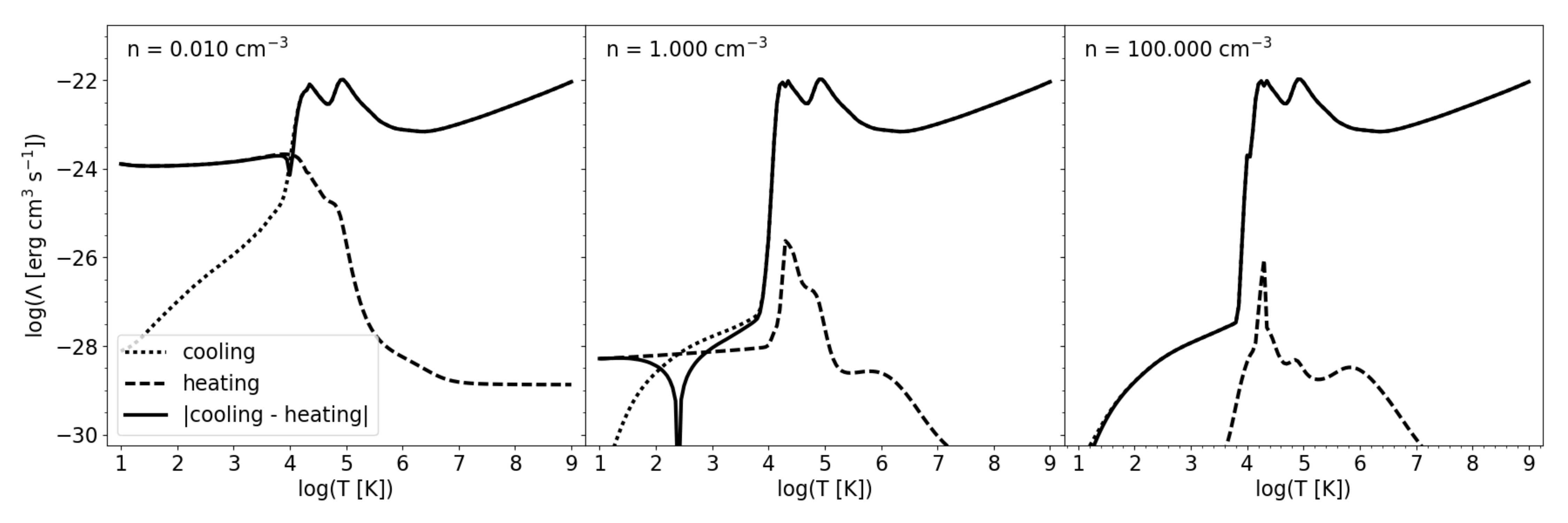}
\caption{The total heating (dotted) and cooling (dashed) rates extracted from the core of a Jeans-length sized cloud irradiated by a \citet{HM2012} metagalactic UV background as modeled in \textsc{Cloudy}. The absolute value of the net cooling or heating
rate is shown with solid lines.}
\label{fig:cooling}
\end{figure*}

The cooling curves used in this work include a correction to a significant conceptual inconsistency one may encounter when using tabulated metal line cooling rates in combination with a self-shielding approximations. \citet{Hu2017} details this inconsistency: metal cooling rates computed under the assumption of an optically thin UV background overestimate the electron fraction in regimes where self-shielding is important ($-0.5 \leq \rm{log(n~[cm^{-3}])} < 2$). This results in an artificially enhanced metal line cooling rate in regions of self-shielding. This can be a significant effect, particularly at higher metallicities. We demonstrate this in Fig.~\ref{fig:cooling comparison}, which gives the absolute value of the net cooling rate for our full model at $Z = 0.1$~Z$_{\odot}$ (discussed below) as compared to a model using the optically thin metal line cooling rates at a range of densities. The effect is most significant at moderate densities, where the net cooling rate can be an order of magnitude higher. In low temperature regions, this can significantly shift the inferred equilibrium temperature and reduce the cooling time of dense gas. Fig.~\ref{fig:cooling} shows the net cooling rate and the individual heating and cooling rates from our generated tables at low metallicity, $Z = 0.01$~Z$_{\odot}$, comparable to that used in our dwarf galaxy. We have made these tables publicly available in the main distribution of \texttt{Grackle} and discuss how they were generated below.

For densities where self-shielding is important, we computed \textsc{Cloudy} models of one-dimensional clouds at each temperature and density pair with a physical size corresponding to the Jeans length. In cases where the Jeans length is very large, we limit the size to 100 pc, a reasonable approximation for the maximum size of a self-gravitating cloud. We then adopt the metal line cooling rates obtained from the center of these clouds, whose outside is exposed to the UV background. These models were computed with a modified version of the \texttt{CIAOLoop}\footnote{https://bitbucket.org/brittonsmith/cloudy\_cooling\_tools (our version: https://github.com/aemerick/cloudy\_tools)} code used in \citet{2008MNRAS.385.1443S}. In computing these tables we had to include the cosmic ray ionization that dominates in optically thick clouds. Without this effect, \textsc{Cloudy} becomes unstable in entirely neutral regions. We adopted a cosmic ray ionization rate of 10\% the Milky Way value, though we note that varying this from 1\% to 100\% the Milky Way value did not have an effect on the extracted metal line cooling rates.

Finally, we note that the cooling and heating rates in our simulation will deviate from these curves, as they depend upon local conditions that cannot be accounted for here. The primordial cooling and heating rates are computed consistently with the non-equilibrium chemistry solver, leading to deviations from the tabulated primordial rates, and the heating rates depend upon the local stellar radiation field, which is not included in these diagrams.

\setcounter{figure}{0}
\section{H$^{-}$ Photodetachment}
\label{appendix:H minus}

\begin{figure}
\centering
\includegraphics[width=0.95\linewidth]{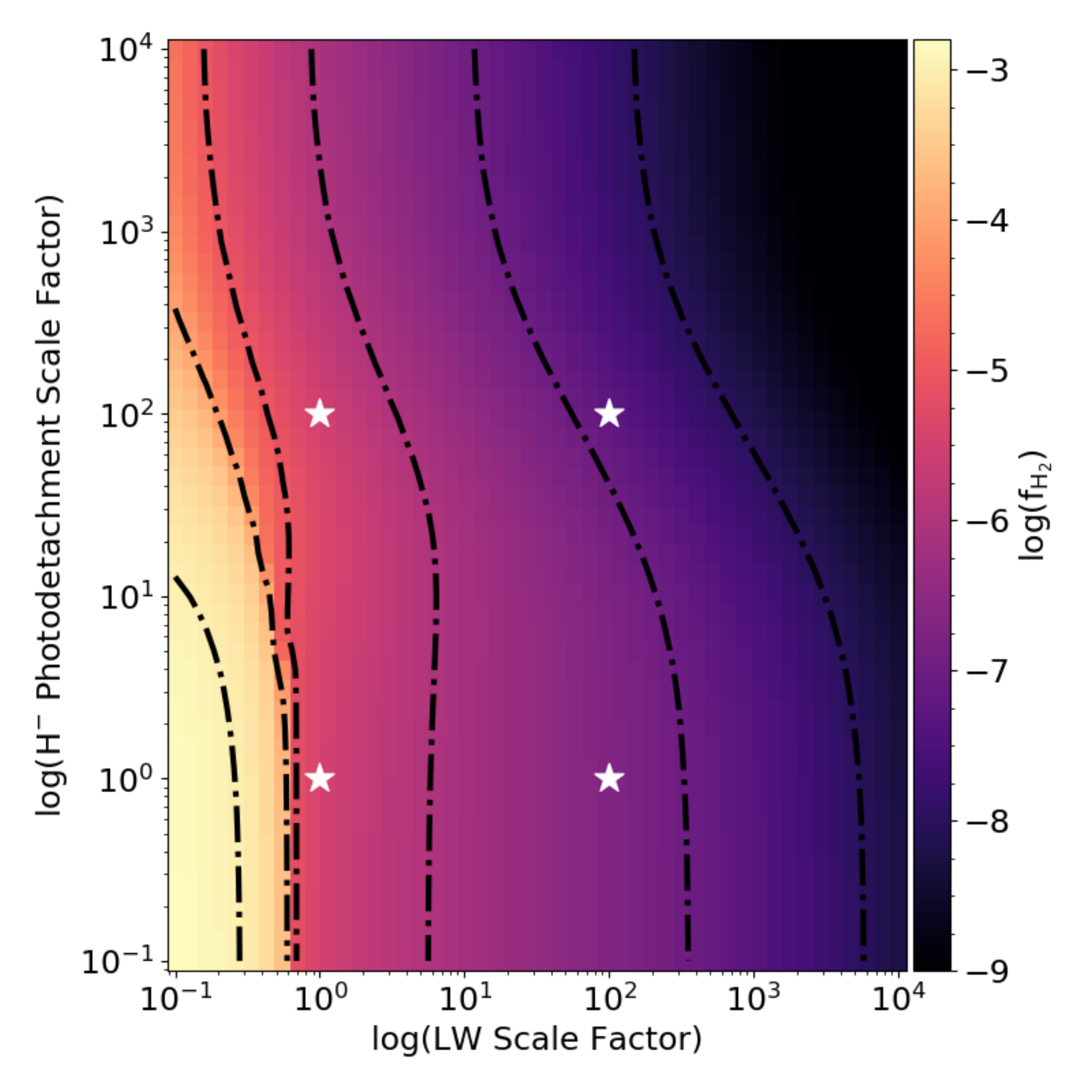}
\caption{The final H$_2$ mass fraction in a grid of one-zone Grackle cooling models run with the same Grackle parameters and metallicity as our full hydrodynamics simulation. Each model is held at a uniform density $n = 10$~cm$^{-3}$. Each model varies either the H$^{-}$ photodetachment rate or the Lyman-Werner background over their respective UVB values by a given scale factor. Contours give factors of ten change in the H$_2$ fraction. We note the regime below 1.0 on each axis is somewhat unphysical, and is never reached in our hydrodynamics simulations.}
\label{fig:Grackle grid}
\end{figure}

As discussed in Section~\ref{sec:molecular gas content}, H$_2$ formation is dominated by the gas-phase H$^{-}$ channel in our simulations. Accounting for the photodetachment of H$^{-}$ from the stellar radiation field, which may dominate over the UVB, can be important for correctly modeling H$_2$ formation in low metallicity gas~\citep{Wolcott-Green2012}. Our current model does not include this contribution. To test the significance of this radiation in determining the final H$_2$ mass we run a grid of one-zone Grackle cooling models.\footnote{See the ``cooling\_cell.py'' example here for more details: https://grackle.readthedocs.io/en/latest/Python.html}  These runs iterate through the non-equillibrium chemistry and cooling solvers with a constant density (10~cm$^{-3}$) and initial temperature of 10$^{4}$~K for 200~Myr. In Fig.~\ref{fig:Grackle grid} we plot the final H$_2$ mass fraction in these cells for variations over five orders of magnitude in the adopted H$^{-}$ photodetachment rate and the Lyman-Werner radiation background. As shown, the H$_2$ mass fraction is much more sensitive to the Lyman-Werner radiation field than the H$^-$ photodetachment rate. In effect, our simulations move left-to-right in this diagram (only) as stars form (right) or die (left). A complete model, including the stellar contribution to H$^{-}$ photodetachment, would move somewhat diagonally.

We follow up on the qualitative conclusions of these models by running four simulations at the four points in the diagram. These runs have identical initial conditions, random SN driving, cooling, heating, and chemistry as our fiducial simulation, but with no star formation. We vary the Lyman-Werner radiation and the H$^{-}$ photodetachment rate by the factors shown, which are comparable to the ratio between the typical ISRF and the UVB in the disk of our galaxy (see Fig.~\ref{fig:ISRF}). Changing the background H$^{-}$ photodetachment rate by a factor of 100 only (top left star) leads to a reduction in the final H$_2$ mass fraction by 3, while changing the Lyman-Werner background by a factor of 100 (bottom right point) decreases the H$_2$ mass by a factor of 6.9. Applying both scale factors gives a reduction of 7.9. Thus, we see that the difference between the more realistic case in which both rates increase by the same factor, and the case assumed in the simulations, is only about 20\%. 

Although the replacement of star formation by random SNe in these models produces qualitatively different galaxy properties, the point of these simulations is to be maximal models of H$_2$ formation. We demonstrate that, even in this case, there is not a dramatic change in the total H$_2$ formation and that the effect of varying the H- photodetachment rate is sub-dominant to other factors. Therefore, we do not expect that ignoring this component of the stellar radiation field will have a significant dynamical impact on our galaxy evolution.

H$^-$ photodetachment arises from lower energy photons, above 0.76~eV, with a peak in effectiveness around 2~eV. For this reason, low mass stars may contribute significantly, if not dominate, the radiation field responsible for this reaction. The Lyman-Werner radiation field is dominated by short-lived massive stars, with some contribution from older stars up to a few hundred Myr, and is thus subject to significant fluctuations, particularly in periods of no star formation. During these times, H$^-$ photodetachment from low energy photons may play a non-negligible role in regulating H$_2$ formation. Indeed there is H$_2$ growth during these periods due to a decrease in the Lyman-Werner field, but it is unclear how much of this would be attenuated by accounting for photons from lower mass stars. Doing so in our current model, applying an optically thin $1/R^2$ field for each star, would be computationally challenging due to the dramatic increase in the number of stars whose radiation would need to be tracked. 

As emphasized in Section~\ref{sec:molecular gas content}, further work modeling low-metallicity dwarf galaxies with a more detailed chemistry model would be useful to clarify the role of various competing phenomena in driving the molecular properties of these galaxies.

\setcounter{figure}{0}
\section{Resolution Study}
\label{appendix:resolution_study}

\begin{figure*}
\centering
\includegraphics[width=0.33\linewidth]{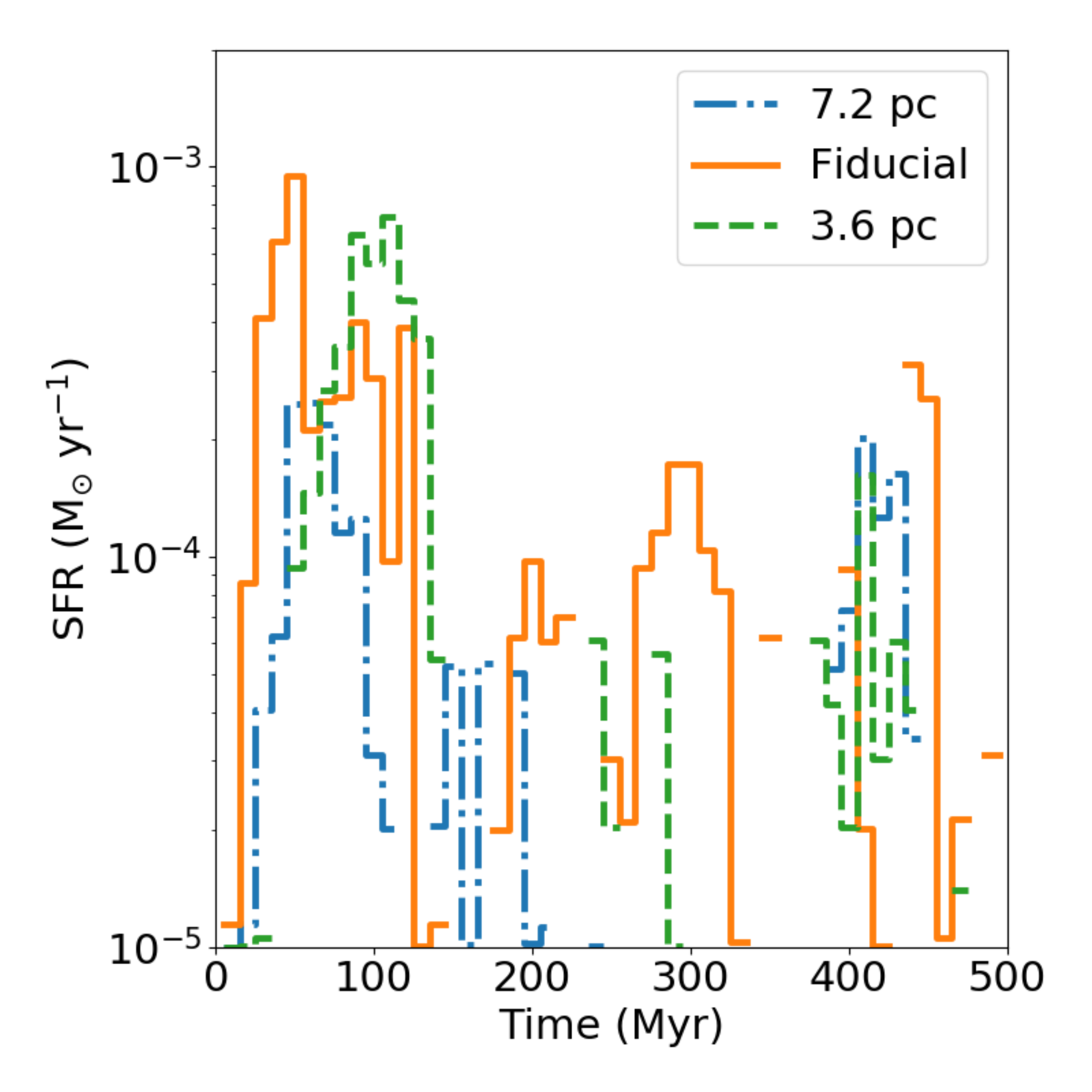}
\includegraphics[width=0.33\linewidth]{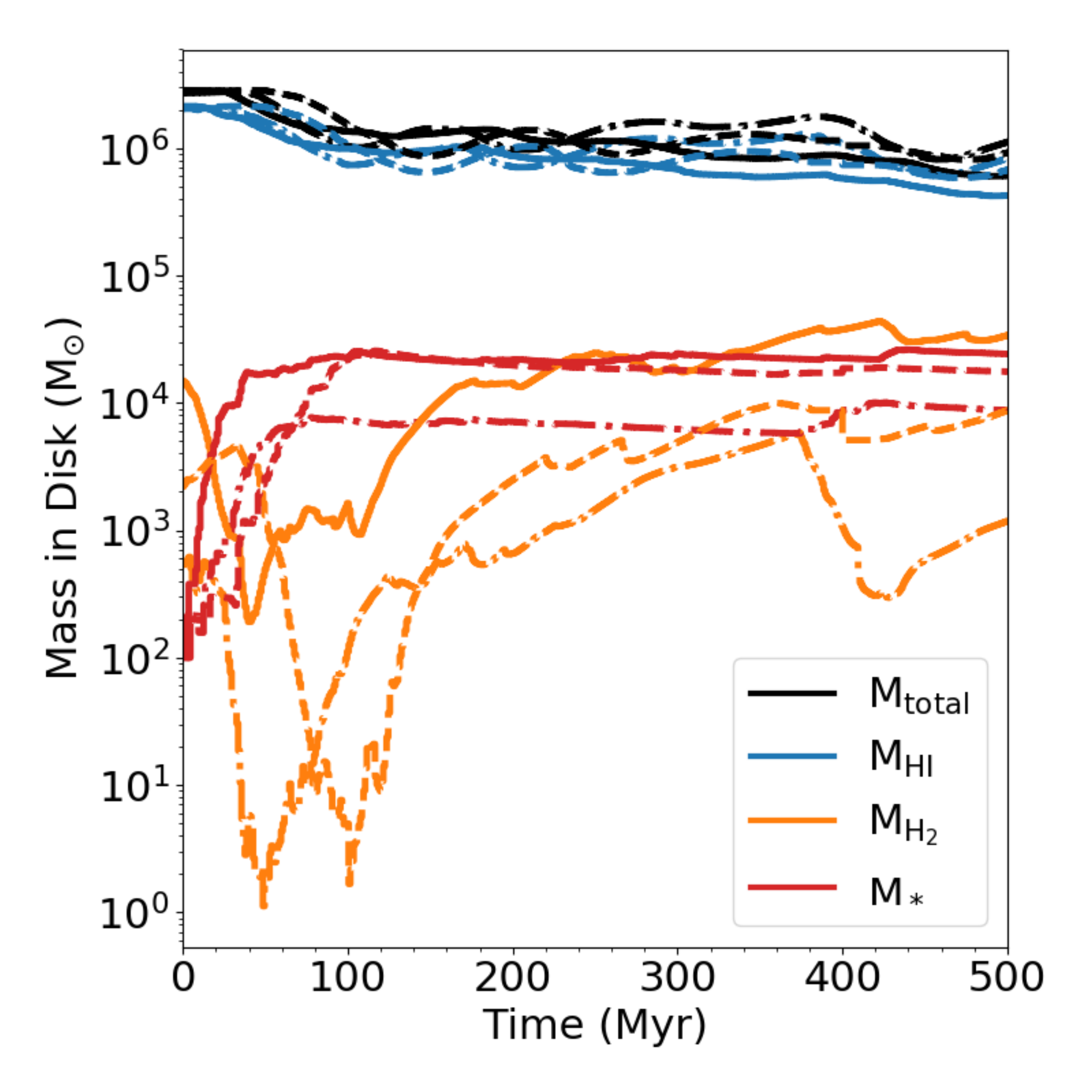}
\includegraphics[width=0.33\linewidth]{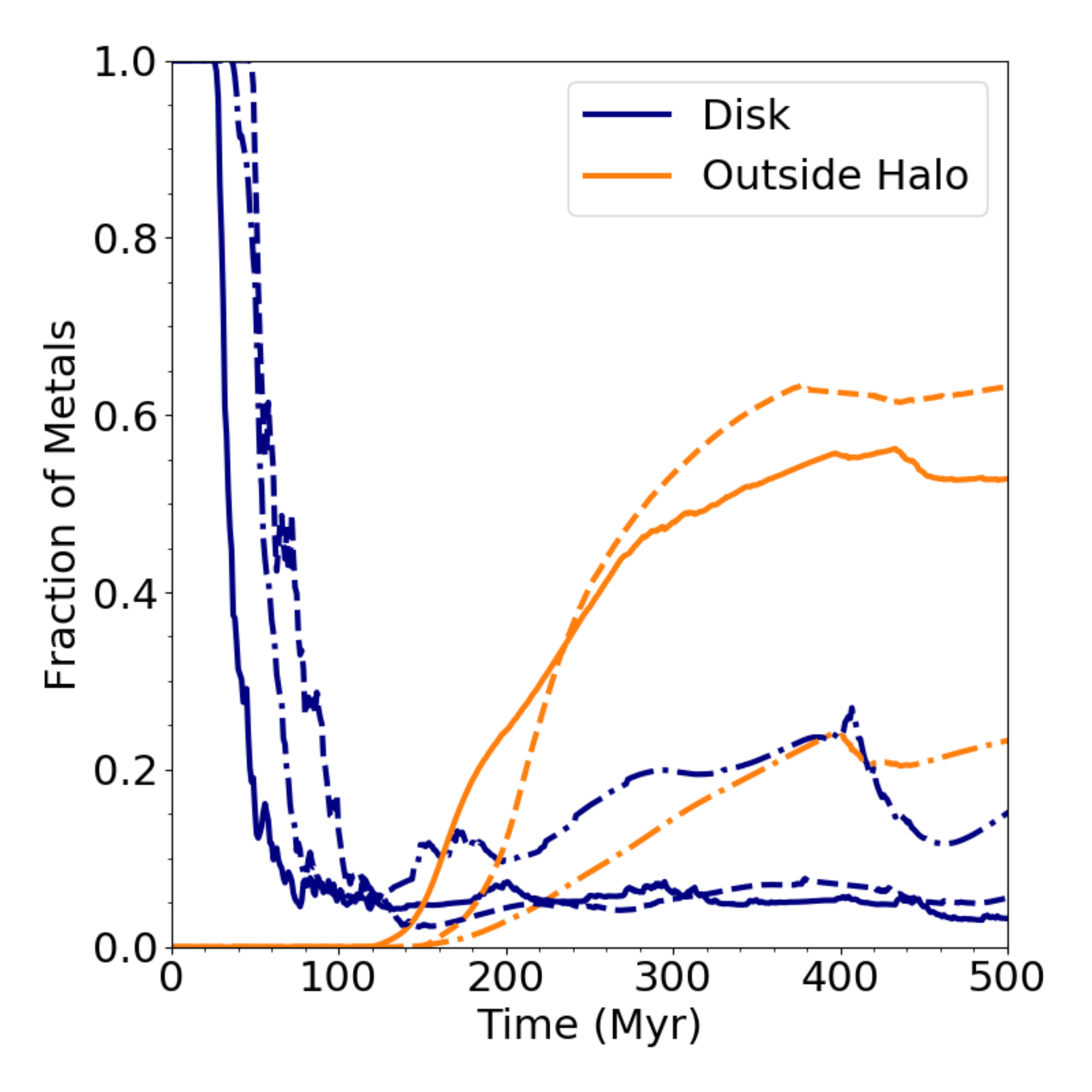}
\caption{A comparison of the SFR (left), disk gas mass (middle), and metal retention (right) evolution for our fiducial run (solid), 3.6~pc run (dotted), 3.6~pc run with a doubled initial supernova driving rate (dashed), and 7.2~pc run (dash-dotted). The line colors in the left panel are for better clarity between the SFRs.}
\label{fig:resolution_study}
\end{figure*}

We demonstrate the effect of varying resolution in the evolution of our galaxy model in Fig.~\ref{fig:resolution_study}. We compare the star formation rate, mass evolution, and metal ejection / retention fractions between our fiducial high-resolution simulation (solid lines) with the same simulation at two lower maximum resolutions, 3.6~pc (dashed lines), and 7.2~pc (dash-dotted lines). The physics in each simulation remains the same with the exception of star formation, which employs a resolution-dependent density threshold. A factor of two decrease in resolution translates to a factor of four decrease in the star formation density threshold, from 200~cm$^{-3}$ in our fiducial runs, to 50~cm$^{-3}$ and 12.5~cm$^{-3}$ in our 3.6~pc and 7.2~pc resolution runs respectively. As a result, the time at which star formation first occurs varies between the simulations, but we again define t = 0 as the time of first star formation in each simulation.

Fig.~\ref{fig:resolution_study} shows that, while decreasing the resolution certainly changes the evolution, the simulations are similar to within a factor of a few. We argue that these differences are within the stochastic variation one would expect from simulations with stochastic star formation. The star formation rate and total stellar mass is correlated with resolution, with our fiducial run having the highest SFR and stellar mass among the three runs. We note that the self-consistent SNe are still resolved in the 3.6~pc runs since these still explode at the low densities seen in Fig.~\ref{fig:SN histogram}. However, although feedback in general is still effective in the 7.2~pc run, SNe are not well resolved at this resolution. 

Generally ineffective feedback leads to increased, not decreased, star formation rates. However, the lower resolution run has the lowest SFR. The source of this difference is likely related to the ability to resolve dense, cold gas in the lower resolution simulations. In particular, the low resolution simulation clearly is unable to resolve the densities at which gas-phase H$_2$ formation becomes efficient, leading to significantly lower H$_2$ fractions (middle panel) and lower cooling rates. In addition, whatever H$_2$ does form at these lower resolutions exists at lower gas densities than in our fiducial run, and is less well protected by self-shielding effects. This is shown clearly in the significant plummet in the H$_2$ mass during the phase of first star formation that is less severe in our fiducial run. In this low metallicity regime, where H$_2$ is an important coolant, the ability to resolve the densities responsible for H$_2$ formation and self-shielding is important. 

\begin{figure*}
\centering
\includegraphics[width=0.475\linewidth]{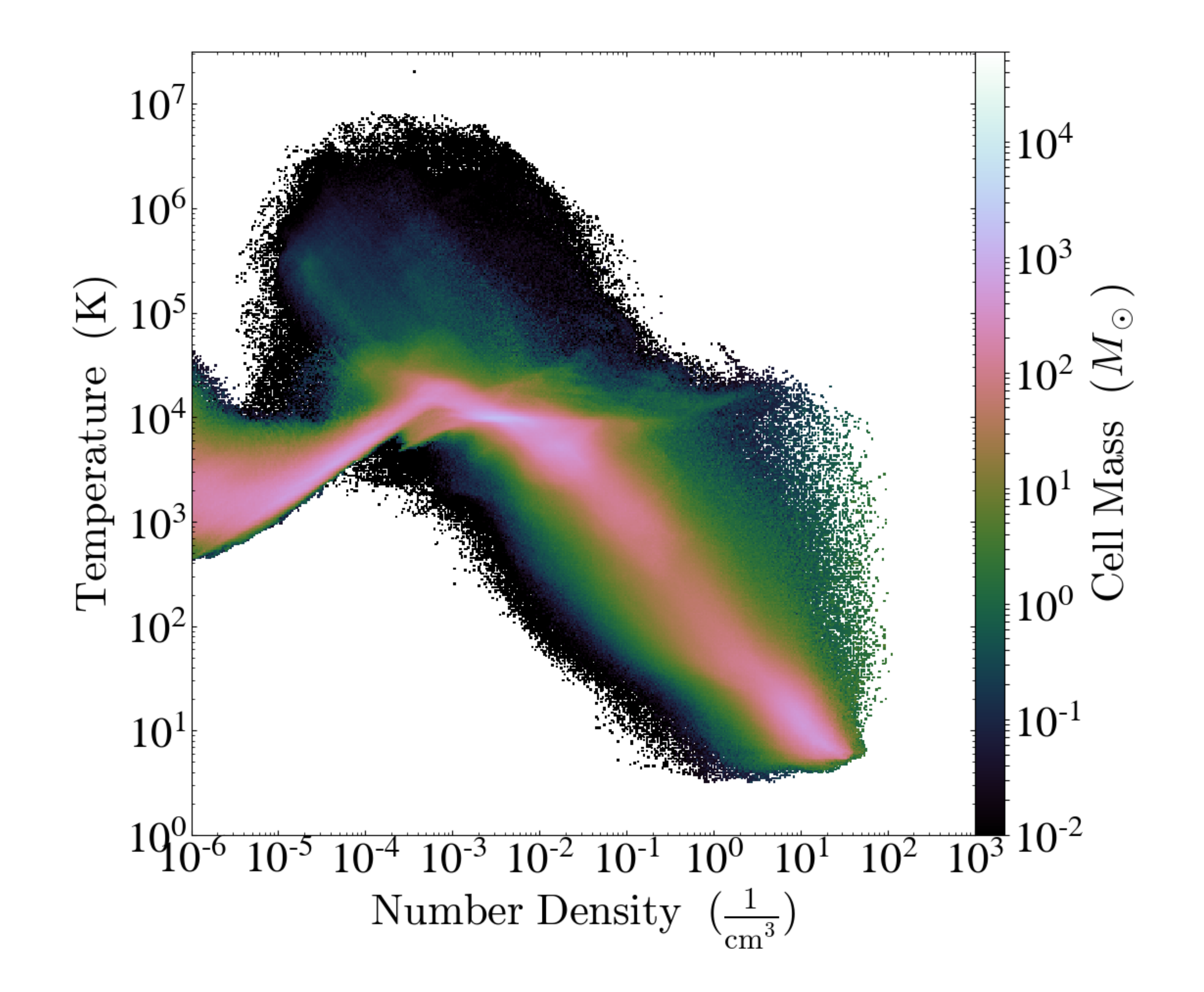}
\includegraphics[width=0.475\linewidth]{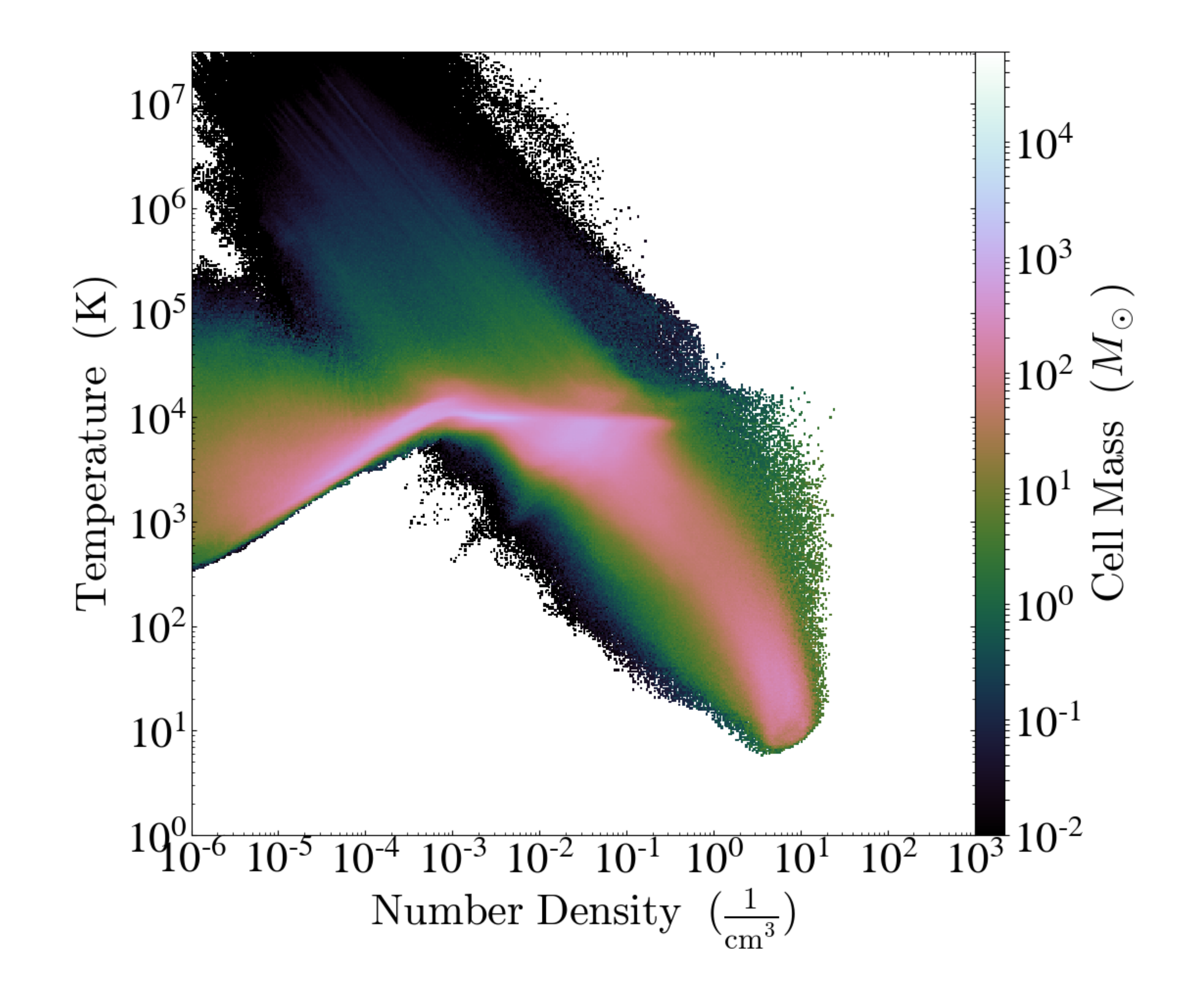}
\caption{Temperature-density phase diagrams for our 3.6~pc run (left) and 7.2~pc run (right), as presented for our fiducial run in Fig.~\ref{fig:phase}.}
\label{fig:phase_resolution}
\end{figure*}

Theses differences in densities are shown clearly by comparing Fig.~\ref{fig:phase} with Fig.~\ref{fig:phase_resolution}. Although the lowest temperature reached in each simulation is comparable, there is less gas in the coldest, densest portion of the phase diagram in each of the lower resolution simulations. In the 7.2~pc run, the gas phases are notably less distinct, with gas more evenly smoothed out in the multi-phase region between cold, dense gas and warm, ionized gas. We note the 7.2~pc simulation, which ran quickly, was conducted without the temperature ceiling used in the 3.6~pc and fiducial runs; we do not expect this to have a significant effect on the comparison.

Finally, the metal retention fraction (right panel) between the fiducial and 3.6~pc runs are quite similar, but is significantly higher for the lowest resolution simulation. It is likely, however, that this is a result of the lower SFR in the 7.2~pc run and is not directly a resolution effect. This also affects the fraction of metals ejected outside the virial radius, with less metals being ejected in the lower resolution runs.

\label{lastpage}

\end{document}